\documentclass[]{article}
\usepackage{emulateapj}
\usepackage{graphicx}
\usepackage[usenames,dvips]{color} 
\def\msun{{\rm ~M}_{\odot}}
\def\rsun{{\rm ~R}_{\odot}}
\def\myr{{\rm ~Myr}}
\def\mdot{\dot M}
\def\mpy{{\rm ~M}_{\odot} {\rm ~yr}^{-1}}
\begin{document}

\title{Compact Object Modeling with the {\tt StarTrack} Population Synthesis Code}

 \author{Krzysztof Belczynski\altaffilmark{1,2}, Vassiliki
         Kalogera\altaffilmark{3}, Frederic A.\ Rasio \altaffilmark{3}, 
         Ronald E.\ Taam\altaffilmark{3}, Andreas Zezas\altaffilmark{4}, 
         Tomasz Bulik\altaffilmark{5}, Thomas J.\ Maccarone\altaffilmark{6,7} 
         and Natalia Ivanova\altaffilmark{8}}

 \affil{
     $^{1}$ New Mexico State University, Dept of Astronomy,
        1320 Frenger Mall, Las Cruces, NM 88003\\
     $^{2}$ Tombaugh Fellow\\
     $^{3}$ Northwestern University, Dept of Physics \& Astronomy,
        2145 Sheridan Rd, Evanston, IL 60208\\
     $^{4}$Harvard--Smithsonian Center for Astrophysics,
     60 Garden St, Cambridge, MA 02138;\\
     $^{5}$ Nicolaus Copernicus Astronomical Center,
     Bartycka 18, 00-716 Warszawa, Poland;\\
     $^{6}$ Astronomical Institute Anton Pannekoek, University of Amsterdam,
     Kruislaan 403, 1098 SJ, Amsterdam, The Netherlands\\
     $^{7}$ School of Physics and Astronomy, University of Southampton,
     Southampton, Hampshire, SO17 1BJ, United Kingdom \\
     $^{8}$ Canadian Institute for Theoretical Astrophysics, University of
     Toronto, 60 St. George, Toronto, ON M5S 3H8, Canada\\
     kbelczyn@nmsu.edu, vicky, rasio, r-taam@northwestern.edu, 
     azezas@head-cfa.cfa.harvard.edu, bulik@camk.edu.pl,
     tjm@phys.soton.ac.uk, nata@cita.utoronto.ca}

 \begin{abstract} 
We present a comprehensive description of the population synthesis code 
{\tt StarTrack}. The original code has been significantly modified and 
updated. Special emphasis is placed here on processes leading to the formation and
further evolution of compact objects (white dwarfs, neutron stars,
and black holes). Both single and binary star populations are considered.   
The code now incorporates detailed calculations of all mass-transfer phases, a full 
implementation of orbital evolution due to tides, as well as the most recent 
estimates of magnetic braking. 
This updated version of {\tt StarTrack} can be used for a wide variety of problems,
with relevance to many current and planned observatories, e.g., 
studies of X-ray binaries (Chandra, XMM-Newton), 
gravitational radiation sources (LIGO, LISA), and gamma-ray burst 
progenitors (HETE-II, Swift). The code has already been used in studies of Galactic 
and extra-galactic X-ray binary populations, black holes in young star clusters, 
Type~Ia supernova progenitors, and double compact object populations.  
Here we describe in detail the input physics, we present the code calibration and
tests, and we
outline our current studies in the context of X-ray binary populations. 
 \end{abstract}


\section{Introduction}
 
The {\tt StarTrack} population synthesis code was initially developed for the 
study of double compact object mergers in the context of gamma-ray burst (GRB) 
progenitors (Belczynski, Bulik \& Rudak 2002b) and gravitational radiation 
inspiral sources (Belczynski, Kalogera \& Bulik 2002c, hereafter BKB02). 
{\tt StarTrack} has undergone major updates and revisions in the last few years. 
With this code we are able to evolve isolated (not dynamically interacting) 
single stars and binaries for a wide range of initial conditions. The input
physics incorporates our latest knowledge of processes governing stellar 
evolution, while the most uncertain aspects are parameterized to allow for 
systematic error analysis.
During the code development, special emphasis was placed on the compact 
object populations:  white dwarfs (WDs), neutron stars 
(NSs), and black holes (BHs). The input physics currently includes all major 
processes important for the formation and evolution of compact objects.
Among other things we have developed fast procedures to treat 
and diagnose various types of mass transfer episodes (including phases 
of thermal timescale and dynamically unstable mass transfer leading to 
common envelopes). We also compute tidal effects on orbital evolution, 
angular momentum losses due to magnetic braking and gravitational radiation, 
as well as mass loss from stellar winds and during mass transfer phases. 
Rejuvenation of binary components is taken into account. The full orbital 
evolution of binaries is also computed, including angular momentum and mass 
loss. Supernovae (SNe) and compact object formation are also treated in detail.

The new version of {\tt StarTrack} presented here has already been tested and 
used in  many applications. Belczynski \& Taam (2004a) studied the formation 
of ultrashort period X-ray binaries and they also demonstrated that the faint 
X-ray Galactic Center population can neither be explained by quiescent NS/BH 
transients nor by hard/faint wind-fed sources (Belczynski \& Taam 2004b). 
Belczynski, Sadowski \& Rasio (2004b) and Belczynski et al.\ (2006) developed 
a comprehensive description of young BH populations, which can also provide 
realistic initial conditions for the dynamical modeling of BHs in star clusters.    
Belczynski et al.\ (2004a) derived for the first time  a synthetic X-ray 
luminosity function which agrees with {\em Chandra\/} observations of NGC 1659, 
and Sepinsky, Kalogera, \& Belczynski (2005) explored the numbers and spatial 
distribution of X-ray binaries formed in young star clusters.
Belczynski, Bulik \& Ruiter (2005b) tested different models of Type~Ia SN
progenitors, arriving at the conclusion that the double degenerate scenario most 
easily reproduces the observed delay times between star formation and Type~Ia SNe.   
Belczynski et al.\ (2005a) used {\tt StarTrack} to study the gravitational 
radiation signal from the Galactic population of double WDs. 
Nutzman et al.\ (2004), O'Shaughnessy et al.\ (2005a,b,c), and 
Ihm, Kalogera, \& Belczynski (2005) studied binary compact object populations  
and derived merger rates and detection rates by ground-based interferometers;
they also examined BH spin magnitudes and studied the eccentricities of double 
neutron stars. {\tt StarTrack} was also incorporated into a simple stellar 
dynamics code, allowing the study of the effects of dynamical interactions on 
binary populations in dense star clusters. In that form it has been used for 
the study of binary fractions in globular clusters (Ivanova et al.\ 2005) and 
an investigation of intermediate-mass BHs in clusters and their connection to 
ultra-luminous X-ray sources (Blecha et al.\ 2005).

Among other things {\tt StarTrack} has been adapted for the study of accretion 
powered X-ray binaries (XRBs). In forthcoming papers we will present the 
synthetic populations of XRBs formed in different stellar environments. We
will start with young starburst galaxies, and move on to spiral and, eventually,
old elliptical galaxies. In the next stage it will be possible to compare the
models with rapidly improving observations of various X-ray point source 
populations. This will offer a new perspective to the study of several uncertain
aspects of binary evolution leading to the formation of XRBs. It may also 
result in an independent diagnostic of star formation rates for nearby 
galaxies, since both the numbers and properties of XRBs are directly connected 
to the star formation history (see e.g., Grimm, Gilfanov \& Sunyaev 2003;
Gilfanov 2004; Kim \& Fabbiano 2004; Belczynski et al.\ 2006, in preparation).  

In this paper we provide a detailed description of the current version of 
{\tt StarTrack}, and we present the results of a number of tests. 
We describe the implementations of single star evolution in \S\,2, 
binary orbit evolution in \S\,3, stellar wind mass loss/accretion in \S\,4, 
Roche lobe overflow calculations in \S\,5, spatial velocities in \S\,6, and the 
assumed distributions of initial parameters in \S\,7.  
In \S\,8 we discuss the validity of various input physics assumptions, and we 
compare {\tt StarTrack} calculations with detailed evolutionary models 
and with various observations. Section \S\,9 is dedicated to the discussion of
X-ray binary modeling. In \S\,10 we conclude with a short summary.

\section{Single Stellar Evolution}

In all subsequent sections we use units of $M_\odot$ for mass, $R_\odot$ for
orbital separations and stellar radii, Myr for time, $L_\odot$ for bolometric
luminosity, unless specified otherwise. We use $R$ and $M$ to denote stellar
radius and mass, while $a, e$ represent the binary orbital parameters: semi-major
axis and eccentricity, respectively. Index $i=1,2$ is used to mark the binary
components (or single stars for consistency), or to denote an accretor and a
donor in mass transfer calculations: $i={\rm acc,don}$. Roche lobe
parameters are indexed with ``lob". The initially more massive (at Zero 
Age Main Sequence) binary component is referred to as primary, while its 
companion as secondary.

\subsection{Overview}

The evolution of single stars and non-interacting binary components 
have remained mostly unchanged since the last published description of the code 
(BKB02) and therefore  we only give a brief outline here. 
However, we do point out the new additions and reiterate the modifications 
to the original formulas which were used as the base for the implementation 
of single star evolution in {\tt StarTrack}. 
 
To evolve single stars from the Zero Age Main Sequence (ZAMS) until remnant
formation (WD, NS, BH, or a remnant-less 
supernova) we employ the analytic formulas of Hurley, Pols \& Tout (2000). 
Each star is followed along an evolutionary track specific for its initial 
mass and metallicity. Various wind mass loss rates that vary with the
stellar evolutionary stage are incorporated into the code and their effect 
on stellar evolution is taken into account. Once the remnant is formed, we 
terminate the calculations but keep track of the numbers, properties and 
formation times of a given type of remnant. Additionally, for white dwarf 
remnants we take into account their subsequent luminosity evolution, and 
follow cooling tracks adopted from Hurley et al.\ (2000).  

\subsection{Stellar types}
We follow Hurley et al.\ (2000) to denote different stages of stellar
evolution with an integer $K_{\rm i}=1..n$, where\\  
0 -- Main Sequence (MS) $M \leq 0.7 \msun$ \\
1 -- MS $M > 0.7 \msun$ \\
2 -- Hertzsprung gap (HG) \\
3 -- Red Giant Branch (RG) \\
4 -- Core Helium Burning (CHeB) \\ 
5 -- Early Asymptotic Giant Branch (EAGB) \\
6 -- Thermally Pulsing AGB (TPAGB) \\
7 -- Helium Main Sequence (HeMS) \\
8 -- Helium Hertzsprung gap (HeHG) \\
9 -- Helium Giant Branch (HeGB) \\
10 -- Helium White Dwarf (He WD) \\
11 -- Carbon/Oxygen White Dwarf (CO WD) \\
12 -- Oxygen/Neon White Dwarf (ONe WD) \\
13 -- Neutron Star (NS) \\
14 -- Black Hole (BH) \\
15 -- massless remnant (after SN Ia explosion) \\
16 -- Hydrogen White Dwarf (H WD) \\
17 -- Hybrid White Dwarf (Hyb WD) \\
In addition to the star types introduced and coded by the numbers $K_{\rm i}=1...15$  
in the original Hurley et al.\ (2000) formulas, we have introduced two new
stellar types $K_{\rm i}=16,17$. $K_{\rm i}=16$ denotes a H-rich white dwarf. 
Only main sequence stars less massive than about 
$0.7 \msun$ can produce such a H-rich remnant through mass loss 
in a close binary
system. These low-mass stars do not process a significant amount of hydrogen
into helium in their cores (even in a Hubble time) and once their mass is 
stripped below the hydrogen burning limit (close to $\sim 0.08 \msun$) they become
degenerate H-rich white dwarfs. These stars, although not frequently
encountered in population synthesis, may become donors in the shortest-period 
interacting binaries.     
$K_{\rm i}=17$ denotes a hybrid white dwarf, with a carbon-oxygen-helium mixture in
the core and a helium envelope. These objects are the remnants of naked Helium main 
sequence stars ($K_{\rm i}=7$) which are stripped of mass below $0.35 \msun$ 
during  Roche lobe overflow (RLOF). At that point, thermonuclear reactions stop and 
the star becomes degenerate (eg., Savonije, de Kool \& van den Heuvel 1986).

\subsection{Modifications}
Several major changes to the original Hurley et al.\ (2000) formulas have been 
implemented  within {\tt StarTrack}.

\subsubsection{Compact object masses} 

The remnant masses of neutron stars and black holes are calculated in a 
different way than originally suggested by Hurley et al.\ (2000). In the present 
version of the code we have further 
revised our prescription presented in Belczynski et al. (2002c) to include the 
more recent calculations of FeNi core masses and allow for the possibility of 
NS formation through electron capture supernovae (ECS). White dwarfs masses are 
calculated with the original formulas of Hurley et al.\ (2000), although ONe WDs 
are formed in a slightly narrower range since we allow for ECS NS formation 
(see below). 

We determine the mass of a NS/BH remnant using information on the final CO and 
FeNi core masses, combined with the knowledge of the pre-supernova mass of the 
star. For a given initial ZAMS mass, the final CO core mass is obtained from 
the original Hurley et al. (2000) formulas, while we use the models of Timmes, 
Woosley \& Weaver (1996) to estimate final FeNi core mass.
The results of Timmes et al.\ (1996, see their Fig.2) show two distinctive FeNi 
core masses (we use models with the addition of Si shell mass), below and above 
initial masses of $M_{\rm zams} \sim 18-19 \msun$. The dichotomy arises from 
different carbon burning (convective versus radiative) the in pre-supernova 
stellar core. For higher mass progenitors ($M_{\rm zams} \gtrsim 20 \msun$) 
there is a slow rise in a FeNi core mass that we approximate with a linear relation. 
The final FeNi core mass for a given CO core mass ($M_{\rm CO}$) is obtained
from:\\ 
$M_{\rm FeNi}=$ \\
\begin{equation}
 \left\{ \begin{array}{ll}
  1.50 & M_{\rm CO} < 4.82\ (M_{\rm zams}<18)\\
  2.11 & 4.82 \leq M_{\rm CO} < 6.31 (18<M_{\rm zams}<25)\\
  0.69 M_{\rm CO} - 2.26 & 6.31 \leq M_{\rm CO} < 6.75 (25<M_{\rm zams}<30) \\ 
  0.37 M_{\rm CO} - 0.07 & M_{\rm CO} \geq 6.75\ (M_{\rm zams} \geq 30)
\end{array}
\right.
\end{equation}
where all masses are expressed in $\msun$.
The above dependence was obtained for solar metallicity models, but we use
it for all compositions considered here ($Z=0.0001-0.3$; the metallicity
range covered by Hurley et al. 2000 in fits to the stellar evolution models).  
Timmes et al. (1996) results for  metallicity 
are very similar for zero metallicity.
For example, in the mass range most important for NS formation 
the core mass changes from $\sim 1.50 \msun$ (solar metallicity) to $\sim 1.58 
\msun$ (zero metallicity). The differences can be larger for BH progenitors, 
but then the exact core mass does not play such an important role on the final 
BH mass, as most BHs form through fall back or direct collapse (see below).    

The effects of  material fallback (ejected initially in the SN explosion) during 
the star's final collapse are included. For the most massive stars 
we also allow for the possibility of a silent collapse (no supernova explosion) 
and direct BH formation. For solar metallicity and standard wind mass loss the 
compact object masses are obtained from 
 \begin{equation}
 M_{\rm rem,bar}=\left\{ \begin{array}{lr} M_{\rm FeNi}& M_{\rm CO} \leq
5\,{\rm M}_\odot\\ M_{\rm FeNi} + f_{\rm fb} (M-M_{\rm FeNi})& 5<M_{\rm
CO}<7.6\\ M& M_{\rm CO} \geq 7.6\,{\rm M}_\odot\\ \end{array}\right.
\label{Mrem}
 \end{equation}
where M is the pre-supernova mass of the star, and $f_{\rm fb}$ is the
{\em fall-back} factor, i.e. the fraction (from 0 to 1) of the stellar envelope 
that falls back. The value of $f_{\rm fb}$ is interpolated linearly 
between $M_{\rm CO}=5 \msun$ ($f_{\rm fb}=0$) and $M_{\rm CO}=7.6 \msun$
($f_{\rm fb}=1$). The regimes of no fall-back ($M_{\rm CO} \leq 5 \msun$),
partial fall-back ($5<M_{\rm CO}<7.6 \msun$) and direct collapse ($M_{\rm CO}
\geq 7.6 \msun$) are estimated from core collapse models of Fryer, Woosley
\& Hartmann (1999) and the analysis of Fryer \& Kalogera (2001). 

We also allow for NS formation through ECS (e.g., Podsiadlowski et al. 2004). 
Following Hurley et al. (2000) we 
use the He core mass at the AGB base to set the limits for the formation of various CO cores. 
If the He core mass is smaller than $M_{\rm cbur1}$ the star forms a  
degenerate CO core, and ends up forming a CO WD. If the core is more massive
than $M_{\rm cbur2}=2.25 \msun$ the star forms a 
non-degenerate CO core with subsequent burning of elements 
until the formation of a FeNi core which ultimately collapses to a NS or a BH. Stars with cores between $M_{\rm cbur1}$ and $M_{\rm cbur2}$ may 
form partially degenerate CO cores. If such a core reaches a critical mass 
($M_{co,crit}=1.08 \msun$, Hurley et al. 2000), it ignites CO off-center and  
non-explosively burns CO into ONe, forming a degenerate ONe core. 
If in subsequent evolution 
the ONe core increases its mass to $M_{\rm ecs}=1.38 \msun$ the core collapses due to electron capture on Mg, and forms NS (we will refer to a NS formed in this  
way as ECS 
NS, as opposed to a regular FeNi core collapse compact object formation). The ECS NSs are assumed to 
have unique masses of $M_{\rm rem,bar}=M_{\rm ecs}$. If the ONe core mass remains 
below $M_{\rm ecs}$ the star forms a ONe WD. 

Hurley et al. (2000) suggested $M_{\rm cbur1}=1.66 \msun$ corresponding to 
$M_{\rm zams}=6.5 \msun$ for $Z=0.02$. Later calculations with the same, 
but updated evolutionary code (Eldridge \& Tout 2004a,b) indicated that ECS may occur 
for higher initial masses ($M_{\rm zams} \gtrsim 7.5 \msun$). For our standard model, we adopt 
$M_{\rm cbur1}=1.83 \msun$ ($M_{\rm zams}=7.0 \msun$), and this results in ECS NS formation above $M_{\rm zams}=7.6 \msun$ (for masses $M_{\rm zams}=7.0-7.6 \msun$ the ONe core does not reach $M_{\rm ecs}$ and a ONe WD is formed). 
It is noted that binary evolution through RLOF, may either decrease the initial the mass of the ZAMS star required to form the core of mass $M_{\rm cbur1}$ (due to rejuvenation) or increase it (due to mass loss). Therefore, binary evolution effectively leads to wider initial progenitor mass for ECS NS formation. Also metallicity, and wind mass loss influences the ECS NS formation range.

\begin{figure}
\includegraphics[width=1.0\columnwidth]{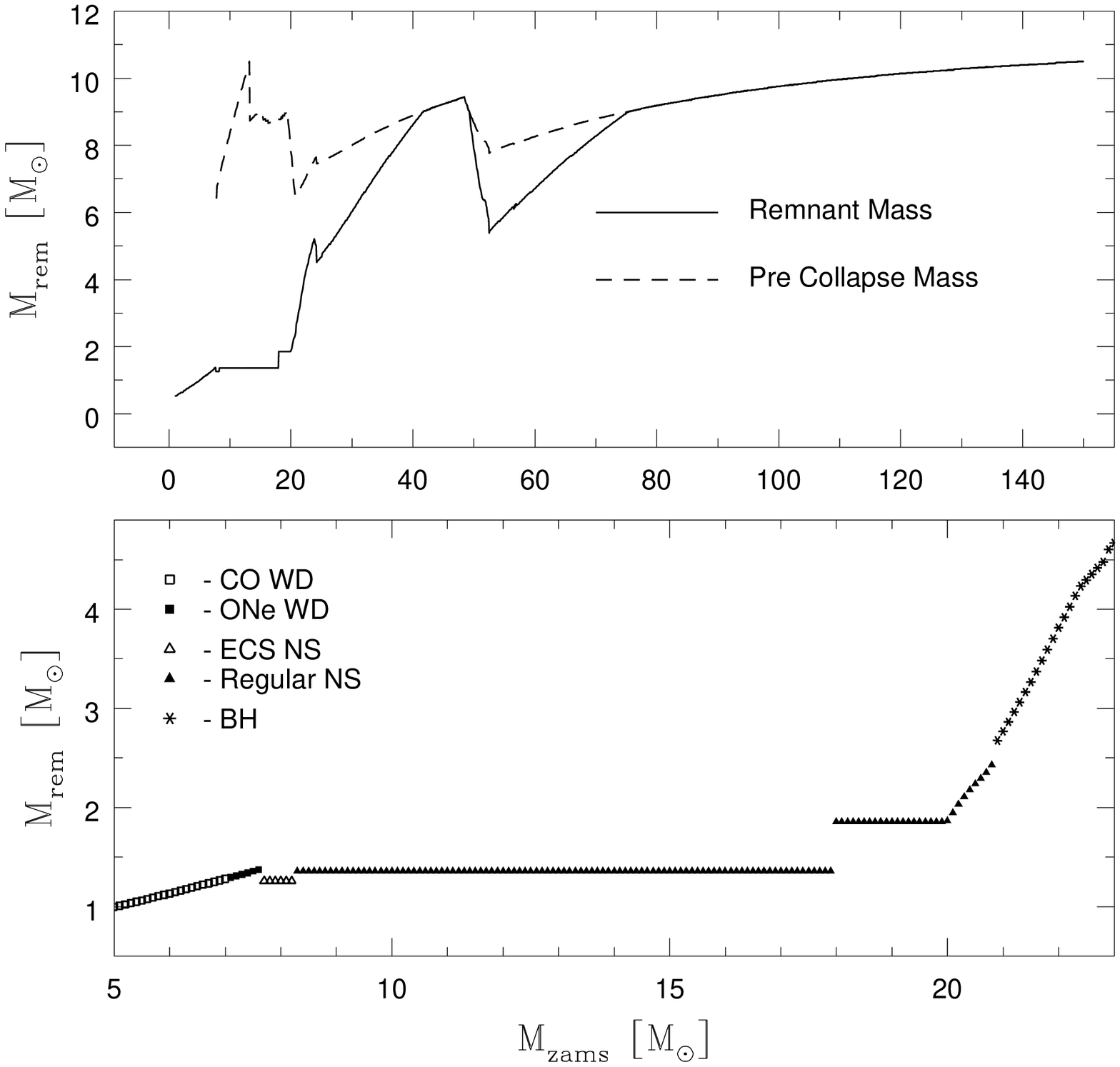}
\caption{Final compact object masses in function of initial mass for 
single star evolution (with solar metallicity and standard winds). 
Top panel shows the full mass range, and indicates pre-collapse mass 
of the progenitor star. Bottom panel shows the mass range important 
for NS formation, with different types of remnants marked on the plot. 
For discussion see \S\,2.3.1.
}
\label{Mfin}
\end{figure}

The remnant masses calculated above express the mass of baryons, and for NS/BH 
remnants we convert them to gravitational masses ($M_{\rm rem}$). We use the quadratic relation 
\begin{equation}
M_{\rm rem,bar} - M_{\rm rem} = 0.075\ M_{\rm rem}^2
\end{equation}
for neutron stars (Lattimer \& Yahil 1989; see also Timmes et al. 1996), while 
for black holes we simply approximate the gravitational mass with 
\begin{equation}
 M_{\rm rem} = 0.9\ M_{\rm rem,bar}
\end{equation}
The resulting remnant mass spectrum covers a wide range of masses and is 
presented in Figure~\ref{Mfin}.

In summary, Helium WDs form from progenitors of initial masses in range $M_{\rm zams} \lesssim 0.8 \msun$, CO WDs from $M_{\rm zams}=0.8-7 \msun$, and ONe WD are formed in range $M_{\rm zams}=7-7.6 \msun$. 
Neutron stars formed through ECS ($M_{\rm zams}=7.6-8.3 \msun$) have mass of 
$M_{\rm rem} = 1.26 \msun$. Regular core-collapse (of FeNi cores) NS are formed 
with mass $M_{\rm rem} = 1.36 \msun$ ($M_{\rm zams}=8.3-18 \msun$); 
$M_{\rm rem} = 1.86 \msun$ ($M_{\rm zams}=18-20 \msun$) and for higher initial 
progenitor masses NSs form up to adopted maximum NS mass: $M_{\rm NS,max}$. 
In our standard model we adopt $M_{\rm NS,max}=2.5 \msun$, and above  
$M_{\rm zams} \gtrsim 21 \msun$ BHs are are formed.  
It is found that single stars may form BHs up to $\sim 11 \msun$ for solar 
metallicity ($Z=0.02$) and $\sim 30 \msun$ for lower metallicities 
($Z=0.001-0.0001$), which is consistent with the current observations of the most massive BHs in Galactic X-ray transients. 
The initial-final mass relation described above and presented in
Figure~\ref{Mfin} holds only for single stellar evolution and for a specific
metallicity ($Z=0.02$). Effects of binary evolution, in particular mass
loss/gain in RLOF, may alter the initial-final mass relation in two ways. 
First, a compact object may have a different mass; higher if a progenitor
(or a compact object itself) accreted mass or smaller if a progenitor
lost mass in RLOF. 
Second, the initial mass limits for formation of a given type of compact 
object are not sharp, since with binary mass loss/gain stars
of various initial masses may form compact objects of a given mass and type.  
These limits are blurred by binary evolution.

The mass estimates of neutron stars in relativistic double neutron star
binaries point to a NS formation mass of $\sim 1.35 \msun$ (Thorsett \& Chakrabarty 
1999). 
It is also 
suggestive that the NS mass in Vela X-1 is high $\sim 1.9 \msun$. Since  
Vela X-1 is a high-mass X-ray binary, system with wind-fed accretion 
(small mass capture efficiency) and a 
massive star donor (short lifetime), the NS probably has not accumulated 
much mass, and the measured mass is close to its formation mass. Our adopted 
model for NS formation masses (Timmes et al. 1996) falls in qualitative agreement 
with these observations.  
Finally we allow compact object masses to increase through accretion in binary systems. 
Accretion and mass accumulation onto WDs is described in detail in \S\,5.7.
For NS we need to adopt a maximum NS mass, over which NS collapses to BH. Such a 
collapse may lead to a short-hard Gamma-ray burst event. Depending on the preferred 
equation of state the maximum NS mass may vary in a wide range ($\sim 2-3 \msun$), 
and in particular may reach $\sim 3 \msun$ if rotation is included (Morrison, Baumgarte
\& Shapiro 2004). 
At the moment 
the highest measured NS mass is $2.1 \pm 0.2 \msun$ for a millisecond pulsar in 
PSR J0751+1807; a relativistic binary with helium white dwarf secondary (Nice 
et. al. 2005). As stated above we adopt $M_{\rm NS,max}=2.5 \msun$ for our 
standard model, but we relax this assumption in parameter studies.

\subsubsection{Wind mass loss} 
The compilation of stellar wind mass loss rates presented in Hurley et al.\
(2000) has been expanded to include mass loss from low- and intermediate-mass 
main sequence stars. We have adopted the formulas of Nieuwenhuijzen \& de 
Jager (1990) to calculate the mass loss rates for main sequence stars below
$\sim 8 \msun$. Although the mass loss from these stars is not large enough
to significantly alter the evolution of a mass--losing star, it may
play an important role in the formation and evolution of wind-accreting close 
binaries. Even with small mass transfer rates characteristic for the low- and 
intermediate-mass main sequence stars, the X-ray luminosities for accreting 
BHs and NSs are high enough to be detected in deep Chandra exposures. 
A number of faint point X-ray sources were discovered in the Galactic center 
with deep exposures (Wang, Gotthelf \& Lang 2002; Muno et al. 2003), some of which may 
be explained in terms of wind-fed close binaries (Pfahl, Rappaport \&
Podsiadlowski 2002a; Bleach 2002; Willems \& Kolb 2003; Belczynski \&
Taam 2004b).  

\subsubsection{Rotational velocities}
A compilation of updated observational data on rotational velocities is used 
to initiate the stellar spins on the ZAMS. The spin evolution is followed as
detailed here for single stars and in \S\,3 for binary components. 
 In order to obtain a functional form of the relation of the
equatorial rotational velocity and stellar mass, we used the
compilation  of rotational velocities of Stauffer \& Hartmann (1986) for stars in open clusters.
 The difference between cluster and field stars is quite small for
massive stars (with a maximum difference of $\sim 10\%$ for
intermediate B-type stars), but can be as high as 40\% for stars later
than F-type,  with  field stars having systematically lower rotational
velocities.

\begin{figure}
\includegraphics[width=0.6\columnwidth]{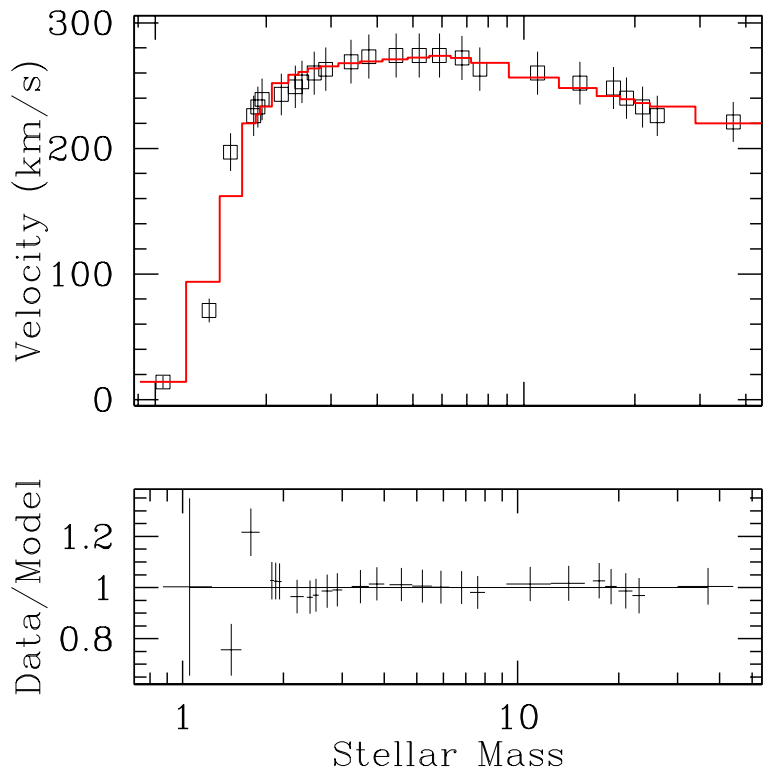}
\caption{
Initial rotational velocities of stars used in {\tt StarTrack} calculations. 
In the top panel we present the fit to the observational data from 
Stauffer \& Hartmann (1986). In the bottom panel we show the ratio
of the data and the model.
}
\label{fig03}
\end{figure}

The mean rotational velocity $\rm{\overline{v}_{rot}}$ was determined
from the projected velocity ($\rm{v_{rot}\,sin\,i}$) assuming a random
distribution of angles with $\rm{sin\,i=\pi/4}$. We fitted
$\rm{\overline{v}_{rot}}$ as a function of stellar mass, and we
obtained the following empirical functional form
                                                                                                       
\begin{equation}
 \rm{\overline{v}_{rot}} = \left\{ \begin{array}{ll}
  \frac{10.0 \, M_{\rm i} ^{-\alpha_{1}}}{c + M_{\rm i} ^{-\beta_1}} & 
   if ~ M_{\rm i} > M_{o} \\
  \frac{13.32 \, M_{\rm i} ^{-\alpha_{2}}}{c + M_{\rm i} ^{-\beta_1}} & 
   if ~ M_{\rm i}  \leq M_{o}
\end{array}
\right.
\end{equation}
where, $\alpha_{1} = -0.035^{+0.06}_{-0.31}$,   $\alpha_{2} =
0.12^{+0.09}_{-0.04}$,    $\beta_1 = 7.95^{+0.33}_{-0.31}$  and $M_{o} =
6.35^{+6.5}_{-2.1}$ (errors are at the $1~\sigma$ level).
We stress that this is only an empirical functional form of the 
equatorial rotational velocity as a function of stellar mass.
In Fig~\ref{fig03} we present the observational data
from Stauffer \& Hartmann (1986), together with the best fit
function. In the bottom panel of this figure we also show the ratio
of the Stauffer \& Hartmann (1986) data and the model.

The spin angular momentum of a star may be expressed as
\begin{equation} 
J_{\rm i,spin} = I_{\rm i} \omega_{\rm i} = k_{\rm i} M_{\rm i} R_{\rm i}^2 
\omega_{\rm i}
\label{sin01}
\end{equation}
where, $\omega_{\rm i}= \rm{\overline{v}_{rot}}/R_{\rm i}$ is the angular rotational 
velocity and the coefficient $k_{\rm i}$ varies as the star evolves and its internal
structure changes (e.g., it is 2/5 for a solid sphere and 2/3 for a
spherical shell). Following Hurley et al.\ (2000) we consider two structural 
components for each star: a core and an envelope. The spins of these two components 
may decouple in the course of evolution, although we keep them coupled in our standard 
model calculations. The spin angular momentum of a star is then 
\begin{equation}
J_{\rm i,spin} = [k_{\rm i,env} (M_{\rm i}-M_{\rm i,c}) R_{\rm i}^2 + k_{\rm
i,core} M_{\rm i,c} R_{\rm i,c}^2] \omega_{\rm i}
\label{sin02}
\end{equation}

We use different values than Hurley at al.\ (2000) for the internal structure
coefficient $k_{\rm i}$.
For stars with no clear core-envelope structure ($K_{\rm i}=0,1,7,10,11,12,13,
14,16,17$) we use simple polytropic models (e.g., Lai, Rasio \& Shapiro 1993) with 
$n=1.5$ and $n=3$ for low-mass and high-mass objects, respectively, giving
\begin{equation}
k_{\rm i,env} = \left\{
               \begin{array}{ll}
                 0.205 & M_{\rm i} < 1 \msun \\
                 0.075 & M_{\rm i} \geq 1 \msun
               \end{array}
             \right.
k_{\rm i,core}=0
\label{ppp1}
\end{equation}
For giants with a clear separation between core and envelope 
($K_{\rm i}=3,4,5,6,9$) and additionally for stars in the Hertzsprung gap 
($K_{\rm i}=2,8$) we use detailed models of giant envelopes (Hurley et al.\ 
2000) and for the core we apply a polytropic model with $n=1.5$ to obtain
\begin{equation}
k_{\rm i,env}  = 0.1,
\ \ \ \ k_{\rm i,core} = 0.205
\label{ppp2}
\end{equation}

Conservation of the spin angular momentum of a star is used then to
determine its rotational velocity. Additional angular  
momentum losses from magnetic
braking (see \S\,3.2) are also taken into account.

\subsubsection{Convective/Radiative envelopes}  
Stars with convective and radiative envelopes respond differently to various
physical processes (e.g., magnetic braking, tidal interactions or mass loss).  
Stars that have a significant convective envelope are: 
low-mass H-rich MS stars ($K_{\rm i}=0,1$) within the mass range of
$0.35 \msun - M_{\rm ms,conv}$, where $M_{\rm ms,conv}$ is the maximum mass
for a MS star to develop a convective envelope; giant-like stars ($K_{\rm i}=
3,5,6$) independent of their mass and evolved low-mass Helium stars 
($K_{\rm i}=9$) below $M_{\rm he,conv}=3.0 \msun$.  
For stars crossing H-rich Hertzsprung gap ($K=2$) and core helium burning stars
($K_{\rm i}=4$ we obtained detailed models using the code described in Ivanova 
\& Taam (2004) to check how far from Hayashi line stars cross the border
between radiative and convective envelopes; stars cooler than $\log(T_{\rm eff}) 
= 3.73 \pm 0.02$ have convective envelopes, while hotter stars have radiative 
envelopes. We have also examined the models presented by Schaller et al.
(1992) for both solar metallicity and $Z=0.001$ and have found that the above 
temperature cut works rather well for these two classes of stars for both 
metallicities.
MS stars with masses below $\sim 0.35$ are fully convective. The value of 
$M_{\rm ms,conv}$) depends strongly on metallicity 
\begin{equation}
M_{\rm ms,conv} = \left\{
               \begin{array}{ll}
                 1.25                    & Z \geq 0.02 \\
                 -1532 Z^2 + 55.73 Z + 0.747 & 0.001 < Z < 0.02 \\
                 0.8                     & Z \leq 0.001 \\
               \end{array}
             \right.
\label{MScon}
\end{equation}
Values of $M_{\rm ms,conv}$ in metallicity range $Z=0.001-0.02$ are obtained 
from a fit to detailed evolutionary calculations (Ivanova 2006). 
All other stars (e.g., $K=8$ -- Helium stars in the Hertzsprung gap) are assumed 
to have radiative envelopes. We use the original Hurley et al.\ (2002) formulas 
to calculate the mass and depth of convective envelopes.

\subsubsection{Helium star evolution}
We assume that low-mass evolved Helium stars ($K_{\rm i}=9$) below 
$M_{\rm he,conv}=3.0 \msun$ (as opposed to $2.2 \msun$ in Hurley et al.\ 2000) 
expand and form deep convective envelopes in their late stages of evolution 
(e.g., Ivanova et al. 2003; Dewi \& Pols 2003). Helium stars with convective
envelopes are subject to strong tidal interactions (convective tides as 
opposed to radiative damping, see \S\,3.3), and if found in an interacting binary, 
they may alter significantly the 
fate of a given system. All helium stars ($K_{\rm i}=7,8,9$) may be subject 
to stable RLOF. However, in dynamically unstable cases we assume a binary
component merger in the case of a HeMS donor ($K_{\rm i}=7$) and a Helium Hertzsprung 
gap donor ($K_{\rm i}=8$; e.g., Ivanova et al. 2003) or we follow a given 
system through a CE phase for evolved He star donors ($K_{\rm i}=9$ 
and test whether the system survives or merges. The examination of RLOF stability 
and development of dynamical instability are described in detail in \S\,5).

The treatment of helium stars is important, for example, in later stages
of evolution leading to double neutron star formation. The immediate 
consequences, leading to the formation of a new class of close double neutron 
stars, were discussed in Belczynski \& Kalogera (2001) and Belczynski, Bulik \& 
Kalogera (2002a). 
The formation of the new class of (ultracompact) double neutron stars involves 
Helium Hertzsprung gap stars initiating CE phase. Now, guided by the better
understanding of CE phase survival (Ivanova et al.\ 2003), we do not allow
survival in such cases, and in our reference model both double neutron star 
merger rates and double neutron star properties have changed (Belczynski et
al. 2007). However, we note that it is still predicted that a significant
fraction ($\sim 40\%$) of close double neutron star binaries form in  
very close orbits (merger times smaller than 100 Myr).
Due to significant updates of the code and new observational results on short 
GRBs with double neutron stars suggested as their progenitors (e.g., Fox et al. 
2005) new {\tt StarTrack} calculations relevant to the double neutron star 
formation have been performed (Belczynski et al. 2007; Belczynski et al., in
prep.).

\section{Binary Orbital Evolution}

Throughout the course of binary evolution we track the changes in orbital 
properties. A number of physical 
processes may be responsible for these changes. 
In the general case of eccentric orbits we numerically integrate a set of four
differential equations describing the evolution of orbital separation, eccentricity
and component spins, which depend on tidal interactions as well as 
angular momentum losses associated with magnetic braking, gravitational
radiation and stellar wind mass losses. 
For circular orbits with synchronized components, we can obtain an exact 
solution for the change of orbital separation using conservation of angular
momentum. Losses of angular momentum and/or mass associated with RLOF 
events, magnetic braking and gravitational radiation are taken into 
account. 

We assume that any system entering RLOF becomes circularized and
synchronized (if it had not already reached this equilibrium state before RLOF). 
In such a case we circularize to the periastron distance 
\begin{equation}
  \begin{array}{l}
    a_{\rm fin}=a_{\rm int} (1-e) \\ 
    e_{\rm fin}=0
  \end{array}
\label{eccrlof}
\end{equation}
where and $int,\ fin$ mark initial and final values), and both components are 
synchronized to the new mean angular orbital velocity.   
Non circular orbits and non synchronous RLOF cases are important for
massive binaries, in which massive stars may remain eccentric/unsynchronized 
as tidal interactions are not as effective as for low mass stars (see \S\,3.3).  
In particular, the vast majority of close (coalescing) NS-NS binary progenitors 
may evolve through such configuration more than once.  It should be  noted that our 
prescription (eq.~\ref{eccrlof}) does not conserve total angular momentum (i.e., 
sum of orbital and spin angular momenta of the components). 

For most of the cases in double compact object progenitor evolution, our 
prescription leads to a moderate ($\sim 20\%$) loss of angular momentum. 
For systems which have not been circularized and synchronized before
entering RLOF there might be substantial mass loss (e.g., Hut \& Paczynski
1984), and although this is not taken into account in our calculations, it
may also lead to some angular momentum loss.

For some systems, consisting of low mass object (e.g., a neutron star) and a
massive star (e.g., a massive CHeB star) eccentricity is induced by
the supernova explosion and once the massive star overfills its Roche lobe we
circularize/synchronize the system, and this may lead to a slight increase 
of total angular momentum ($\sim 10\%$). However, it needs to be stressed
that there is no solution that conserves total angular momentum in such
cases, as systems evolve toward a Darwinian unstable state (see \S\,3.3) and eventually evolve towards a common
envelope phase. Since the orbital shrinkage of a system with a massive star 
donor during the common envelope phase is usually dramatic ($\sim 2$ orders of magnitude), 
the moderate change of orbital separation prior CE phase (eq.~\ref{eccrlof}) and
its specific magnitude (in the prescription used) does not play an important role. 

Violent processes like SN explosions or common
envelope phases are taken into account in binary orbital evolution. 
Also nuclear evolution of components (expansion/contraction affecting 
stellar spins) is considered. In what follows we describe the elements 
used to calculate the orbital evolution.
The orbital angular momentum of the binary and its mean angular velocity are 
expressed as 

\begin{equation}
J_{\rm orb} = {M_1 M_2 \sqrt{a G (M_1+M_2)} \over M_1+M_2} \sqrt{1-e^2}
\label{jorb}   
\end{equation}

\begin{equation}
w_{\rm worb} = \sqrt{G (M_1+M_2)} a^{-1.5}
\end{equation}
where $G$ is the gravitational constant.

\subsection{Gravitational radiation} 
Binary angular momentum loss due to gravitational radiation is estimated
for any type of binary following Peters (1964)

\begin{equation}
\,d J_{\rm gr}/\,d t = - {32 \over 5} {G^{7 \over 2} {M_1}^2 {M_2}^2
\sqrt{M_1+M_2} \over c^5 a^{7 \over 2} (1-e^2)^2} (1+{7
\over 8} e^2)
\label{jgr}
\end{equation}
where, $c$ is the speed of light. Emission of gravitational radiation causes
orbital decay as well as circularization, both taken into account
during the evolution of a binary system. For any given system, the merger time may be easily estimated (e.g., see eq.14 in BKB02).

\subsection{Magnetic Braking}

Each binary component may be subject to magnetic braking, causing the
decrease of the component's rotation. 
In the case of a detached binary configuration magnetic braking is applied 
directly to the component spins, while during RLOF the effects of magnetic 
braking are applied to the orbit, since the components are then kept in 
synchronism. 
Three different prescriptions for magnetic braking are incorporated within
the {\tt StarTrack} code and may be used interchangeably for parameter
studies. In what follows we provide a detailed description of the specific 
braking laws adopted.

Magnetic braking is applied to stars with a significant convective envelope, 
i.e., for low-mass H-rich MS stars, H-rich giant-like stars and cool HG and CHeB 
stars (see \S\,2.3.4 for details) with the exception of 
low-mass evolved Helium stars for which there is not much known about magnetic
fields. For fully convective MS stars ($K_{\rm i}=0$, $M<0.35 \msun$) 
magnetic braking may also operate, although it has been hypothesized that 
the braking is suppressed (Rappaport, Verbunt \& Joss 1983; Zangrilli, Tout 
\& Bianchini 1997) in order to provide an explanation of the observed period 
gap for cataclysmic variables. Therefore we assume that magnetic braking is 
not operative for fully convective stars, independent of the prescription 
used. Since massive core helium burning stars, more massive H-rich MS 
stars, and He-rich MS stars have radiative envelopes, we assume that magnetic 
braking does not operate in these stars.  The prescription for the loss of 
angular momentum associated with magnetic braking $\,d J_{\rm i,mb}/\,d t$ 
takes several forms. Historically, most studies have adopted the form 
suggested by Rappaport et al.\ (1983) where 
\begin{equation}
\,d J_{\rm i,mb}/\,d t = - 5.8 \times 10^{-22} {M_i} {R_i}^{\gamma} {\omega_i}^3 
\label{jmb1}   
\end{equation}
with parameter $\gamma=2$ in our model calculations.  However, studies 
based on the observations of rapidly rotating stars show that the Skumanich 
relation ($\dot J \propto \omega^3$) is inadequate in this regime and point to 
a weakening of magnetic braking due to saturation of the dynamo (Andronov, 
Pinsonneault \& Sills 2003).  In this case, the angular momentum loss rate 
takes the form 
\begin{equation}
\,d J_{\rm i,mb}/\,d t = -8.88 \times 10^{-22} \sqrt{R_i/M_i} \left\{ \begin{array}{ll}
   {\omega_i}^3 & {\omega_i} \leq \omega_{\rm crit} \\
   {\omega_i} \omega_{\rm crit}^2 & {\omega_i} > \omega_{\rm crit} \\
\end{array} \right.
\label{jmb2}
\end{equation}
where, $i$ denotes the component for which magnetic braking is operating, 
$\omega_{\rm i}$\ [$\myr^{-1}$] is angular velocity, and  $\omega_{\rm crit}$ stands 
for a critical value of angular velocity above which the angular momentum loss 
rate enters the saturated regime. If the latter law is used, the saturation is 
applied only for MS stars and $\omega_{\rm crit}$ is interpolated from Table~1 
of Andronov et al.\ (2003). 

In addition, we also include the form of magnetic braking from the results of a study 
by Ivanova \& Taam (2003).  In this latter study, 
an intermediate form of the angular momentum loss rate was derived ($\dot J \propto 
\omega^{1.3}$) based on a two component coronal model as applied to the observational 
data relating stellar activity to stellar rotation.  Specifically, we adopt 

\begin{equation}
\,d J_{\rm i,mb}/\,d t = -619.2 {R_i}^4 \left\{ \begin{array}{ll}
   (\omega_i/9.45 \times 10^7)^3 & {\omega_i} \leq \omega_{\rm x} \\
   10^{1.7} (\omega_i/9.45 \times 10^7)^{1.3} & {\omega_i} > \omega_{\rm x} \\
\end{array} \right.
\label{jmb3}
\end{equation}
with $w_{\rm x}=9.45 \times 10^8 \myr^{-1}$. This law is used for the 
{\tt StarTrack} standard model calculations.

\subsection{Tidal Evolution}

The evolution of the orbital parameters ($a,e$) as well as component  
spins ($\omega_{\rm i}$, $i=1,\,2$) driven by tidal interactions of
binary components is computed in the standard equilibrium-tide, weak-
friction approximation (Zahn 1977, 1989), following the formalism of 
Hut (1981)\footnote{Note that upon entering RLOF any binary 
system is instantly synchronized and circularized.}. This formalism 
allows us to treat binaries with arbitrarily large eccentricities. 
We assume that the only sources of
dissipation are eddy viscosity in convective envelopes and radiative
damping in radiative envelopes. Specifically, 
we integrate numerically the following differential equations in  
parallel with the stellar evolution

\noindent
\begin{eqnarray}
\left( {\,d a \over \,d t} \right)_{\rm tid} = & - 6 F_{\rm tid}
\left( {k \over T} \right)_{\rm i}
q_{\rm i} (1+q_{\rm i}) \left( {R_{\rm i} \over a} \right)^8 {a \over  
(1-e^2)^{15/2}} \nonumber \\
& \times \left( f_1(e^2) - (1-e^2)^{3/2} f_2(e^2) {\omega_{\rm i} \over
\omega_{\rm orb}} \right)
\label{tid1}
\end{eqnarray}

\noindent
\begin{eqnarray}
\left( {\,d e \over \,d t} \right)_{\rm tid} = & - 27 F_{\rm tid}
\left( {k \over T} \right)_{\rm i}
q_{\rm i} (1+q_{\rm i}) \left( {R_{\rm i} \over a} \right)^8 {e \over  
(1-e^2)^{13/2}} \nonumber \\
& \times \left( f_3(e^2) - {11 \over 18} (1-e^2)^{3/2} f_4(e^2)  
{\omega_{\rm i}
\over \omega_{\rm orb}} \right)
\label{tid2}
\end{eqnarray}

\noindent
\begin{eqnarray}
\left( {\,d \omega_{\rm i} \over \,d t} \right)_{\rm tid} = & 3 F_
{\rm tid}
\left( {k \over T} \right)_{\rm i}
{q_{\rm i}^2 \over r_{\rm i,gyr}^2} \left( {R_{\rm i} \over a} \right)
^6 {\omega_{\rm orb} \over
(1-e^2)^6} \nonumber \\
& \times \left( f_2(e^2) - (1-e^2)^{3/2} f_5(e^2) {\omega_{\rm i}  
\over \omega_{\rm orb}}
\right)
\label{tid3}
\end{eqnarray}

\noindent
where\\
$f_1(e^2)=1+{31 \over 2}e^2+{255 \over 8}e^4+{185 \over 16}e^6+{25  
\over 64}e^8$\\
$f_2(e^2)=1+{15 \over 2}e^2+{45 \over 8}e^4+{5 \over 16}e^6$\\
$f_3(e^2)=1+{15 \over 4}e^2+{15 \over 8}e^4+{5 \over 64}e^6$\\
$f_4(e^2)=1+{3 \over 2}e^2+{1 \over 8}e^4$\\
$f_5(e^2)=1+3e^2+{3 \over 8}e^4$\\
and $r_{\rm i,gyr}$ is the gyration radius and is defined by $I_{\rm i}
\equiv M_{\rm i} (r_{\rm i,gyr} R_{\rm
i})^2$, with $I_{\rm i}$ denoting the moment of inertia of a given  
binary
component. Here the mass ratio is defined as follows,
\begin{equation}
q_{\rm i} = \left\{
               \begin{array}{ll}
                 M_2 / M_1 & i=1 \\
                 M_1 / M_2 & i=2
               \end{array}
             \right.
\label{tid4}
\end{equation}

The quantity $(k/T)_{\rm i}$ is the ratio of the apsidal motion 
constant $k$ (which depends on the interior structure of the star) 
over the timescale $T$ of tidal dissipation. Following Hurley, Tout \& Pols 
(2002), we calculate that constant for either the equilibrium 
tide with convective damping ($(k/T)_{\rm i}=(k/T)_{\rm i,con}$) 
or the dynamical tide with radiative damping ($(k/T)_{\rm i}=(k/T)_{\rm 
i,rad}$).  
Radiative damping is applied to stars with radiative envelopes: MS stars 
with mass above $M_{\rm ms,conv}$, CHeB stars with mass above $7 \msun$,
massive evolved He stars and all He MS stars. For all other stars, 
convective damping is applied (see \S\,2.3.4 for details on
convective/radiative envelopes). We do not calculate tides on stellar 
remnants, e.g., on WDs ($K_{\rm i} \geq 10$).

The constant for convective damping is obtained from 
\begin{equation}
\left( {k \over T} \right)_{\rm i,con} = {2 \over 21} {f_{\rm i,conv} \over
\tau_{\rm i,conv}} {M_{\rm i,env} \over M_{\rm i}}\ {\rm yr}^{-1}
\label{tid5}
\end{equation}
where $M_{\rm i,env}$ is the mass contained in the convective  
envelope of component $i$.
The eddy turnover time $\tau_{\rm i,conv}$ is calculated as
\begin{equation}
\tau_{\rm i,conv}=0.431 \left[
{M_{\rm i,env} R_{\rm i,env} (R_{\rm i} - {1 \over 2} R_{\rm i,env})  
\over 3 L_{\rm i}}
\right]^{1/3} {\rm yr}
\label{tid6}
\end{equation}
with $R_{\rm i,env}$ denoting the depth of the convective envelope and
$L_{\rm i}$ the bolometric luminosity of a given component (Rasio et  
al.\ 1996).

The numerical factor $f_{\rm i,conv}$ is defined as
\begin{equation}
f_{\rm i,conv}=min \left[ 1, \left( P_{\rm i,tid} \over 2 \tau_{\rm  
i,conv}
\right)^2  \right]
\label{tid7}
\end{equation}
with the tidal pumping timescale $P_{\rm i,tid}$ defined as
\begin{equation}
{1 \over P_{\rm i,tid}} = \left| {1 \over P_{\rm orb}} - {1 \over
P_{\rm i,spin}} \right|
\label{tid8}
\end{equation}
where $P_{\rm orb}$ and $P_{\rm i,spin}$ are the binary orbital period 
and the spin period of component $i$, respectively. This factor 
represents the reduction in the effectiveness of eddy viscosity when 
the forcing period is less than the turnover period of the largest 
eddies (Goldreich \& Keeley 1977)

The constant for radiative damping is calculated from
\begin{equation}
\left( {k \over T} \right)_{\rm i,rad} = 1.9782 \times 10^4 
\sqrt{M_{\rm i} R_{\rm i}^2 \over a^5} (1+q_{\rm i})^{5/6} E_2\ {\rm yr}^{-1}
\label{tid9}
\end{equation}
where a second-order tidal coefficient $E_2=1.592 \times 10^{-9} 
M_{\rm i}^{2.84}$ was fitted (Hurley et al. 2002\footnote{Note that there 
is a typo in their eq.(42); missing $\sqrt{}$. However, their binary code 
utilizes the proper formula. Jarrod Hurley, private communication.}) to 
values given  by Zahn (1975).  

Finally, we have introduced an additional scaling factor $F_{\rm tid}$ 
in the evolution equations (eq.~\ref{tid1},~\ref{tid2},~\ref{tid3}) 
which we normally set to: 
\begin{equation}
F_{\rm tid} = \left\{
               \begin{array}{ll}
                 F_{\rm tid,con}=50  & {\rm convective\ env.} \\
                 F_{\rm tid,rad}=1  & {\rm radiative\ env.} \\
               \end{array}
             \right.
\label{Ftid}
\end{equation}
and the distinction between the stars with convective and radiative
envelopes is given in \S\,2.3.4. This factor makes tidal 
forces (both in case of convective and radiative damping) more 
effective than predicted by the standard Zahn theory. The choice of 
this specific value of $F_{\rm tid}$ is a result 
of our calibration against the cutoff period for circularization of 
binaries in M67 and from the orbital decay of the high mass X-ray 
binary LMC X-4 (for details see \S\,8.2).

The orbital angular momentum change associated with tides is 
calculated from
\noindent
\begin{eqnarray}
\,d J_{\rm i,tid}/\,d t = & 3 F_{\rm tid} I_{\rm i} \left( {k \over T} 
                          \right)_{\rm i} {q_{\rm i}^2 \over r_{\rm i,gyr}^2} 
                          \left( {R_{\rm i} \over a} \right)^6 {\omega_
                          {\rm orb} \over (1-e^2)^{6}} \nonumber \\  
                          & \times \left( f_2(e^2) - (1-e^2)^{3/2} f_5(e^2) 
                          {\omega_{\rm i} \over \omega_{\rm orb}} \right) 
\label{tid10}
\end{eqnarray}
and the change of binary parameters is calculated with
eqs.~\ref{tid1},~\ref{tid2},~\ref{tid3}.

{\em Pre--main sequence tidal synchronization and circularization.}
We also allow for pre-MS tidal interactions. Since we do not
follow pre-MS evolution, all binaries with orbital periods shorter
than $4.3\,$d (Mathieu et al.\ 1992) are simply assumed to have  
circularized and all binary components to have synchronized by the 
time they reach the ZAMS. For binaries with longer orbital periods 
we apply our assumed distribution of initial eccentricities 
(see \S\,7) and initial rotational velocities for binary components 
(see \S\,2.3.3).

{\em Darwin instability.} 
One important consequence of tidal interactions in massive binaries is 
the possible occurrence of the Darwin instability (e.g., Lai et al.\
1993). When the more massive component is spinning slowly compared to the
orbital rate of its companion, tidal forces will tend to spin it up,
leading to loss of orbital angular momentum (orbital decay).
Usually this orbital decay will stop when synchronization is established.
However, if, in the synchronized state, more than a third of the total 
binary angular momentum would be in the component spins, then synchronization 
can never be reached and the components will continue to spiral in.
We follow this process until one of the binary components overflows its Roche 
lobe. In all cases
RLOF is found to be dynamically unstable (\S\,5.1 and \S\,5.2) and the system 
goes through a CE phase leading either to a merger, or further orbital decay 
with envelope ejection (\S\,5.4).

\subsection{Mass and Angular Momentum Loss from Binaries}

Mass lost from the binary components in stellar winds carries angular 
momentum, in turn affecting the orbit through tidal coupling. 
Similarly, during RLOF, some of the transferred 
material and its associated angular momentum may be lost from the system. In this section 
we consider the amount of angular momentum loss associated both with 
stellar winds and RLOF phases. However, for RLOF we only consider here 
dynamically stable phases, while the change of the orbit following unstable RLOF
(common envelope events) is described in \S\,5.4. 

For stellar winds we assume spherically symmetric mass loss, which carries 
away the specific angular momentum of the mass--losing component (Jeans mode mass loss). The corresponding change of the orbit (Jeans-mode mass loss) is calculated from 
\begin{equation}
a(M_1+M_2)\ =\ {\rm const.}
\label{mw01}
\end{equation}
The above approach holds for circular orbits, however the change in
binary separation $a$ is similar for eccentric orbits (Vanbeveren, Van
Rensbergen \& De Loore 1998, see p.\ 124).

In the case of stable RLOF with compact accretors (WD, NS, BH; $K_{\rm acc}=
10,11,12,13,14,16,17$)  we limit (although this assumption may be relaxed) 
accretion to the Eddington critical rate
\begin{equation}
\mdot_{\rm edd} = 2.088 \times 10^{-3} {R_{\rm acc} \over \epsilon (1+X)}
\mpy
\label{dMedd}
\end{equation}
and the corresponding critical Eddington luminosity may be expressed as 
\begin{equation}
L_{\rm edd} = \epsilon {G M_{\rm acc} \mdot_{\rm edd} \over R_{\rm acc}}
\label{Ledd}
\end{equation}
where $R_{\rm acc}$ denotes the radius at which the accretion onto compact
object takes place (a NS or a WD radius, and three Schwarzschild radii 
for a BH), $X$ denotes the composition of accreted material (0.7 for the
H-rich material, and 0.0 for all other compositions), and $\epsilon$ gives 
the conversion efficiency of gravitational binding energy to radiation 
associated with accretion onto a WD/NS (surface accretion $\epsilon=1.0$) and 
onto a BH (disk accretion $\epsilon=0.5$).
We also note that above some critical (very high) accretion rate, nuclear 
burning will start on the WD surface. This can be much more radiatively 
efficient than the gravitational energy release and the above relations 
break down.
If the mass transfer rate is higher than $\mdot_{\rm edd}$ we expect the excess
material to leave the system from the vicinity of the accreting object and thus 
to carry away the specific angular momentum of the accretor. The angular momentum 
loss associated with a given systemic mass loss in a RLOF phase is obtained 
from 
\begin{equation}
\,d J_{\rm RLOF}/ \,d t = R_{\rm com}^2 w_{\rm orb} (1-f_{\rm a}) \mdot_{\rm don} 
\label{jmt1}
\end{equation}
where $R_{\rm com} = a M_{\rm don} / (M_{\rm don} + M_{\rm acc})$
is the distance between the accretor and binary center of mass, and 
$\mdot_{\rm don}$ is the mass transfer rate (donor RLOF rate, see 
eq.~\ref{mt10a}).
The $f_{\rm a}$ fraction of material transferred from the donor is accreted 
on the compact object. If mass transfer is sub-Eddington then $f_{\rm a}=1$ 
(conservative), otherwise it is $f_{\rm a} = \mdot_{\rm edd} /  \mdot_{\rm don}$ 
(non-conservative evolution). 
Here we assume that the radiative efficiency is not a function 
of the mass transfer rate.  Some work has suggested that at high transfer
rates, flows onto black holes may become radiatively inefficient as photons are
trapped in the flow and advected into the black hole (see e.g., Abramowicz et al. 
1988), or that substantially super-Eddington accretion may be possible in 
non-spherical accretion flows (e.g., Begelman 2002). In the current version of the
code, we do not consider these possibilities.

For all other, non-degenerate accretors ($K_{\rm acc}=0,1,2,3,4,5,6,7,8,9$) 
we assume a non-conservative evolution through stable RLOF, with part of the 
mass lost by the donor accreted onto the companion ($f_{\rm a}$), and the rest 
($1-f_{\rm a}$) leaving the system with a specific angular momentum 
$j_{\rm loss}$ expressed in units of ${2 \pi a^2 / P_{\rm orb}}$ (see 
Podsiadlowski, Joss \& Hsu 1992). The angular momentum loss is then estimated 
from
\begin{equation}
\,d J_{\rm RLOF} / \,d t = j_{\rm loss} {J_{\rm orb} \over M_{\rm don} + 
                           M_{\rm acc}} (1-f_{\rm a}) \mdot_{\rm don}
\label{jmt2}
\end{equation}
For our standard model calculation we adopt $j_{\rm loss}=1$, and $f_{\rm a}=0.5$ 
(half of the transferred mass lost from system, e.g. Meurs \& van den Heuvel 1989). 
However, we note that the amount of mass loss (as well as the specific angular 
momentum with which the mass is lost from the binary) may change from case to 
case. Ideally, one would like to know the details of mass transfer/loss for all 
potential binary configurations, and change $j_{\rm loss}$ and $f_{\rm a}$
according to the types of interacting stars as well as binary properties. 
Since, such predictions or understanding are not available, we treat $j_{\rm loss}$ and $f_{\rm a}$ as parameters, which are applied evenly to all the stars in a given simulation.

\section{Wind Mass Loss/Accretion Implementation}

We adopt the compilation of mass loss rates from Hurley at al.\ (2000). 
We have further extended the original formulas to include winds from low- and 
intermediate-mass MS stars. The structure of the star (and its subsequent 
evolution) in response to stellar wind mass loss is self-consistently 
taken into account with the Hurley et al.\ (2000) evolutionary formulas. 
The most important effects include possible removal of the H-rich 
envelope of a massive star or a more gradual nuclear evolution with
decreasing mass.
The effects of wind mass loss from binary components on the orbital
parameters are also accounted for (see \S\,3.4). 

The effects of mass increase of binary components due to accretion
from the companion winds are neglected. Either the wind accretion rates are
very low or the high wind accretion phases do not last for long, which does
not translate into significant mass increase of a companion star. 
However, we estimate the wind accretion rates onto NSs and BHs since it may 
give rise to bright X-ray emission (see \S\,9). 
 
The wind accretion rate is calculated in the general case of
eccentric orbits, i.e. we obtain accretion rate (and accretion
luminosity) for a specified position on the orbit, or we integrate over
a specific part of the orbit (e.g., corresponding to the exposure time of 
given observations).  This may be of importance for eccentric wind-fed
binaries, e.g., high mass X-ray binaries (see \S\,9.2). We have also implemented 
an orbital-averaged solutions. The two solutions may be adopted as required 
for a given project or analysis.

\subsection{General Eccentric Orbit Case}

We follow the Bondi \& Hoyle (1944) accretion model to calculate the accretion 
from stellar wind. As an approximation we may express (Boffin \& Jorissen 1988) 
the accretion rate as
\begin{equation}
{\mdot_{\rm acc,wind}} = \alpha_{\rm wind} { 2 \pi (G M_{\rm acc})^2 \over 
                         (V^2_{\rm rel} + c^2_{\rm wind})^{3/2}} \rho 
\label{wind01}
\end{equation}
where $\alpha_{\rm wind}=1.5$ is the accretion efficiency in the Bondi--Hoyle 
model, although it may be as low as 0.05 in some specific cases 
(e.g., see hydrodynamical simulations of Theuns, Boffin \& Jorissen 1996 for 
Barium star formation), $c_{\rm wind}$ is the wind sound speed, and 
$V_{\rm rel}$ is the relative velocity of the wind with respect to the accreting 
star. The local (undisturbed) density of the wind matter $\rho$ in the 
vicinity of the accreting object may be calculated in a steady 
spherically symmetric case from 
\begin{equation}
{\mdot_{\rm don,wind}}= - 4 \pi r^2 \rho V_{\rm wind}
\label{wind02}
\end{equation}
where ${\mdot_{\rm don,wind}}$ is the wind mass loss rate form the donor, $r$ 
is the instantaneous distance between the two stars, and $V_{\rm wind}$ is the wind 
velocity. We assume that the wind flow is supersonic ($V_{\rm rel} \gg c_{\rm wind}$)
so that $c^2_{\rm wind}$ may be dropped from eq.~\ref{wind01}. We introduce 
$\rho$ (expressed through eq.~\ref{wind02}) into eq.~\ref{wind01} to obtain
\begin{equation}
{\mdot_{\rm acc,wind}} = - \alpha_{\rm wind} { (G M_{\rm acc})^2 \over
                         2 V^3_{\rm rel} V_{\rm wind} r^2} \mdot_{\rm don,wind} 
\label{wind03}
\end{equation}
The accretion rate calculated with eq.~\ref{wind03} varies as the accreting
object moves in its orbit around the mass--losing star. The relative distance $r$ 
of the two stars is obtained through the Kepler equation for a given orbit.   
Obviously $r$ is a function of orbital position. 
The vector of the relative velocity $\vec V_{\rm rel}$ is defined as 
\begin{equation}  
\vec V_{\rm rel} =  \vec V_{\rm acc,orb} + \vec V_{wind}
\label{wind04}
\end{equation}
where $\vec V_{\rm acc,orb}$ denotes the instantaneous velocity of the accretor 
on the orbit relative to the mass losing star, and is readily obtained for a 
given position through the Kepler equation. 
The direction of the wind velocity vector $\vec V_{wind}$ follows the vector
pointing toward the accretor on its relative orbit around the mass--losing star.  
We set the wind velocity proportional to the escape velocity from the surface 
of the mass--losing star 
\begin{equation}
V^2_{\rm wind}=2 \beta_{\rm wind} {G M_{\rm don} \over R_{\rm don}}, 
\label{wind05}
\end{equation}
and vary $\beta_{\rm wind}$ with the spectral type of the mass--losing star. 
For extended ($R_{\rm don}>900 \rsun$) H-rich giants 
($K_{\rm don}=2,3,4,5,6$) slow winds are assumed $\beta_{\rm wind}=0.125$. For 
the most massive MS stars ($> 120 \msun$) $\beta_{\rm wind}=7$, for low mass 
MS stars ($< 1.4 \msun$) $\beta_{\rm wind}=0.5$ and the value of $\beta_{\rm
wind}$ is interpolated in-between. For He-rich stars ($K_{\rm don}=7,8,9$); 
$\beta_{\rm wind}=7$ for $M_{\rm don}>120 \msun$, $\beta_{\rm wind}=0.125$ for 
$M_{\rm don} < 10 \msun$, and is interpolated in-between.
The values of $\beta_{\rm wind}$ follow from the observations of wind 
velocities for different type of stars (Lamers, Snow \& Lindholm 1995; Kucinskas 
1999) and are adopted from the discussion of wind properties in Hurley et al. 
(2002).

\subsection{Orbit-averaged Case}

We use eq.~\ref{wind03} to obtain the orbit-averaged accretion rate. 
The wind velocity vector is assumed to be perpendicular to the orbital 
speed vector (as on a circular orbit), i.e., $V^2_{\rm rel}= V^2_{\rm acc,orb}+V^2_{wind}$. 
The wind velocity is taken from eq.~\ref{wind05}.
The orbital velocity of the accretor is taken to be constant and is obtained from the
circular orbit approximation  $V^2_{\rm acc,orb}= G(M_{\rm acc}+M_{\rm
don})/a$. Finally, $1/r^2$ is substituted in eq.~\ref{wind03} with its mean 
value over one orbital revolution, i.e., $1/(a^2\sqrt{1-e^2})$ to obtain 
\begin{equation}
{\mdot_{\rm acc,wind}} = - {F_{\rm wind} \over \sqrt{1-e^2}} 
                     \left( {G M_{\rm acc} \over V^2_{\rm wind}} \right)^2
                     {\alpha_{\rm wind} \over 2 a^2} 
                     { \mdot_{\rm don,wind} \over (1+V^2)^{3/2}} 
\label{wind06}
\end{equation}
where $F_{\rm wind}$ is a parameter (see below) and 
$V^2={V^2_{\rm acc,orb}/V^2_{\rm wind}}$.

For highly eccentric orbits, the averaged (over one orbit) accretion rate
calculated with the eq.~\ref{wind06} may exceed the companion mass loss
rate. This is a direct result of the orbital averaging used above. 
To avoid this we follow Hurley et al. (2002; \S\,2.1) and 
adopt $F_{\rm wind}$ such that ${\dot M_{\rm acc,wind}}$ never exceeds $0.8 
{\dot M_{\rm don,wind}}$.

\section{Roche Lobe Overflow Calculations}

Different physical processes may be responsible for driving RLOF. In the 
following we describe the treatment of mass loss and mass accretion in our 
model.

\subsection{Mass Transfer/Accretion Rate}

For any binary system during RLOF phases with a non-degenerate donor
($K_{\rm don}<10$) we calculate the radius mass exponents for the donor and 
its Roche lobe
\begin{equation}
\zeta_{\rm don} = {\partial \ln R_{\rm don} \over \partial \ln M_{\rm don}}
\label{mt1}   
\end{equation}
\begin{equation}
\zeta_{\rm lob} = {\partial \ln R_{\rm don,lob} \over \partial \ln M_{\rm don}}
\label{mt2}   
\end{equation}
and we estimate the change of donor radius with time due to its nuclear
evolution as
\begin{equation}
\zeta_{\rm evl} = {\partial \ln R_{\rm don} \over \partial t}
\label{mt3}    
\end{equation} 
The above derivatives are calculated numerically with the use of the analytic
single star formulas of Hurley et al. (2000). 
The time derivative of the stellar radius ($\zeta_{\rm evl}$) is obtained directly 
from the single star formulas, since the radius of a given star is tracked in time. 
To obtain response of the donor to mass loss ($\zeta_{\rm don}$) we
calculate the response of the star to an instantaneous (over a timestep of only  
1\,yr, a time interval unimportant for stellar evolution) mass loss 
through (artificially) increased wind mass loss. Finally, the Roche lobe 
exponent is obtained by removing 1\% of the donor mass, part of which is 
transferred to the accretor and the rest is lost with the specific angular
momentum of the accretor from the binary (see \S\,3.4); a new Roche lobe radius 
and the numerical derivative are then readily calculated.  

RLOF may be driven by different physical processes; angular momentum losses
connected to magnetic braking (applied directly to the orbit, given the 
assumption of synchronism during RLOF) and gravitational radiation or expansion due to 
nuclear evolution. We do not include tides as we assume that binary is
circular and synchronized during RLOF.   
The timescales for magnetic braking, and gravitational radiation 
are calculated from
\begin{equation}
\tau_{\rm mb} = - {J_{\rm orb} \over \,d J_{\rm don,mb}/\,d t + \,d J_{\rm
acc,mb}/\,d t } 
\label{mt4}   
\end{equation}
\begin{equation}
\tau_{\rm gr} = - {J_{\rm orb} \over \,d J_{\rm gr}/\,d t}
\label{mt6}
\end{equation}
where expressions for $\,d J_{\rm gr}/\,d t,\ \,d J_{\rm i,mb}/\,d t$, are given 
in \S\,3.1, and \S\,3.2 respectively.

If RLOF is driven by the combination of angular momentum losses 
changing the orbit and nuclear evolution of the donor we then 
calculate the mass transfer rate from  
\begin{equation}
\mdot_{\rm eq} =  - {\zeta_{\rm evl}+{2 \over \tau_{\rm mb}} + {2 \over \tau_{\rm gr}}  
                    \over \zeta_{\rm don} - \zeta_{\rm lob}} M_{\rm don}
\label{mt7}   
\end{equation}
and the corresponding mass transfer timescale
\begin{equation}
\tau_{\rm eq} = - {M_{\rm don} \over \mdot_{\rm eq}}
\label{mt8}
\end{equation}

Additionally we estimate the thermal timescale for the donor following 
Kalogera \& Webbink (1996) from 
\begin{equation}
\tau_{\rm th} = {30 \times {M_{\rm don}}^2 \over R_{\rm don} L_{\rm don}}
\label{mt9}
\end{equation}
and the mass transfer rate on the thermal timescale
\begin{equation}
\mdot_{\rm th} = - {M_{\rm don} \over \tau_{\rm th}}
\label{mt10}
\end{equation}

In the case of stable RLOF, $\tau_{\rm eq}>\tau_{\rm th}$, a donor is in 
thermal equilibrium,  and we use eq.~\ref{mt7} to calculate the mass transfer 
rate. Otherwise, for $\tau_{\rm eq} \leq \tau_{\rm th}$, RLOF proceeds on 
the thermal timescale and we evolve a given system calculating the mass transfer 
rate from eq.~\ref{mt10}. We follow the timescales of the donor as it evolves 
through RLOF, and apply the appropriate mass loss rate. For example, a massive 
donor may be transferring mass on the thermal timescale at first, but once it 
loses a fraction of its mass, the mass transfer becomes stable and RLOF proceeds 
on the timescale defined by $\tau_{\rm eq}$.
However, in some cases the RLOF is so rapid that it may eventually lead 
to a dynamical instability. Once $\mdot_{\rm eq}$ changes sign and 
becomes positive, the donor loses its 
equilibrium, and the 
system evolves either on the thermal or dynamical timescale. In this case a 
special {\em diagnostic diagram} is used (see below) to decide which
of the two timescales is relevant. 
We also allow for the development of a delayed dynamical instability, which 
may occur for stars with a radiative envelope, but with a deep convective 
layer.
Dynamical instability during RLOF leads to a spiral-in of the binary components
and  common envelope evolution (CE). 
We follow the CE phase to determine whether the binary survives (ejection of 
the envelope at the expense of orbital energy) or if a merger of the binary 
components (single star formation) occurs.  

The following summarizes the calculation of the RLOF mass transfer rates
\begin{equation}
\mdot_{\rm don}= \left\{ \begin{array}{ll}
 CE/merger & M_{\rm don} > q_{\rm ddi} \times M_{\rm acc}\\  
 \mdot_{\rm eq} &  \mdot_{\rm eq}<0\ and\ \tau_{\rm eq}>\tau_{\rm th} \\
 \mdot_{\rm th} &  \mdot_{\rm eq}<0\ and\ \tau_{\rm eq}<=\tau_{\rm th} \\
 \mdot_{\rm th}/CE/merger & diagnostic\ diagram \\
\end{array}\right.
\label{mt10a}
\end{equation}
where we additionally assume that above some critical mass ratio ($q_{\rm ddi}
\equiv M_{\rm don} / M_{\rm acc} $)  
the binary system will evolve toward delayed dynamical instability (Hjellming 
\& Webbink 1987), leading to rapid inspiral and CE evolution. 
For H-rich stars Hjellming (1989) gives a range $q_{\rm ddi}=2-4$ depending 
on the evolutionary state of a donor, while Ivanova \& Taam (2004) obtain $q_{\rm 
ddi}=2.9-3.1$. In our standard model calculations we adopt $q_{\rm ddi}=3$
for H-rich stars ($K_{\rm i}=0,1,2,3,4,5,6$). For He-rich stars we adopt
critical mass ratios from Ivanova et al. (2003); $q_{\rm ddi}=1.7$ for HeMS 
stars ($K_{\rm i}=7$), while $q_{\rm ddi}=3.5$ for evolved He stars 
($K_{\rm i}=8,9$). We note that the study of Ivanova et al. (2003) 
was targeted for He stars with NS accretors only. However, we adopt their 
values for systems with He star donors and arbitrary accretors, since detailed 
models for arbitrary accretors are not available.  
Also, dynamical instability may be encountered if the trapping radius of the 
accretion flow exceeds the Roche lobe radius of the accretor (\S\,5.4).   
Additionally, we consider the case of spiral--in in the case of the Darwin instability, 
where the components' spin angular momentum is comparable to the 
orbital angular momentum (\S\,3.3).

For the donor stars without a well defined core-envelope structure ($K_{\rm
don}=0,1,7,10,11,12,16,17$) we assume that dynamical instability during RLOF 
{\em always} leads to a merger. The same is assumed for the donors in the
Hertzsprung gap ($K_{\rm don}=2,8$) as there is no clear entropy jump at the 
core-envelope transition (Ivanova \& Taam  2004; Belczynski et al. 2007). 
In the case of a merger a single stellar object is formed. However, we do not 
follow its evolution here, as the chemical composition and structure of merged 
remnants is not well understood and certainly is different than normal stars. 
This may lead to an underestimate of our synthetic supernovae rate, since 
potentially some merger products are massive enough to evolve and explode as 
Type II or Ib/c SNe. 
For H-rich and He-rich giant-like donors ($K_{\rm don}=3,4,5,6,9$) we follow 
CE evolution, and assuming ejection of the entire donor envelope, we calculate 
the most probable outcome with conservation of energy (see \S\,5.4).
If RLOF is encountered for a system with an evolved Helium star donor 
($K_{\rm i}=8,9$), then it is found that for low donor masses ($\lesssim 4-5
\msun$) RLOF is stable (although it may proceed at very high rates) while 
for higher donor masses it leads to a CE phase (e.g., see Ivanova et al. 2003). 
The survival of the binary then depends on the donor properties (e.g.,
stellar structure, envelope binding energy, its mass, binary separation) and
in particular for He stars in the Hertzsprung gap ($K_{\rm don}=8$) CE phase 
always leads to merger. 

The mass accretion rate in a dynamically stable RLOF is calculated from
\begin{equation}
\mdot_{\rm acc}= f_{\rm_a} \mdot_{\rm don}
\label{mt10b}
\end{equation}
where $\mdot_{\rm don}$ is the donor RLOF mass transfer rate (see
eq.~\ref{mt10a}). The parameter $f_{\rm_a}$ denotes the fraction of the 
transferred mass which is accreted, while the rest ($1-f_{\rm_a}$) 
is ejected from the system (see \S\,3.4).
Mass accretion in dynamically unstable cases (CE events) is calculated 
only for NS and BH accretors, since only then significant accretor 
mass gain may be expected in spite of the very short timescales (for details 
see BKB02).

\subsection{Diagnostic Diagram for Rapid Mass Transfer}
\begin{figure}
\includegraphics[width=1.1\columnwidth]{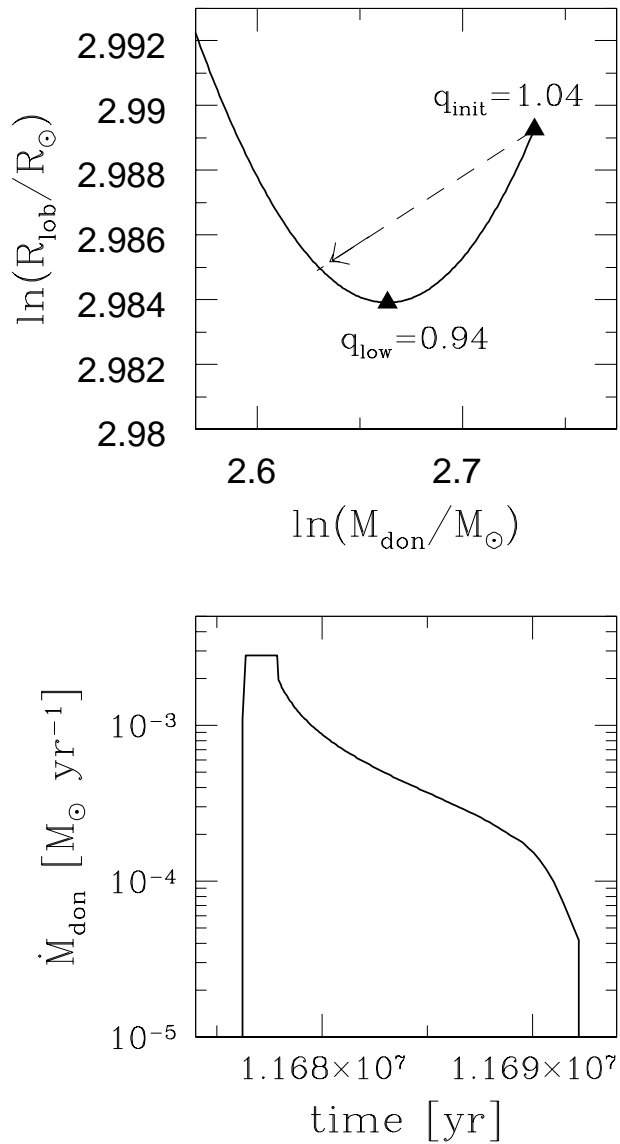}
\caption{
The diagnostic diagram (top panel) used to decide whether a binary should
be evolved on a thermal timescale or rather RLOF is dynamically unstable
(leading to CE evolution and a potential merger).
If the mass ratio at the onset of RLOF ($q_{\rm int}$) is much greater 
than the mass ratio at the moment when the orbit starts expanding 
($q_{\rm low}$) then the system is dynamically unstable, otherwise 
RLOF on a thermal timescale is assumed. The arrow represents the partial 
derivative of donor radius (equal to the Roche lobe radius) with
respect to its mass, and points to the place where the 
donor is expected to regain thermal equilibrium. 
The bottom panel shows a specific system: a $16 \msun$ Hertzsprung 
gap donor with a $15 \msun$ MS companion in an 8-day orbit, for which the 
diagnostic diagram is plotted. The mass transfer begins on a thermal 
timescale (flat part) and then evolves on a slower nuclear timescale 
(decline). For more details see \S\,5.2.
}
\label{fig01}
\end{figure}

The aforementioned diagnostic diagram is shown in Fig~\ref{fig01}. 
Once RLOF proceeds on the thermal timescale, and the donor is no longer in thermal 
equilibrium, we do not have proper stellar models to use and calculate the donor 
properties (e.g., RLOF rate).  
Therefore, we use an approximate method and calibrate it based on the results from 
detailed stellar evolutionary and mass transfer calculations, which are not limited 
to stars in thermal equilibrium. When the donor loses its equilibrium, we use the 
stellar and binary properties to predict whether the system will evolve through the
phase of thermal mass transfer and the donor will regain its equilibrium, or the
RLOF will become dynamically unstable and will eventually lead to CE evolution. 
We plot the donor Roche lobe radius versus decreasing donor mass under the 
assumption that mass transfer is non-conservative and proceeds on the thermal
timescale (see eqs.~\ref{mt9} and ~\ref{mt10}). For NS/BH
accretors the accretion rate is limited by the Eddington rate, while for all
other accretors, a fraction $f_{\rm a}$ of transferred material is accreted. 
The associated specific angular momentum loss is described in \S\,3.4.
As the mass of the donor decreases with mass transfer the Roche lobe 
first shrinks and then at some critical mass ratio ($q_{\rm low}$), it starts expanding again (see the solid line on the top panel, Fig.~2). 
If the mass ratio at the moment the star loses its equilibrium $q_{\rm int}$ is 
not greatly different than $q_{\rm low}$ we expect that the donor may regain 
the equilibrium when the system is expanding.  The dashed line arrow in
Figure~2 shows the expected behavior of the donor when it loses its
equilibrium. If the system does not evolve into a CE phase then we expect the donor to
regain its equilibrium at the position indicated by the arrow. Of course this is
just an approximation, since, as the donor evolves,
 the radius-mass exponent changes. We use a number of published (Tauris \& 
Savonije 1999; Wellstein \& Langer 1999; Wellstein, Langer \& Braun 2001; 
Dewi \& Pols 2003) and unpublished (N.\ Ivanova 2004, private communication) 
detailed calculations to calibrate the diagnostic diagram. Based on these studies 
we find that a CE phase ensues if 
\begin{equation}
 CE\ \left\{ \begin{array}{ll}
        q_{\rm int} \geq 1.2\ q_{\rm low} & K_{\rm don}=2,3,4,5,6\\
        q_{\rm int} \geq 2.0\ q_{\rm low} & K_{\rm don}=0,1,7,8,9\\
 \end{array} \right.
\label{mt15}
\end{equation}
Otherwise the system is evolved through RLOF on the donor's thermal timescale.

\subsection{Thermal Timescale Mass Transfer}  

Once a binary is identified as a thermal timescale RLOF system, we assume that 
the mass transfer rate remains constant throughout the entire episode. We calculate 
the rate using eq.~\ref{mt10} where we use properties corresponding to the time the 
donor loses its thermal equilibrium. This may be justified by the following: {\em (i)} 
thermal mass transfer rates have been shown to be rather constant within a factor of 
$\sim 2-3$ (Paczynski 1971), {\em (ii)} since the rates are calculated at the time the 
star loses equilibrium, it is a good approximation (and the best possible with only 
equilibrium stellar models being available) for the short lived phase of thermal mass 
transfer that follows. 

In the bottom panel of Figure~\ref{fig01} we show an example calculation through a 
thermal RLOF phase, followed with a slower (driven by nuclear evolution) RLOF 
period after the donor has regained its thermal equilibrium. 
The specific system  was chosen to match the RLOF calculation of Wellstein
et al.\ (2001) for a $16 \msun$ and $15 \msun$ binary with an initial period of
8 days. The RLOF starts when the primary evolves off the main sequence and crosses 
the Hertzsprung gap. Mass transfer initially proceeds on a thermal timescale
at a very high rate ($\sim 2.8 \times 10^{-3} \mpy$), then the star regains its
equilibrium and the RLOF rate decreases with time by more than order of
magnitude ($\sim 10^{-4} \mpy$). 
Our calculation can be directly compared to Wellstein et al.\ (2001): see their Figure~4,
left panel. Their detailed stellar evolution calculation shows a thermal RLOF rate of 
$\sim 10^{-3} \mpy$, followed by a slower RLOF phase characterized by
rates of $\sim 10^{-4} \mpy$, very similar to what we find with our simplified 
prescription. Our RLOF phase lasts about twice as long as that of 
Wellstein et al.\ (2001), who in contrast to our calculation assumed 
conservative evolution and did not include effects of tidal spin-orbit 
interactions. We choose not to modify our standard model assumptions 
(e.g., neglect tidal interactions) for comparisons, and therefore emphasize
some differences with previous calculations. 
More comparisons of RLOF sequences are presented in \S\,8.1.

\subsection{Dynamical Instability and Common Envelopes} 

Dynamically unstable mass transfer may be encountered in a number of ways. Most often 
it is the direct consequence of stellar expansion during rapid nuclear evolution phases. However,
loss of orbital angular momentum (e.g., via magnetic braking, gravitational
radiation, or tides) may also lead to dynamical instability. 

Additionally, we allow a system to evolve into a CE phase if the 
trapping radius of the accretion flow exceeds the Roche lobe radius of the
accretor. 
The trapping radius is defined as (Begelman 1979)   
\begin{equation}
R_{\rm trap} = {{\dot M_{\rm don}} \over {\dot M_{\rm edd}} }
               {R_{\rm acc} \over 2}.
\end{equation}
Following King \& Begelman (1999) and Ivanova et al. (2003) we check whether the 
mass transfer rate exceeds a critical value above which the system is engulfed 
in a CE  
\begin{equation}
{\dot M_{\rm trap}} = 2 \times {\dot M_{\rm edd}} 
{R_{\rm acc,lob} \over R_{\rm acc}}
\label{trapping}
\end{equation}
where $R_{\rm acc,lob}$ is the accretor Roche lobe radius, and $\dot M_{\rm
edd}$ is the Eddington critical accretion rate (see eq.~\ref{dMedd}).

Below we present two different implementations of the orbital contraction 
calculation during CE that are incorporated in {\tt StarTrack}.

{\em Standard Energy Balance Prescription.}~
If dynamical instability is encountered a binary may enter a CE phase. 
We use the standard energy equation (Webbink 1984) to calculate the 
outcome of the CE phase
\begin{equation}
 \alpha_{\rm ce} \left( {G M_{\rm don,fin} M_{\rm acc} \over 2 A_{\rm fin}}
   - {G M_{\rm don,int} M_{\rm acc} \over 2 A_{\rm int}} \right) =
{G M_{\rm don,int} M_{\rm don,env} \over \lambda R_{\rm don,lob}}
 \label{ce}
\end{equation}
where, $M_{\rm don,env}$ is the mass of the donor envelope ejected from the binary, 
$R_{\rm don,lob}$
is the Roche lobe radius of the donor at the onset of RLOF, and the indices ${\rm int,\ fin}$ denote 
the initial and final values, respectively. The parameter $\lambda$ is a measure of the
central concentration of the donor (de Kool 1990; Dewi \& Tauris 2000).
The right hand side of equation~(\ref{ce}) expresses the binding energy of
the donor's envelope, the left hand side represents the difference
between the final and initial orbital energy, and $\alpha_{\rm ce}$ is
the CE efficiency with which orbital energy is used to unbind the
stellar envelope. If the calculated final binary orbit is too small to
accommodate the two post-CE binary components then a merger occurs. 
In our calculations, we combine $\alpha_{\rm ce}$ and $\lambda$ into one 
CE parameter, and for our standard model, we assume that 
$\alpha_{\rm ce}\times\lambda = 1.0$. This is for all but evolved naked Helium stars
($K_{\rm i}=9$) for which we adopt $\alpha_{\rm ce} = 1.0$ and $\lambda = 
0.3R_{\rm i}^{-0.8}$, where $R_{\rm i}$ is radius of Helium star in solar radii. 
The relation for $\lambda$ was obtained with Ivanova's (2003) evolutionary code. 
If a compact object spirals in the common envelope it may
accrete significant amounts of material because of hyper-critical accretion 
(Blondin 1986; Chevalier 1989, 1993; Brown 1995). We have incorporated  a numerical scheme to include the effects of  hyper-critical accretion on NSs
and BHs in our standard CE prescription (for details see BKB02).  Compact
objects gain, on average, several tenths of solar mass in CE if hyper-critical
accretion is allowed. 
However, we also allow for evolution with no hyper-critical accretion 
following recent results of accretion flow calculations with geometry
specific for compact object moving through common envelope. These
calculations indicate that accretion can be limited to only $0.01 \msun$ 
(E.Ramirez-Ruiz, private communication).

{\em Alternative Angular Momentum Prescription}
In addition to the standard prescription for common envelope
evolution based on comparing the binding and orbital energies (see above), 
we investigate the alternative approach proposed by Nelemans \& Tout (2005), 
based on the non-conservative mass transfer analysis by Paczynski \& 
Ziolkowski (1967), with the assumption that the mass loss reduces the 
angular momentum in a linear way. This leads to reduction of the orbital 
separation
\begin{equation}
{A_{\rm fin}\over A_{\rm int}}=\left( 1 -\gamma {M_{\rm don,env}\over M_{\rm tot,int}}\right)
         {M_{\rm tot,fin}\over M_{\rm tot,int}} \left({M_{\rm don,int} 
         M_{\rm acc,int} \over M_{\rm don,fin} M_{\rm acc,fin}}\right)^2
\end{equation}
where $M_{\rm don,env}$ is the mass of the donor envelope lost by the system, 
$M_{\rm tot,int}$, $M_{\rm tot,fin}$ are the total masses of the system before 
and after CE, and $\gamma$ is a scaling factor. Following Nelemans \& Tout
(2005) we use $\gamma=1.5$ and note that hyper-critical accretion is not
included in this prescription.

The two above prescriptions are extended (e.g., BKB02) to the case where 
both stars lose their envelopes, which happens if the stars have giant-like 
structure ($K_{\rm i}=3,4,5,6,9$) at the onset of CE phases (see Bethe
\& Brown 1998).

\subsection{Mass Transfer from Degenerate Donors}
Degenerate donors ($K_{\rm don}=10,11,12,16,17$), are also 
considered. The RLOF is assumed to be driven by gravitational radiation 
only 
\begin{equation}
\mdot_{\rm don} = M_{\rm don} D^{-1} {\,d J_{\rm gr}/\,d t \over J_{\rm orb}}
\label{mt13}  
\end{equation}
with 
\begin{equation}
D={5 \over 6}+ {1 \over 2} \zeta_{\rm don}-{1-f_{\rm a} \over 3 (1+q)}-
{ (1-f_{\rm a}) (1+q) \beta_{\rm mt}+f_{\rm a} \over q} 
\label{mt17}
\end{equation}
where the mass ratio is defined as $q={M_{\rm acc} / M_{\rm don}}$, $f_{\rm a}$
denotes the fraction of transferred material that is accreted by the 
companion
(defined and evaluated in \S\,3.4), and 
$\beta_{\rm mt}= M_{\rm don}^2 / (M_{\rm don}+M_{\rm acc})^{2}$.

\subsection{Effects of Mass Transfer on Stellar Evolution} 

Mass loss/gain changes the subsequent evolution of stars. We implement RLOF mass 
loss/gain by adding an extra term in the original Hurley et al.\ (2000) stellar 
evolution formulae. In case of mass loss we increase the wind mass loss rate to 
match the combined effects of wind and RLOF mass loss. To treat mass gain and 
potential accretor rejuvenation, we add the RLOF mass accretion rate, as calculated 
in \S\,5.6, to the accretor wind mass loss rate (they have opposite signs).
Additionally, in case of rejuvenation evolutionary timescales and stellar ages are modified as 
suggested by Tout et al. (1997) and Hurley et al. (2002) by the relative
mass change (see the following equation) to calculate the 
net effect on the star.   
For main sequence stars we can calculate the change of the age of a given star 
(due to accretion or mass loss) from  
\begin{equation} 
t_{\rm age,fin}= f_{\rm rej} {\tau_{\rm ms,fin} \over \tau_{\rm ms,int}}  t_{\rm age,int}
\label{rejuv}
\end{equation}
where $\tau_{\rm ms}$ is the main sequence lifetime, and indices $int,\ fin$ mark
the state before and after some amount of mass is transferred, respectively. 
The factor $f_{\rm rej}$ is unity for all mass losing stars and for hydrogen MS stars 
($K_{\rm i}=1$) with radiative cores ($0.3<M_{\rm i}<1.25 \msun$), while it is 
$f_{\rm rej}=M_{\rm i,int}/M_{\rm i,fin}$
for hydrogen MS stars with convective cores and helium MS stars ($K_{\rm i}=7$; 
they have convective cores) and it reflects the effects of additional 
fuel in the core. For HG stars ($K=2$) we change the timescales using 
$f_{\rm rej}=1$ as for MS stars with radiative cores.
In this way we ensure that the subsequent evolution of the donor is followed consistently,
i.e., evolutionary timescales and physical properties of mass losing/gaining stars
(e.g., core masses) are changed in agreement with stellar models. Our wind mass loss 
formulae are implemented the same way as in the original Hurley et al.\ (2000, see their \S\,7.1) 
and the above scheme allows for appropriate changes of evolutionary timescales
and core masses, both in cases of mass loss and gain.

For simplicity, we assume that the composition of the accreted material matches 
that of the accretor, although this may not always be the case.  
Only in the case of accretion onto white dwarfs we take into account
the composition of accreted material (see \S\,5.7).

\subsection{Mass Accumulation onto White Dwarfs}  

A number of important phenomena, e.g., novae and Type Ia SN explosions or
accretion-induced collapses, are associated with mass accretion onto WDs.  
We incorporate the most recent results to estimate the accumulation efficiencies 
on WDs. 
In particular we consider accretion of matter of various compositions 
onto different WD types. We also include the possibility that NS 
formation can occur via accretion induced collapse (AIC) of a massive 
ONe white dwarf (e.g., Bailyn \& Grindlay 1990; Belczynski \& Taam 2004a).

In this section we discuss the accumulation of material and growth of 
the WD 
mass in binary systems. Only during dynamically stable RLOF 
phases can the mass accretion onto WDs be sustained for a prolonged 
period of time and hence affect the evolution of accreting WDs. 
During dynamically unstable cases (i.e., CE evolution) we assume that 
the WDs do not accrete any material. 

If dynamical instability is encountered for a binary with two white 
dwarfs we assume that a merger occurs.  Based on the results of 
Saio \& Nomoto (1998) mergers of a ONe WD with 
any type of WD companion and two CO WDs lead to either AIC and NS formation 
(if total merger mass $M_{\rm merger}$ is above $M_{\rm ecs}=1.38 \msun$) or 
the formation of the single ONe WD (with new mass equal to $M_{\rm merger}$). 
For mergers of CO WD and He WD, we assume a Type Ia SNa explosion; either 
sub-Chandrasekhar ($M_{\rm merger}< 1.44 \msun$) (see Woosley, Taam, \& 
Weaver 1986; Woosley \& Weaver 1994) or Chandrasekhar mass SN 
Ia ($M_{\rm merger}>1.44 \msun$. 
Mergers of other types of WDs have total mass below the Chandrasekhar mass, and
in particular for CO WD and H WD we assume formation of single CO WD, while 
for He WD and H WD we assume formation of single He WD with masses equal to
$M_{\rm merger}$.   
 
During a phase of sustained mass accumulation the massive ONe WD ($K=12$)
may eventually collapse to a NS. We include AIC in our standard model
calculations since it naturally follows from the adopted accumulation
physics (see below). Since little is known about potential asymmetries of
the collapse, we either apply no natal kick (standard model) or a full natal 
kick (parameter studies) obtained from Arzoumanian, Chernoff, \& Cordes 
(2002) or Hobbs et al. (2005, see also \S\,6.2). However, we also allow 
for the possibility of SN Ia explosion instead of AIC in parameter studies. 
It is also worth noting the difference between accretion and accumulation. 
The calculation of accretion rate during stable RLOF was described in \S\,5.1, 
and this rate could be used to calculate, for example, the accretion luminosity 
of the system 
(mostly in the UV part of spectrum for WD accretors). However, it is believed 
that in many cases (see below) not all of the accreted material remains on the  
surface of the accreting WD. Mass is lost either in shell flashes (nova-like 
explosions) or through optically thick winds from the surface of 
accreting WDs. To calculate the actual WD mass growth through the RLOF phase 
the accumulation efficiency, $\eta_{\rm acu}$, which is defined as
\begin{equation}
\mdot_{\rm acu} = \eta_{\rm acu} \mdot_{\rm acc}
\label{wd00}
\end{equation}
must be known. Here,  
$\mdot_{\rm acu}$ is the mass accumulation rate on the surface of WD 
and the mass accretion rate ($\mdot_{\rm acc}$) is given by eq.(~\ref{mt10b}). 
In what follows we discuss the accumulation efficiency in various 
evolutionary scenarios.  

{\em Accretion onto Helium and Hybrid white dwarfs.}
It is assumed that if the mass accretion rate $\mdot_{\rm acc}$ from the 
H-rich donor ($K_{\rm don}=0,1,2,3,4,5,6,16$) is smaller than
some critical value $\mdot_{\rm crit1}$, an unstable hydrogen shell flash 
will occur in the accreted layer.  In response, the envelope will expand 
beyond the Roche lobe of the white dwarf.  We shall assume no material is 
accumulated, and that the accumulation efficiency is $\eta_{\rm acu}=0.0$, 
i.e. the entire accreted material is lost from the binary. 
If $\mdot_{\rm acc} > \mdot_{\rm crit1}$ then the material piles up on the
WD leading to mass loss from the system. Assuming that a contact binary 
configuration is not formed in this case, the system will eventually 
undergo an inspiral. For giant-like 
donors we assume the system evolves through CE to examine if the system
survives; for all other donors we call it a merger and halt binary
evolution. The critical accretion rate is calculated as
\begin{equation}
\mdot_{\rm crit1} = l_0 M_{\rm acc}^{\lambda} (X*Q)^{-1} \mpy
\label{wd01}
\end{equation}
where, $Q=6\times 10^{18}\ {\rm erg\ g}^{-1}$ is an energy yield of hydrogen 
burning, $X$ is the hydrogen content of accreted material. For Population I 
stars (metallicity $Z>0.01$) we use $X=0.7, l_0=1995262.3, \lambda=8$, 
while for Population II stars ($Z \leq 0.01$) we use $X=0.8, l_0=31622.8, 
\lambda=5$ (Ritter 1999, see his eq. 10,12 and Table~2).
                                                                                                       
If the mass accretion rate from the He-rich donor ($K_{\rm don}=7,8,9,10,17$) 
is higher than $\mdot_{\rm crit2} = 2\times 10^{-8} \mpy$ all the material 
is accumulated ($\eta_{\rm acu}=1.0$) until the accreted layer of material 
ignites in a helium shell flash. At this point degeneracy is lifted, a  
main sequence helium star ($K_{\rm acc}=7$) is formed and further accretion 
on the helium star is then taken into account. Following the calculations of 
Saio \& Nomoto (1998) we estimate the maximum mass of the accreted shell at which 
the flash occurs as 
\begin{equation}
\Delta M = \left\{ \begin{array}{ll}
  -7.8 \times 10^4 \mdot_{\rm acc} + 0.13 & \mdot_{\rm acc} < 1.64 \times 10^{-6}  \\
  0 {\rm (instantaneous\ flash)} & \mdot_{\rm acc} \geq 1.64 \times 10^{-6}
\end{array}
\right.
\label{wd02}
\end{equation}
where $\mdot_{\rm acc}$ is expressed in $\mpy$.

The newly formed helium star may overfill its Roche lobe, in which case
either a single helium star is formed (He or Hyb WD companion, $K_{\rm 
don}=10,17$), a helium contact binary is formed (HeMS companion, $K_{\rm
don}=7$) which we assume leads to a merger or the system goes through CE 
evolution (evolved helium star companion, $K_{\rm don}=8,9$\footnote{For
$K_{\rm don}=8$ merger is assumed in such a case.}).
                                                                                                       
For accretion rates lower than $\mdot_{\rm crit2}$, accumulation
is also fully efficient ($\eta_{\rm acu}=1.0$). However, the SN Ia  
occurs at a sub-Chandrasekhar mass
\begin{equation}
M_{\rm SNIa} = -4 \times 10^8 \mdot_{\rm acc} + 1.34 \msun ,
\end{equation}
where $\mdot_{\rm acc}$ is expressed in $\mpy$.
For mass accretion rates close to $\mdot_{\rm crit2}$, the above
extrapolations from the results of Hashimoto et al. (1986) yield
masses smaller than the current mass of the accretor, and we
assume an instantaneous SN Ia explosion. We note that above explosions 
disrupt the accreting WD, and although possibly subluminous, they 
appear as Type Ia SNe (no Hydrogen).  We do not
consider the accumulation of heavier elements since they could only
originate from more massive WDs (e.g., CO or ONe WDs), which would 
have smaller radii and could not be donors to lighter He or Hyb WDs.
                                                                                                       
{\em Accretion onto Carbon/Oxygen white dwarfs.}
In this case we adopt the prescription from Ivanova \& Taam (2004). For 
H-rich donors and mass accretion rates lower than $10^{-11} \mpy$
there are strong nova explosions and no material is accumulated
($\eta_{\rm acu}=0.0$). In the range $10^{-11} < \mdot_{\rm acc} <
10^{-6} \mpy$ we interpolate for  $\eta_{\rm acu}$ from Prialnik \&
Kovetz (1995, see their Table 1). For rates higher than $10^{-6} \mpy$
all accreted material burns into helium ($\eta_{\rm acu}=1.0$).
Additionally we account for the effects of strong optically thick winds
(Hachisu, Kato \& Nomoto 1999), which eject any material accreted over the
critical rate
\begin{equation}
\mdot_{\rm crit3} = 0.75 \times 10^{-6} (M_{\rm acc}-0.4)  \mpy.
\end{equation}
This corresponds to $\eta_{\rm acu} = \mdot_{\rm crit3} / \mdot_{\rm acc}$
for $\mdot_{\rm acc} \geq \mdot_{\rm crit3}$. The accretor is allowed to
increase in mass up to $1.4 \msun$, and then explodes as a Chandrasekhar mass
SN Ia. 
In the case of He-rich donors, if the mass accretion rate is higher
than $\mdot_{\rm crit4}$ helium burning is stable and contributes 
to the accretor mass ($\eta_{\rm acu}=1.0$). For rates in the range
$\mdot_{\rm crit4} \div \mdot_{\rm crit5}$ accumulation is calculated from\\

\noindent
$\eta_{\rm acu}^{0.8} = -0.35 (\log\mdot_{\rm acc}+6.1)^2+1.02\ \ [-6.5\div-6.34]$ \\

\noindent
$\eta_{\rm acu}^{0.9} = -0.35 (\log\mdot_{\rm acc}+5.6)^2+1.07\ \ [-6.88\div-6.05]$ \\

\noindent
$\eta_{\rm acu}^{1.0} = -0.35 (\log\mdot_{\rm acc}+5.6)^2+1.01\ \ [-6.92\div-5.93]$ \\

\noindent
\[ \eta_{\rm acu}^{1.1-1.2} = \left \{ \begin{array}{l}
                           0.54 \log\mdot_{\rm acc}+4.16\ \ [-7.06\div-5.95] \\
                          -0.54 (\log\mdot_{\rm acc}+5.6)^2+1.01\ [-5.95\div-5.76] \\
                           \end{array} \right. \]

\noindent
$ \eta_{\rm acu}^{1.3} = -0.175 (\log\mdot_{\rm acc}+5.35)^2+1.03\ \ [-7.35\div-5.83]$ \\

\noindent
\begin{equation}
\eta_{\rm acu}^{1.35}  = -0.115 (\log\mdot_{\rm acc}+5.7)^2+1.01\ \ [-7.4\div-6.05] 
\label{wdacc10}
\end{equation}
and represents the amount of material that is left on the surface of
the accreting WD of a specific mass (denoted by a superscript on $\eta_{\rm acu}$
in $\msun$) after the helium shell flash cycle (Kato \& Hachisu 1999, 2004).
Logarithms of critical mass accretion rates for a given specific WD mass are given 
in square brackets: $[\log  (\mdot_{\rm crit5}/\mpy) \div \log (\mdot_{\rm crit4}/\mpy)]$. 
To obtain the accumulation rate for CO WD within the mass range $0.7-1.4 \msun$ 
we incorporate results of the closest (by mass) model from the set of 
eqs.~\ref{wdacc10}.
If the WD mass drops below $0.7 \msun$ we use $\eta_{\rm acu}=1.0$ and we set 
$\log \mdot_{\rm crit4}= \log \mdot_{\rm crit5} = -7.6$ (see Kato \& Hachisu 2004).
The mass of the CO WD accretor is allowed to increase up to $1.4 \msun$, and 
then a Chandrasekhar mass SN Ia takes place in the two above He-rich accretion
regimes. If mass accretion rates drop below $\mdot_{\rm crit5}$, the
helium accumulates ($\eta_{\rm acu}=1.0$) on top of the CO WD and once the 
accumulated mass reaches
$0.1 \msun$ (Kato \& Hachisu 1999), a detonation follows and ignites the CO
core leading to the disruption of the accretor in a sub-Chandrasekhar mass
SN Ia (e.g., Taam 1980; Garcia-Senz, Bravo \& Woosley 1999). If the mass of the
accreting WD has reached $1.4 \msun$ before the accretion layer has reached
$0.1 \msun$ then the accretor explodes in a Chandrasekhar mass SN Ia.
Carbon/Oxygen accumulation takes place without mass loss ($\eta_{\rm acu}=1.0$) 
and leads to SN Ia if Chandrasekhar mass is reached.
                                                                                                       
{\em Accretion onto Oxygen/Neon/Magnesium white dwarf.}
Accumulation onto ONe WDs is treated the same way as for CO WD accretors.
The only difference arises when an accretor reaches the Chandrasekhar mass.
In the case of ONe WD this leads to an AIC and NS formation, and binary 
evolution continues (see Belczynski \& Taam 2004a, 2004b).

\section{Spatial Velocities}

\subsection{Overview}

All stars (single and binary systems) may be initialized with 
arbitrary velocities appropriate for a given environment. For example, a  
galactic rotation curve may be used for a field population of a given
galaxy, or a velocity dispersion can be applied for a cluster population. 
The velocities of stars are then followed throughout their 
evolution. Single stars and binary systems are subject to recoil (change 
of spatial velocity) in SN explosions. Additionally, binary systems may be 
disrupted as a result of an especially violent explosion. We account for both 
mass/angular momentum losses as well as for SN asymmetries (through natal 
kicks that NSs and BHs receive at their formation; see below). The detailed description 
of SN explosion treatment is given in BKB02. Here, we only list the new additions 
to {\tt StarTrack}. The most important modification allows us to 
trace velocities of disrupted binary components after a SN explosion. 
For the first time, a full general approach with explosions taking place on orbits 
of arbitrary eccentricity (in contrast to circular orbits only) is applied to follow 
the trajectories of disrupted components. First population synthesis results are 
presented in Belczynski et al. (2006).

\subsection{Natal Kick Distribution}

At the time of birth,  NSs and possibly BHs receive
a natal kick, which is connected to asymmetries in SN
explosions. We use the distributions inferred from observed velocities of
radio pulsars. 
We have replaced the natal kick distribution used in BKB02 
(Cordes \& Chernoff 1998) with two more recent alternatives.  
One presented by Arzoumanian et al. (2002) is a bimodal distribution with a 
weighted sum of two Gaussians, one with $\sigma=90$ km sec$^{-1}$ (40\%) and 
another with $\sigma=500$ km sec$^{-1}$ (60\%).
The other was derived by Hobbs et al. (2005) and is a single Maxwellian
with $\sigma=265$ km sec$^{-1}$. According to this most recent study there is 
no indication of a bimodal (low- and high-velocity) kick distribution
claimed in earlier studies (e.g., Fryer, Burrows \& Benz 1998; Cordes \& 
Chernoff 1998; Arzoumanian et al. 2002). If this is indeed true, then some
theoretical models built in support of the bimodal kick distribution (e.g., 
Pfahl et al.\ 2002b and Podsiadlowski et al.\ 2004 model of high mass X-ray binaries) 
may need to be revised. Motions of many hundreds of pulsars are expected to be 
measured in the next few years. These measurements will provide better constraints 
on the natal kick distribution (Hobbs et al. 2005). Until then we will use both
distributions to assess the associated uncertainties in {\tt StarTrack}
calculations.

Compact objects formed without any fall back receive full kicks drawn from one of the above distributions; this case includes most of NSs (see \S\,2.3.1). The only exception are 
NS formed through ECS, for which we adopt either no natal kicks (standard model) or 
full kicks (as part of parameter studies). During accretion induced collapse of WD to NS in an 
accreting binary system, the NS is formed through the same ECS process and the same 
prescription is adopted. The recent numerical simulations of AIC and NS formation 
through ECS point towards significantly lower energies of explosion than for regular 
core collapse SNe (Dessart et al. 2006), and although the kick mechanism is not yet 
identified, these results may be also indicative of lower kicks in ECS NS formation.   
For compact objects formed with partial fallback, heavy NSs and light BHs, kicks are 
lowered proportionally to the amount of fallback associated with NS/BH formation
\begin{equation}
V_{\rm kick} = (1-f_{\rm fb}) V
\label{kick001}
\end{equation} 
where $V$ is the kick magnitude drawn from either Arzoumanian et al. (2002) 
or Hobbs et al. (2005) distribution, and $f_{\rm fb}$ is a fallback 
parameter, i.e., the fraction (from 0 to 1) of the stellar envelope that
falls back (see also \S\,2.3.1). 
For the most massive BHs, formed silently (no SN explosion) in a direct 
collapse ($f_{\rm fb}=1$) of a massive star to a BH, we assume that no 
natal kick is imparted. 
The adopted natal kick distribution and kick scaling for NSs and BHs can be 
readily changed for parameter studies (e.g., full BH kicks).

\subsection{Supernova disruptions}

Just prior to the SN explosion, the two components of the
binary move with velocities $\vec v_1^I$ and $\vec v_2^I$, which, in the 
center of mass (CM) system of coordinates, denoted here with the 
superscript $I$, satisfy
\begin{equation}
M_{\rm 1,int}\vec v_1^I + M_{\rm 2,int}\vec v_2^I =0 
\end{equation}
where $M_{\rm 1}$ denotes the SN component and  $M_{\rm 2}$ its companion. 
Subscripts ${\rm int,\ fin}$ stand for initial and final values. 

We make no assumptions about the orbit; it can have an arbitrary eccentricity,
in contrast to the derivation by Tauris \& Takens (1998), who assumed that
the orbit is circular prior to the explosion.
At the moment of a supernova explosion the orbital separation is $r_0 \vec n$. 
The exploding star loses its envelope, its mass becomes $M_{\rm 1,fin}$ and 
receives a kick $\vec w$, so now its velocity in the coordinate system $I$ is
\begin{equation}
\vec v_{\rm 1,int}^I = \vec v_1^I + \vec w\, .
\end{equation}

The secondary star may be  affected by the expanding shell 
and may receive an additional velocity $\vec v_{\rm imp}$, however it has 
been shown (Kalogera 1996) that the effect of this velocity is small, 
unless the pre-supernova orbital separation is smaller than $\simeq 3$\,R$_\odot$. 
We assume that the velocity of the companion is not affected by 
the impact of supernova ejecta. We also assume that the velocity of the 
shell is large and the shell leaves the system 
quickly, i.e. $v_{\rm shell} >> r_{\rm o} P$, where $P$ is the orbital period
of the system prior to the explosion.

In order to calculate the final velocity of the two stars we
first transform the velocities to the CM system of the two post-SN stars.
The velocity of this system, denoted as $II$, in relation to system $I$ is
\begin{equation}
 {\vec v}^{II}_{\rm CM} = { M_{\rm 1,fin}{\vec v}^I_{\rm 1,int}+ M_2 {\vec 
                          v}^I_{\rm 2,int} \over M_{\rm 1,fin}+ M_2} 
 \end{equation}
The relative velocity of the two stars in this system is
\begin{equation}
{\vec v}^{II} = {\vec v_1^I} - {\vec v_2^I}+{\vec w}-{\vec v}_{\rm imp}
\end{equation}
while the initial direction between the two stars 
remains the same as in the coordinate system $I$, $\vec n^{II}=\vec n^{I}$.
In this new system the relative motion of the stars is a hyperbola
in the plane perpendicular to the angular momentum vector:
\begin{equation}
\vec J = \mu r_0 {\vec n}^{II} \times {\vec v}^{II}\, ,
 \end{equation}
where $\mu = M_{\rm 1,fin}M_2/(M_{\rm 1,fin}+M_2)$ is the reduced mass of 
the system. It is convenient now to introduce a third coordinate system
$III$ in which  the angular momentum $\vec J$ lies 
along the z-axis. The transformation from $II$ to $III$ is a rotation
$\cal R$:  $v^{III}={\cal R}v^{II}$,$n^{III}={\cal R}n^{II}$ .
The orbit in $III$ is described by 
\begin{equation}
 r = {p \over 1 + \epsilon \cos\phi} \label{traj}
\end{equation}
where 
\begin{eqnarray}
p={J^2\over \alpha\mu} \ \ {\rm and} \ \
\epsilon=\sqrt{1+{2EJ^2\over\alpha^2\mu}},
\end{eqnarray}
with $E=\mu (v^{II})^2/2-\alpha/|r_0|$ is the (positive) 
energy of the system, and  $\alpha=G M_{\rm 1,fin}M_{2}$.
The final velocity, at $r\rightarrow \infty$, follows from
energy conservation:
\begin{equation} |\vec v^{III}_{\rm fin}| = \sqrt{2E\over \mu}.
\end{equation}

In order to find the direction of the final velocity we note that
conservation of angular momentum implies that at infinity 
($r\rightarrow \infty$): the final relative $\vec v^{III}_{\rm fin}$ 
is parallel to the direction between the stars $\vec n^{III}_{fin}$. 
The initial position of the two stars on the trajectory described by 
eq.~\ref{traj} is
\begin{equation}
 \cos\varphi_{\rm int}= {1\over\epsilon}\left({p\over r_0 - 1 }\right) .
 \end{equation}
The sign of the angle $\varphi_{\rm int}$ is negative if 
the two stars initially lie on the descending branch of the hyperbola
$\vec v^{III}_{\rm int} \vec r^{III}_0>0$ and positive  if they are on 
the ascending one $\vec v^{III}_{\rm int} \vec r^{III}_0<0$. 
In the first case, when the two stars are initially 
on the descending branch, we need to compare the distance of closest
approach on the orbit $r_{\rm min}=p/(1+\epsilon)$ with 
the radius of the companion star to examine whether the two stars
collide instead of escaping to infinity.

We obtain the  final position on the trajectory from 
\begin{equation}
 \cos\varphi_{\rm fin}=-{1\over \epsilon}
\end{equation}
and $\varphi_{\rm fin} > 0$. Thus the final direction between
the two stars   at 
$r=\infty$ is $\vec n^{III}_{\rm fin}=T(\varphi_{\rm fin}-
\varphi_{\rm int})\vec n^{III}$, where $T(\phi)$ is the matrix of rotation 
around the z-axis, and their relative velocity is:
\begin{equation}
\vec v^{III}_{\rm fin}=\sqrt{2E\over \mu} \vec n^{III}_{\rm fin}.
\end{equation}
We now have to transform quantities from system $III$ back to system $I$ to obtain 
the final velocities of the two disrupted binary components in the initial (pre-SN) 
CM system:
\begin{eqnarray}
v^{I}_{\rm 1,fin}={\cal R}^{-1}\left({-M_2 v^{III}_{\rm fin}\over M_{\rm 1,fin}+M2}
\right) + v^{II}_{\rm CM}\\
v^{I}_{\rm 2,fin}={\cal R}^{-1}\left({M_{\rm 1,fin} v^{III}_{\rm fin}\over
M_{\rm 1,fin}+M2}\right) + v^{II}_{\rm CM}.
\end{eqnarray}

\section{Distributions of Initial Parameters}

Each binary system is initialized by four parameters, which are assumed to be
independent: the primary mass $M_1$ (the initially more massive component), 
the mass ratio $q={M_2 / M_1}$, where $M_2$ is the mass of the secondary,
the semi-major axis $a$ of the orbit, and the orbital eccentricity $e$. 

For both single stars and binary system primaries, we use the initial
mass function adopted from Kroupa, Tout \& Gilmore (1993) and Kroupa \& Weidner 
(2003),
\begin{equation}
 \Psi(M_1) \propto \left\{ 
              \begin{array}{ll}
                {M_1}^{-1.3} & 0.08 \leq M_1 < 0.5 \msun \\
                {M_1}^{-2.2} & 0.5 \leq M_1 < 1.0 \msun \\
                {M_1}^{-\alpha_{\rm imf}} & 1.0 \leq M_1 < 150 \msun \\
              \end{array}
            \right.
\label{init01}
\end{equation}
where parameter $\alpha_{\rm imf}=2.35-3.2$, with our standard choice being 2.7
for field populations and 2.35 for cluster populations. 
Stars are generated within an initial mass range: $M_{\rm min} - M_{\rm
max}$, and the range is chosen accordingly based on the targeted stellar 
population. For example, NS studies would require
evolution of single stars within range $\sim 8-25 \msun$ while the 
formation of WDs would require an initial range $\sim 0.08-8 \msun$. 
Binary evolution, due to mass transfer events (both mass accretion and
mass loss) may significantly broaden any of the ranges mentioned above.  

We assume a flat mass ratio distribution,
 \begin{equation}
 \Phi(q) = 1 
 \end{equation} 
in the range $q=0-1$ in agreement with recent observational results of Kobulnicki, 
Fryer \& Kiminki (2006). However, it should be noted that massive binaries may form 
with components of comparable mass (Pinsonneault \& Stanek 2006), and we will test 
this alternative mass ratio distribution in our parameter studies.
Given the value of the primary mass
and the mass ratio, we obtain the mass of the secondary $M_2= q M_1$. 
 
The distribution of initial binary separations is assumed to be 
flat in the logarithm (Abt 1983),
 \begin{equation}
 \Gamma(a) \propto {1\over a},
 \end{equation} 
where $a$ ranges from a minimum value, such that the primary fills 
at most 50\% of its Roche lobe at ZAMS, up to $10^5 \, {\rm R}_\odot$.

Finally, we adopt the thermal-equilibrium eccentricity distribution for initial
binaries,
 \begin{equation}
 \Xi(e) = 2e,  
 \end{equation} 
in the range $e = 0-1$ (e.g., Heggie 1975; Duquennoy \& Mayor 1991).

\section{Calibrations and Comparisons}

\subsection{Mass Transfer Sequences}

In the following subsections we present {\tt StarTrack} 
mass transfer calculations and compare them to published and 
unpublished results based on the use of stellar evolution and mass transfer codes.

\subsubsection{Case B Mass Transfer: MS+HG binary}

We choose this RLOF sequence from Wellstein et al.\ (2001, their Model B) 
and start with a $16 \msun$ + $15 \msun$ ZAMS binary in a 8 day circular orbit. 
RLOF starts after the primary evolves off the MS.
The system at the onset of RLOF ($t=11.5$ Myr since ZAMS) is characterized by: 
$K_1=2,\ K_2=1,\ P_{\rm orb}=7^d.9,\ e=0, M_1=15.6 \msun,\ M_2=14.7
\msun,\ R_1=19.9 \rsun,\ {\rm and}\ R_2=11.4 \rsun$. The evolution of the
system during the RLOF phase is shown in Figure~\ref{case1}.

The RLOF phase proceeds on the thermal timescale of the donor, which is rapidly
expanding while crossing the Hertzsprung gap.  First, there is a phase 
characterized by very high mass transfer rates ($\sim 3 \times 10^{-3} 
\mpy$), until the mass ratio is reversed and the donor becomes the less massive 
binary component. Shortly thereafter, the transfer rate slowly decreases ($\sim
10^{-3} - 10^{-4} \mpy$). The mass ratio is reversed just right after the
onset of RLOF and the orbit expands. 

Our assumption is that all (100\%) of the transferred material is accreted by 
the companion (conservative evolution is adopted for a better comparison with 
Wellstein et al. 2001 calculation). RLOF terminates when the 
envelope of the donor is nearly exhausted and its radius contracts below the 
Roche lobe radius, thereby, causing the system to become detached.   
The primary loses most of its mass and becomes a core helium burning star 
($M_1=3.99 \msun,\ K_1=4$), while the secondary gains mass and is
rejuvenated ($M_2=26.2 \msun,\ K_2=1$). The orbital period increases to reach 
$\sim 80$ days at the end of the RLOF phase. Both stars continue to evolve in a
detached configuration.  Eventually, the primary becomes a naked  helium star 
($M_1=3.98 \msun,\ K_1=7$) that evolves and loses some mass ($M_1=3.5 \msun,\ 
K_1=8-9$) until the final explosion and forms a neutron star ($M_1=1.45 \msun,
\ K_1=13$). Since the mass of the primary is
quite large during the helium burning stage it does not significantly expand and in
particular it does not initiate another RLOF. The helium star primary reaches
a maximum radius of $R_1 \sim 10 \rsun$ just prior the explosion, while its Roche
lobe radius is $R_{\rm 1,lob} \sim 60 \rsun$. After the explosion the system is
disrupted due to a large natal kick the neutron star receives (e.g., Hobbs et al.
2005). At this point the secondary is still on the main sequence ($M_2=25.7 \msun,
\ K_2=1$), and will eventually (in $\sim 2$ Myr) form a single black hole 
($M_2=6.4 \msun,\ K_1=14$). 

The calculation of Wellstein et al.\ (2001) shows similar behavior during the first 
RLOF phase in terms of the duration, mass transfer rate,  and orbital period. 
The final donor and accretor masses in both simulations are almost the same. 
Therefore, our calibration and rates used for thermal and then nuclear
timescale RLOF are in very good agreement with the detailed evolutionary
calculation of Wellstein et al.\ (2001). 

However, Wellstein et al.\ (2001) find a second RLOF phase, when the primary expands again, once it becomes a helium giant ($K_1=8-9$) after it loses significant amount of mass in a stellar wind: $M_1=3.8,\ 2.8 \msun$ at helium star formation and at the start of the second (case BB) mass transfer phase, respectively. 
Their system evolves through the, so called, case BB mass transfer phase. The RLOF stops when the primary loses a significant part of its helium envelope. The system becomes wider, 
and eventually the primary explodes in type Ib/c supernova forming a neutron star.      
The difference in the evolution into a second RLOF phase is explained by the difference in modeling the helium star evolution, and in particular the stellar winds. The helium star primaries in both calculations start with about the
same mass ($M_1 \sim 4 \msun$) and their lifetimes are similar ($\sim 1$
Myr). However, in the case of Wellstein et al.\ (2001) the helium star loses
significant amount of mass: $\sim 1 \msun$, while our star loses only half
of that: $\sim 0.5 \msun$. Therefore, our helium star remains massive 
and does not significantly expand to initiate the second RLOF as it is 
found by Wellstein et al.\ (2001). The reason for this difference in wind mass loss from helium stars is that we use the downward revised empirical wind mass loss rates (Hamann \& Koesterke 1998; see also Hurley et al. 2000) that 
take into account wind clumping 
and predict rates lower by factor of $\sim 2$ than previously estimated (Hamann
Schonberner, \& Heber 1982)\footnote{Wellstein et al.\ (2001) refer to 
Wellstein \& Langer (1999) for the employed mass loss rates. In Wellstein 
\& Langer (1999) we find that both old and revised mass loss rates for helium 
stars are presented. However, it is apparent from the difference in the results
that Wellstein et al.\ (2001) evolution of their Model B binary employs the
old (high) rates.}.

\subsubsection{Case A Mass Transfer: MS+MS binary}

This RLOF sequence is selected from Wellstein et al.\ (2001, their Model A). 
We start with a $12 \msun$ + $7.5 \msun$ ZAMS binary in a 2.5 day circular orbit. 
RLOF starts while the primary still evolves through the MS phase.
The system at the onset of RLOF ($t=14.8$ Myr since ZAMS) is characterized by: 
$K_1=1,\ K_2=1,\ P_{\rm orb}=2^d.3,\ e=0, M_1=11.9 \msun,\ M_2=7.5 
\msun,\ R_1=8.3 \rsun,\ {\rm and}\ R_2=4.0 \rsun$. The evolution of the
system during the RLOF phase is shown in Figure~\ref{case2}.
In this calculation we invoke conservative evolution ($f_{\rm a}=1.0$; all mass 
lost from donor is accreted by the companion) to match the assumption 
in Wellstein et al.\ (2001). 

{\em First phase.}
At first, the RLOF proceeds on the thermal timescale with a mass transfer rate of $\sim
5 \times 10^{-4} \mpy$, through the so called rapid case A transfer phase. The 
transfer rate then rapidly decreases by more than 2 orders of magnitude until the 
component masses are nearly equal. Subsequent evolution proceeds on the much 
slower nuclear timescale of the donor with transfer rates below $10^{-6} \mpy$. 
RLOF continues until the final stages of the donor MS lifetime, when the primary 
contracts and detaches from its Roche lobe. The evolution of the orbital period 
is characterized by an initial small decrease and then (after the thermal timescale 
phase has ended) a slow but rather constant increase up to 3.5 days. At that point 
the primary mass is $\sim 6 \msun$ and the secondary mass $\sim 13 \msun$.

{\em Second phase.}
After $\sim 0.5$ Myr the primary starts expanding as it enters the Hertzsprung
gap and RLOF restarts. This mass transfer phase is much more rapid and
is driven by expansion of the primary. This phase is characterized by high 
mass transfer rates ($10^{-4} - 10^{-5} \mpy$) and the envelope of the primary 
is soon ($\sim 0.3$ Myr) exhausted, ending the second RLOF phase. During this relatively
short phase, the orbit expands significantly (final orbital period $\sim 260$
days), while the primary loses most of its mass ($M_1 \approx 1 \msun$) while the secondary, 
still on its MS, gains mass ($M_2 \approx 18 \msun$) and is rejuvenated.
The dramatic orbit expansion is an effect of the rather extreme mass ratio for this 
system at the time of the second RLOF. 
For both RLOF phases conservative evolution was applied.
The evolution of this system ends when the massive secondary evolves off MS, initiating 
a CE phase while crossing the Hertzsprung gap. This phase leads to inspiral and 
merger.

The calculation of Wellstein et al.\ (2001) shows a qualitatively similar system 
behavior during both RLOF phases; initial high mass transfer phase, then a
slower one, short break in RLOF followed by another rapid phase while the donor
evolves off the MS. Also the mass transfer rates are comparable as it is
duration of the second RLOF phase. 

However, there is a difference in the duration of the first RLOF phase, our calculation 
showing factor of $\sim 3$ longer RLOF phases than that of Wellstein et al. (2001). The 
initial rapid (on a thermal timescale) phase of RLOF lasts longer in the Wellstein et al. (2001) 
calculation, and the mass transfer rates are slightly different than
in our approximations. These result in a somewhat different mass evolution of the binary components (e.g., our secondary has reached $13.5 \msun$ at the end of first RLOF phase while in  Wellstein et al. (2001) calculation it ends up with $5.5 \msun$), and therefore different mass ratio of the system that in turn influences the subsequent mass transfer calculations and period evolution.
Additionally, we include spin-orbit coupling in our calculations. 
The period of our system after the second RLOF is longer (260 days) than $\sim 100$ days found by Wellstein et al. (2001). 

We note that Wellstein et al. (2001) account for the rejuvenation effects in detail, given that they use a stellar structure code and do not have to assume full rejuvenation as it is adopted by Tout et al.\ (1997), Hurley et al.\ (2002), and our code implementation.  Wellstein et al. (2001) remark that had full rejuvenation been assumed in their calculation the system would have ended in a CE merger during the expansion phase of the secondary after MS evolution in agreement with our findings.

\subsubsection{BH-MS binary} 

This calculation starts with a $10 \msun$ BH + $5 \msun$ ZAMS star. We let
the secondary evolve through about half of its MS lifetime before bringing 
the system into contact at $t=51.3$ Myr (counted from the secondary ZAMS) . 
The system at the onset of RLOF is characterized by:
$K_1=14,\ K_2=1,\ P_{\rm orb}=1^d.0,\ e=0,\ M_1=10 \msun,\ M_2=5
\msun,\ R_1=0.000042 \rsun,\ {\rm and}\ R_2=3.5 \rsun$.  The evolution of the 
system during the RLOF phase is shown in Figure~\ref{case3}.

{\em First phase.}                                                                                                                  
RLOF is stable and proceeds on the nuclear timescale of the secondary with a mass
transfer rate of $\sim 2 \times 10^{-8} \mpy$. Since this rate is
sub-Eddington we allow all the transferred material to be accreted onto the
primary BH, which increases its mass to $\sim 11.5 \msun$, while the
secondary mass decreases to $\sim 3.5 \msun$. During this phase, the period
increases from 1 to 2 days. The phase ends when the secondary begins 
contraction at the end of its MS life.

{\em Second phase.}
RLOF restarts when the secondary crosses the Hertzsprung gap with mass transfer
proceeding at the high rate ($\sim 10^{-6} \mpy$) corresponding 
to rapid expansion of the star on its thermal timescale during that phase. 
At some point the donor starts ascending along the red giant branch, and 
the transfer rate drops by about an order of magnitude to $\sim 3 
\times 10^{-7} \mpy$.   
Since the transfer rate is super-Eddington throughout this entire phase 
we limit accretion onto the BH to the Eddington rate, allowing 
the rest of the material to leave the binary with the specific orbital 
angular momentum of the BH. In the end the BH has increased its mass to 
$12.6 \msun$ and the mass of the donor has decreased to $0.6 \msun$. The 
orbit expands significantly ($\sim 300$ days) during this rapid RLOF phase.
 
The RLOF phase ends at the point when the donor, due to the loss of its almost 
entire H-rich envelope, stops its expansion. The system ends its life as a wide 
BH-WD binary. 

The same RLOF sequence was calculated with the detailed stellar evolution 
code of Ivanova et al.\ (2003; also see Ivanova \& Taam 2004). The comparison of the two 
phases of RLOF shows overall qualitative agreement with the {\tt StarTrack} 
calculation.  The mass transfer rates are virtually the same: $\sim 
2 \times 10^{-8}\ {\rm and} \sim 10^{-6} \mpy$, for first and second phase,
respectively. However, the detailed calculation with the evolution code 
shows a longer duration (by a factor $\sim 2$) for the first RLOF phase.

\subsubsection{BH-RG binary}

This calculation starts with a $7 \msun$ BH + $2 \msun$ ZAMS star. We let
the secondary evolve through about one third of its red giant lifetime before bringing 
the system into contact at $t=1180.4$ Myr (counted from the secondary ZAMS). 
The system at the onset of RLOF is characterized by:
$K_1=14,\ K_2=3,\ P_{\rm orb}=4^d.8,\ e=0,\ M_1=7 \msun,\ M_2=2 
\msun,\ R_1=0.000030 \rsun,\ {\rm and}\ R_2=7.1 \rsun$. The
evolution of the system during RLOF phase is shown in Figure~\ref{case4}. 

RLOF is stable and proceeds through the entire RG phase ($K_2=3$) on the nuclear
timescale of the donor. The mass transfer rate is sub-Eddington and thus the 
material transferred to the BH is entirely accreted. In the end the mass of the 
BH is increased to $8.4 \msun$ while the mass of the donor is decreased to $0.6 
\msun$. As the donor expands, ascending the RG branch, the orbit expands as 
well, and finally the RLOF phase terminates at an orbital period of $\sim 90$ 
days. The phase ends when the donor contracts upon igniting helium in its core. 
The system eventually forms a wide BH-WD binary. 

This RLOF sequence was also calculated with the detailed stellar evolution code
of Ivanova et al.\ (2003). The mass transfer rates found
in both cases are similar ($\sim 10^{-7} - 10^{-8} \mpy$) and in this case
the {\tt StarTrack} timescales are shorter, but do not differ by more than 50\%.

\subsubsection{Short period NS-RG binary}

This RLOF sequence is chosen from Tauris \& Savonije (1999, their example 2b).
We start with a $1.3 \msun$ NS + $1.6 \msun$ ZAMS star in a 3 day circular orbit.
RLOF starts while the secondary is on the RG branch ($t=2321.4$ Myr since 
secondary ZAMS) and the binary is described by:
$K_1=13,\ K_2=3,\ P_{\rm orb}=2^d.8,\ e=0,\ M_1=1.3 \msun,\ M_2=1.6 
\msun,\ R_1=0.000014 \rsun,\ {\rm and}\ R_2=4.7 \rsun$. The
evolution of the system during the RLOF phase is shown in Figure~\ref{case5}.

At first the RLOF proceeds on a thermal timescale with a highly super-Eddington 
mass transfer rate ($\sim 10^{-6} \mpy$). After the donor becomes less
massive than its accretor, the mass transfer is driven by the expansion of 
the red giant donor (on its nuclear timescale) at a much smaller rate of
$\sim 10^{-8} \mpy$. As the mass transfer rate decreases, the NS starts to 
accrete efficiently and its mass increases to $1.9 \msun$.  
Eventually, after $\sim 65$ Myr of RLOF, the RG secondary loses most of its 
mass ($M_2=0.28 \msun$) and contracts, leaving a remnant helium WD. At this point 
the RLOF phase ends (orbital period 60 days), and further evolution leads to 
the formation of wide binary, with a recycled pulsar. 
 
Comparison with the detailed evolutionary calculation of Tauris \& Savonije 
(1999) shows good agreement between the results. The detailed calculations show
an initial rapid RLOF phase followed by a sub-Eddington mass transfer phase, eventually
leading to the formation of NS-He WD binary. Final component masses (NS and
donor: $2.05$ and $0.29 \msun$, respectively) are very similar to the ones 
obtained with {\tt StarTrack}. The final orbital period of 42 days obtained 
by Tauris \& Savonije (1999) is shorter than in our calculation (60 days). 
In addition, there is a difference in the duration of RLOF phase, lasting 123 Myr 
in the Tauris \& Savonije (1999) model, as compared to 60 Myr in our 
calculations.
This may be understood in terms of a different treatment of binary interactions
(tides, magnetic braking, winds) as well as the difference in stellar models 
which may lead to a different starting point of RLOF.

\subsubsection{Long period NS-RG binary}

This RLOF sequence is taken from Tauris \& Savonije (1999, their example 2c).
We start with a $1.3 \msun$ NS + $1.0 \msun$ ZAMS star in a 60 day circular
orbit. RLOF starts while the secondary is on the RG branch ($t=12312.5$ Myr since 
secondary 
ZAMS) and the binary is described by:
$K_1=13,\ K_2=3,\ P_{\rm orb}=60^d.708033,\ e=0,\ M_1=1.3 \msun,\ M_2=0.98
\msun,\ R_1=0.000014 \rsun,\ {\rm and}\ R_2=30.5 \rsun$. The 
evolution of the system during RLOF phase is shown in Figure~\ref{case6}.
                                                                       
RLOF is highly super-Eddington and driven by the expansion of the
donor on a nuclear timescale.
Only shortly before the system detaches as a result of the exhaustion 
of the donor's envelope, the transfer rate becomes sub-Eddington. As a result, 
the donor loses most of its mass ($M_2=0.4 \msun$) while the NS hardly accretes 
any 
material ($M_1=1.43 \msun$). The orbit expands throughout this phase with the 
orbital period increasing to over 300 days. The system eventually forms 
a wide NS-He WD binary, with a potential recycled pulsar (the NS has accreted $\sim 0.1 
\msun$).

The above results are very similar to the calculations of Tauris \& Savonije
(1999), who obtain a $1.5 \msun$ NS with a $0.4 \msun$ NS-He WD binary in 
a 382 day orbit. The mass transfer rates and duration of the RLOF phases are 
similar in both calculations.

\subsubsection{Long period NS-He star binary}

This RLOF sequence follows from Dewi \& Pols (2003, see their Fig.~1).
We start with a $1.4 \msun$ NS + $2.8 \msun$ ZAMS He star in a 10 day 
circular orbit.  RLOF starts while the secondary is already an evolved He 
star ($t=2.9$ Myr) and the binary is described by:
$K_1=13,\ K_2=8,\ P_{\rm orb}=9^d.804617,\ e=0,\ M_1=1.4 \msun,\ M_2=2.5
\msun,\ R_1=0.000014 \rsun,\ {\rm and}\ R_2=15.4 \rsun$. The 
evolution of the system during RLOF phase is shown in Figure~\ref{case7}.

RLOF proceeds on the donor's thermal timescale throughout the entire phase. 
The very high mass transfer rate ($6 \times 10^{-3} \mpy$) makes this phase
very short  and RLOF stops after the envelope of He star is exhausted. Since
the mass transfer rate is highly super-Eddington, the NS hardly accretes
any material while the donor loses its entire He-rich envelope ($M_2=1.65 \msun$). 
The orbital period at first decreases to reach a minimum at 8.9 days, and then  
increases to 9.5 days at the end of RLOF phase.
After the phase of RLOF the secondary core explodes in SN Ic leading to 
double neutron star formation (provided that a natal kick does not disrupt
the binary). This result was presented also in Ivanova et al.\ (2003). 

Dewi \& Pols (2003) calculated mass transfer rates spanning the range: $10^{-4}
- 10^{-2} \mpy$. Our rate is constant and close to the high end of the Dewi
\& Pols (2003) range.
We have adopted a constant mass transfer rate following Paczynski (1971) who pointed 
out that thermal timescale mass transfer rates do not vary by more than factor of 2-3 
(for details see \S\,5.3). This system may appear as an X-ray binary during 
this phase. However, the chances of catching it at this phase are very small, since 
the thermal timescale mass transfer is very short. Besides, in this case the X-rays 
may be significantly degraded because of high optical depths (material shed out of 
the system). On the other hand, some of these sources might appear to be soft 
$\gamma$-ray emitters (i.e. $20-100$ keV range, tail of X-ray emission) with high 
intrinsic absorption, and the discovery of objects 
with these broad characteristics (see e.g., Dean et al. 2005) lends some hope for
detecting this phase of binary evolution.  
The results from Dewi \& Pols (2003) reveal a different period evolution than 
in our simulation; RLOF starts at higher value (10.46 days), and then decreases to 
10.37 days. However, the period changes in both calculations are rather small, and 
are probably related to our consistently high mass transfer rate throughout the RLOF 
phase.  This leads to higher mass and angular momentum loss from the binary which
determines the orbit evolution. Additionally, we include tidal
interactions between binary components (see \S\,3.3 and \S\,8.2 ).  
These differences between models for low mass helium stars were already 
noted by Dewi \& Pols (2003).

\subsubsection{Short period NS-He star binary}

We choose this RLOF sequence from Dewi \& Pols (2003, see their Fig.~3).
We start with a $1.4 \msun$ NS + $3.6 \msun$ ZAMS He star in a 0.6 day 
circular orbit.
RLOF starts while the secondary is already an evolved He star ($t=2.0$ Myr) 
and the binary is described by:
$K_1=13,\ K_2=8,\ P_{\rm orb}=0^d.59,\ e=0,\ M_1=1.4 \msun,\ M_2=3.2 
\msun,\ R_1=0.000014 \rsun,\ {\rm and}\ R_2=2.4 \rsun$. The
evolution of the system during RLOF phase is shown in Figure~\ref{case8}.

The RLOF proceeds on the donor's thermal timescale with a mass transfer rate of $\sim
10^{-3} \mpy$ until the envelope of He star is almost exhausted. Since the 
mass transfer rate is highly super-Eddington, the NS does not accrete
much material while the donor loses most of its He-rich envelope ($M_2=2 \msun$).
The orbital period decreases from 0.6 to $\sim 0.4$ days at the end of 
RLOF phase.
After the RLOF phase ceases the secondary explodes in SN Ib/Ic leading to a
close double neutron star system (again provided that a natal kick does not 
disrupt the binary).

The Dewi \& Pols (2003) RLOF sequence for this case is very similar to our
calculation. They find a period decrease (from 0.65 to 0.47 days) and
a high constant mass transfer rate of a few $\times 10^{-4} \mpy$. The inspiral
phase and CE is not expected in this case, and therefore further evolution
may lead to a close double neutron star formation.

\subsection{Tidal Evolution Calibration}

Whenever coeval binary populations in nearby clusters are observed to  
constrain the circularization rate, it is found that standard tidal
dissipation theories do not match the data (see Meibom \& Mathieu  
2005 for a recent review).
In all cases an increase in the tidal dissipation rate appears necessary
(Claret \& Cunha 1997; Terquem et al.\ 1998). Depending on which  
theory is used, the increase needed in the overall efficiency of tidal 
dissipation is by a factor $\sim 10 - 100$.

We have used {\tt StarTrack} models to calibrate our theoretical  
treatment by comparing them against observations of
{\em (i)} the cutoff period  for circularization in a population of  
MS binaries (in M67), and {\em (ii)} the orbital decay accompanying 
tidal  synchronization in a high mass X-ray binary (LMC X-4).
The results, presented in two following subsections, confirm
that tidal dissipation, at least in case of convective stars, is more 
effective than predicted by our simple theory. Therefore, in all our 
standard model calculations, we will use an increased rate of tidal 
dissipation for convective stars, corresponding to $F_{\rm tid,con}=50$,
while using standard dissipation for radiative stars $F_{\rm tid,rad}=1$, 
but we will also allow for even more effective tidal dissipation rates 
in our parameter studies (all the way to $F_{\rm tid,con}=100$ and 
$F_{\rm tid,rad}=100$). See \S\,3.3 for our implementation of tidal 
dissipation theory and the definition of $F_{\rm tid}$.

\subsubsection{Cutoff Period for M67}

Open star clusters have often been used to test tidal interaction theories
(Mathieu et al.\ 1992; Meibom \& Mathieu 2005). Observations of single-
and double-line spectroscopic binaries allow estimates of the periods and
eccentricities for a number of systems within clusters. It was expected and
then confirmed that the cutoff period ($P_{\rm cut}$, the longest  
period of a circular binary) should increase with the age of the cluster. The  
tidal dissipation depends strongly on the orbital separation and therefore 
the wider, longer-period binaries will take a greater time to circularize.
In principle, with knowledge of the initial conditions in a given cluster, 
the observed value of the cutoff  period may be used to calibrate the 
efficiency of tidal interactions. In practice, binaries within a given 
cluster form with eccentricities, separations and angular momenta which are 
not precisely known. In addition, the observed samples may suffer from small 
number statistics (the observed cutoff periods are only lower limits), 
rendering such a calibration quite uncertain. However, we can use the 
cutoff-period observations to provide at least an order of magnitude 
estimate for the factor by which any standard theoretical estimate must be 
increased.

M67 is an old open cluster with an age of $3.98\,$Gyr and observed cutoff
period of $10-12\,$d (Mathieu, Latham \& Griffin 1990; Mathieu et al.\ 1992) 
and a solar metallicity stellar population (Janes \& Phels 1994). The period 
was estimated for a  sample of MS binaries with components close to the 
cluster turnoff mass. Recently Meibom \&  
Mathieu (2005) proposed a new way to estimate the point of transition from  
circular to eccentric systems. Instead of a simple cutoff period, they use a  
new estimator called the ``tidal circularization period.'' This period is
found from fitting a special function which mimics the tidal circularization 
isochrone of the most frequently occurring eccentric binary orbits for a
given cluster. They find that the tidal circularization period for M67 is 
$12.1\,$d.

Several calculations, with different efficiencies of tidal dissipation,
were performed to try to reproduce the binary population of the open  
cluster M67. In each calculation we have evolved $10^4$ binaries at solar  
metallicity with component masses in the $0.7-1.4 \msun$ range, requiring 
that the mass ratio be greater than 0.5. The limits are somewhat arbitrary, 
but chosen to include the population of bright MS stars observed in M67. 
Most of these stars have convective envelopes, and therefore we try to
calibrate the scaling factor for convective envelopes ($F_{\rm tid,con}$) while
keeping the one for radiative stars constant ($F_{\rm tid,rad}=1$). 
The initial distributions were chosen as in our  standard
evolutionary model (see \S\,5.7), but with IMF exponent
$\alpha_{\rm imf}=2.35$, which is more appropriate for clusters (Kroupa
\& Weidner 2003).

In Figure~\ref{m67} we show synthetic binary MS populations
in the period--eccentricity plane corresponding to an evolution with
different efficiencies for the tidal interaction.
As expected we see that the cutoff period increases for more  
efficient tidal
interactions, $P_{\rm cut} \simeq 4, 7, 10\,$d for $F_{\rm tid,con}=1, 10,
100$, respectively. It is found that only for significantly increased 
dissipation ($F_{\rm tid,con} \gtrsim 10-100$) the the predicted cutoff period 
approach the observed value of 10-12 days. An additional calculation 
with $F_{\rm tid,con}=1000$ results in a cutoff period of $\sim 16$ days, now  
clearly higher than the observed value.

\subsubsection{Orbital decay of LMC X-4}

Levine, Rappaport \& Zojcheski (2000) measured an orbital period decay
for the high mass X-ray binary (HMXB) LMC X-4. The system consists of
a $1.3 \msun$ NS and a massive $15.6 \msun$ companion in a 1.4-day circular 
or almost circular orbit (Woo et al.\ 1996; van der Meer et al.\ 2005).
The X-ray emission in HMXBs is believed to arise from wind accretion
onto the compact object; however it was also suggested that some systems 
may be in an atmospheric RLOF phase (e.g., Kaper 2001).
For wind-fed detached systems, the orbital decay may be directly
connected to the tidal interaction of the HMXB components.
The secondary is 
a massive star, and the source of tidal dissipation is radiative damping. 
Therefore, we use LMC X-4 to check the efficiency of tidal 
interactions for radiative envelopes ($F_{\rm tid,rad}$).  The rotation 
of the massive component decreases with time as it expands during its
evolution. On the other hand, the tidal forces act to synchronize the massive 
component, resulting in loss of orbital energy and angular momentum,
i.e., decay of the orbit. 

If, in fact, LMC X-4 is a wind-fed system and not in RLOF, then the
massive star must be smaller than its Roche lobe $R_{\rm roche} = 8
\rsun$. A $15.6 \msun$ star exceeds that size, while still on MS, after
about $10.5\,$Myr of evolution (from the ZAMS). Subsequent RLOF is
dynamically unstable (extreme mass ratio) and leads to a rapid merger
of the binary components, terminating the HMXB phase.
We perform a set of calculations for a synthetic binary similar to
the LMC X-4 using our standard model parameters, with a metallicity
appropriate for the LMC ($Z=0.007$). We {\em assume} that the binary 
configuration is detached and we calculate the rate of orbital decay. 
The orbital decay rate depends crucially on the current relative radius 
of the massive component of LMC X-4 ($\propto (R/a)^8$, see eq.~\ref{tid1}).

The radius of the $15.6 \msun$ 
star ($Z=0.007$) increases from $R_2=4.5 \rsun$ on ZAMS to $R_2=13 \rsun$ 
at the end of MS phase, which takes $\sim 13$ Myr. Based on the 
Roche lobe radius of secondary for 1.4 day orbit  
the secondary fills its Roche lobe at $\sim 11$ Myr, 
which is close to the end of MS phase. Since the primary has already 
evolved and has formed a NS, a significant amount of 
time must have elapsed  since the binary formation. For example a $30-35 \msun$ star takes 
$\sim 6$ Myr to form compact object, and such a massive star would have 
formed a NS only if stripped of a significant part of its mass in RLOF 
episode. For more massive primaries, the evolution would be slightly
faster ($\sim 4-5$ Myr), but they would more likely have formed BHs. 
So on one hand the secondary cannot be older than $\sim 11$ Myr 
($R_2=8 \rsun$ and $R_2/R_{\rm 2,lob}=1$) and most likely it is not younger 
than $\sim 6$ Myr ($R_2=6 \rsun$ and $R_2/R_{\rm 2,lob}=0.75$). We conclude that 
the secondary is in the late stage of its evolution on the main sequence and 
probably close to filling its Roche lobe (see also Levine et al. 2000). 

We perform the set of calculations for different radii of the massive
component of LMC X-4 ($R_2/R_{\rm 2,lob}=0.75-0.9$) for various 
efficiencies of tidal interactions ($F_{\rm tid,rad}=1,10$). 
The results are shown in Figure~\ref{lmcx4}. 
The orbital decay rate increases with time as the massive component 
expands along the MS and approaches its Roche lobe. The time to reach 
contact (at which point calculations are stopped) decreases with increasing
effectiveness of tidal forces. For comparison we show the
observed orbital decay for LMC X-4, which falls within the model with
standard tidal interactions efficiency ($F_{\rm tid,rad}=1$).
We conclude that in the case of LMC X-4 there is no need for the increased
efficiency of tidal interactions, and therefore we adopt $F_{\rm tid,rad}=1$ 
for massive stars with radiative envelopes for our standard model value.

\section{X-ray modeling}

\subsection{X-ray luminosity calculations}

In our study we consider only accreting binaries with NS and BH primaries, 
which are brighter than some X-ray luminosity cut $L_{\rm x,cut}$. This cut 
may correspond to a detection limit of a particular set of observations. 
Typical $L_{\rm x,cut}$ values for most current {\em Chandra} observations
are in the range $10^{34}-10^{36}$ erg s$^{-1}$. 
At these high luminosities in the Chandra sensitivity range ($\sim 0.3-7$ keV)
the only WD accretors will be supersoft sources, which are easily
identifiable from their X-ray spectra and are thought to have most of their
X-ray emission coming from nuclear burning rather than gravitational energy
release (see Kuulkers et al. 2003 for a review of the X-ray properties of
WD accretors).
Although, for some deep Galactic exposures {\em Chandra} has reached levels  
of $\sim 10^{30}$ erg s$^{-1}$ (e.g., Galactic Center image of Muno et al.\ 
2003) and a contribution from cataclysmic variables may also become important.  
The calculation of X-ray luminosities of systems with WD accretors is described 
in a separate study (Ruiter, Belczynski \& Harrison 2006).

Binary companions to NS/BHs may lose material either through a stellar wind or via 
RLOF. In the latter case, the donors transfer all the material toward the 
accretor, whereas for the wind-fed systems only a fraction of the material 
is captured by the compact object.  We calculate the bolometric luminosity 
($L_{\rm bol}$) arising from the accretion onto a compact object. The 
accretion rate is based on the secular averaged mass accretion rate. If a 
system is detached then we use the wind mass accretion rate (eq.~\ref{wind02}), 
and if system is semi-detached the RLOF accretion rate is used 
(eq.~\ref{mt10b}). 
We do not calculate X-ray luminosities arising from the accretion in 
dynamically unstable phases, since the timescales are very short and 
additionally X-ray emission would be highly absorbed due to large optical 
depths in the CE. The $L_{\rm bol}$ is calculated from 
\begin{equation}
L_{\rm bol} = \epsilon {G M_{\rm acc} \mdot_{\rm acc} \over R_{\rm acc}}
\label{lx01}
\end{equation}
where the radius of the accretor is 10 km for a NS and 3 Schwarzschild radii 
for a BH, and $\epsilon$ gives a conversion efficiency of gravitational 
binding energy to radiation associated with accretion onto a NS (surface 
accretion $\epsilon=1.0$) and onto a BH (disk accretion $\epsilon=0.5$). 

For RLOF-fed systems we make a distinction between persistent and 
transient X-ray sources. All wind-fed systems are considered as persistent 
X-ray sources. The issue of the wind-fed XRBs with massive B$e$ companions 
and their outburst behavior is discussed in \S\,9.2.
 
RLOF-fed systems are subject to a thermal disk instability and may appear 
either as persistent or transient X-ray sources depending on the mass 
transfer rate. A system becomes a transient X-ray source when the RLOF 
rate falls below a certain critical value $\mdot_{\rm disk}$. We use 
the work of  Dubus et al.\ (1999) for H-rich disks (see their eq.30) 
and the study of Menou, Perna, \& Hernquist (2002) for disks with heavier 
elements (see their eqs.1--4)

\begin{equation}
 \mdot_{\rm disk} = \left\{ \begin{array}{ll}

1.5 \times 10^{15} M_{\rm acc}^{-0.4} R_{\rm disk}^{2.1} 
C_1^{-0.5} g s^{-1} & H-rich\\
&\\
5.9 \times 10^{16} M_{\rm acc}^{-0.87} R_{\rm disk}^{2.62}
\alpha_{0.1}^{0.44}\ g s^{-1} & He-rich \\ 
&\\
1.2 \times 10^{16} M_{\rm acc}^{-0.74} R_{\rm disk}^{2.21}
\alpha_{0.1}^{0.42}\ g s^{-1} & CO-rich \\
&\\
5.0 \times 10^{16} M_{\rm acc}^{-0.68} R_{\rm disk}^{2.05}
\alpha_{0.1}^{0.45}\ g s^{-1} & O-rich, \\

\end{array} \right.
\label{lx03}
\end{equation}
where $M_{\rm acc}$ is accretor mass in $M_\odot$, $R_{\rm disk}$ is a
maximum disk radius (2/3 of accretor Roche lobe radius) in $10^{10}$
cm. Constants are: $C_1=C/(5 \times 10^{-4})$, with C being radiation
parameter of typical value $5 \times 10^{-4}$; $\alpha_{0.1}=\alpha/0.1$,
with $\alpha$ being a viscosity parameter. Following Menou et al.\ (1999)  
we adopt $\alpha=0.1$ for all types of donors since there is empirical
evidence from dwarf nova outbursts that this is the right order of magnitude 
for the viscosity parameter. The same value of $\alpha$ is used to derive the
critical mass transfer rate for H-rich disks (Dubus et al.\ 1999). H-rich 
donors are the stars with types $K_{\rm i}=0,1,2,3,4,5,6,16$, He-rich donors are 
$K_{\rm i}=7,8,9,10,17$, CO-rich donors are $K_{\rm i}=11$, while we apply 
formulas for O-rich type donors to ONe WDs ($K_{\rm i}=12$).

We adopt a semi-empirical approach to calculate quiescent X-ray luminosities of 
transient NS RLOF-fed sources, since little is known about the emission 
mechanism 
during quiescence. It is not certain if the emission arises from a low level
accretion or a deep crustal heating (for a detailed discussion see Belczynski 
\& Taam 2004b, and references therein).   
Using X-ray studies of Galactic transient systems with NS accretors (e.g., 
Tavani \& Arons 1997; Rutledge et al. 2001; Campana \& Stella 2003; Jonker,
Wijnands \& van der Klis 2004; Tomsick et al.\ 2004; Campana 2004), we adopt 
$10^{31}$ erg s$^{-1}$ 
as a lower limit for the hard X-ray luminosity, above 2 keV. However, it was 
shown that the average luminosity level can be higher $\gtrsim 10^{32}$ erg 
s$^{-1}$ (e.g., Rutledge et al.\ 2002; Jonker et al.\ 2004). We adopt an X-ray 
luminosity level of $10^{31} - 10^{32}$ erg s$^{-1}$ above 2 keV. Furthermore, 
we assume that the quiescent NS transient X-ray luminosities are evenly 
distributed (in $\log L_{\rm x}$) in the above range.

The quiescent emission from BH transient systems 
is likely related to a low level of mass accretion. Recent observations
of BH transients in their quiescent states (Tomsick et al.\ 2003) reveal rather 
hard spectra that are not well described by a black body. The observed luminosities
are found in the range $\sim 10^{30} - 10^{33}$ erg s$^{-1}$ with a 
median luminosity $\simeq 2 \times 10^{31}$ erg s$^{-1}$. For BH systems
we also use a semi-empirical approach, and we assume that most (80\%) of the
quiescent BH transient X-ray luminosities above 2 keV are evenly distributed 
in the $10^{30} - 10^{32}$ erg s$^{-1}$ range, while the rest (20\%) of the 
systems are slightly brighter: luminosities evenly distributed in the $\sim 
10^{32} - 10^{33}$ erg s$^{-1}$ range  (see Fig.\,3  of Tomsick et al.\ 2003).
Both of the above distributions are uniform in $\log L_{\rm x}$.
There are some indications that the highest quiescent luminosities are
found in the longest period systems (e.g. Garcia et al. 2001), but we do not
implement this effect until confirmed by more observations. 

RLOF-fed transient systems in outburst reach high (close to Eddington) X-ray 
luminosities. We introduce a factor $\eta_{out}$ describing the fraction of the 
critical Eddington luminosity a given system has reached. 
The long period systems, with orbits that are sufficiently extensive for  
a large accretion disk to be formed, are usually found to emit at the 
Eddington luminosity ($L_{\rm edd}$) during outburst, while the outburst 
luminosities of short period systems are lower by about an order of magnitude. 
The correction factor to an X-ray luminosity at outburst corresponding to 
$\eta_{\rm out}=0.1$ and $\eta_{\rm out}=1$ for the short and long period 
systems is applied respectively. The critical periods, above which the 
Eddington luminosity is adopted, are taken to be 1 day and 10 hrs for NS and 
BH transients in outburst, respectively (Chen, Shrader \& Livio 1997; Garcia et 
al. 2003; see also appendix A1 in Portegies Zwart, Dewi \& Maccarone 2004). 

In order to decide if a given transient system is in an active (outburst)
state or inactive (quiescent) state the disk duty cycle
($DC_{\rm disk}$; the fraction of a time a given system spends in the 
outburst) must be known. However, the disk instability theory cannot provide 
a reliable estimate of $DC_{\rm disk}$.  Empirically it is thought that 
$DC_{\rm disk} \lesssim 1\%$ (e.g., Taam, King \& Ritter 2000). We adopt 
$DC_{\rm disk}=1\%$ (probability of finding a system in outburst) in our 
calculations and use Monte Carlo to decide the state of a transient system. 
In practice when we study a stellar population the information for all
X-ray binaries is extracted at some specified time (time slice). Once 
a given system is identified as a transient (see eq.~\ref{lx03}) a random number 
(flat probability distribution) is drawn from the range 0--1. If the number is 
smaller than 0.01 (1\% probability) the system is then in outburst,
otherwise it is in quiescence. The appropriate X-ray luminosity is then
assigned to the system (see eq.~\ref{lx05}). 
Alternatively, we use a phenomenological model for the duty cycle
developed by Portegies Zwart et al.\ 2004. The model is based on the
observations available for the Galactic BH transient systems. In particular 
comparison of the recurrence time and the decay time combined with the
observed peak outburst energy allows to calculate the time in which system 
is brighter than a certain critical X-ray luminosity. Specific application 
of that model will be discussed in the forthcoming paper on the evolution 
of X-ray luminosity function in starburst galaxies (Belczynski et al.\ 2006, 
in preparation).

Finally, the bolometric accretion luminosity is converted to an X-ray
luminosity in a specific energy range. We perform the conversion to the 
0.3 -- 7 keV range, which may be used directly for comparison with {\em 
Chandra} observations.  
For all the persistent RLOF-fed sources, all wind-fed sources and the 
transients in the outburst stage, where accretion is the dominant
contributor to the observed luminosity, we apply a bolometric correction 
($\eta_{\rm bol}$). For all quiescent transients the bolometric correction 
is not needed since we adopted their X-ray luminosities directly from 
observations. For different types of systems we estimate the correction 
to be:

\begin{equation}
 \eta_{\rm bol} = \left\{ \begin{array}{ll}
        0.15 &  {\rm NS:\ wind:\ all} \\
        0.55 &  {\rm NS:\ RLOF:\ pers.,\ outburst\ trans.}\\
        0.8  &  {\rm BH:\ wind:\ all} \\
        0.8  &  {\rm BH:\ RLOF:\ pers.,\ outburst\ trans.}\\
 \end{array} \right.
\label{lx04}
\end{equation}
Corrections were obtained from: La Barbera et al. (2001) for wind-fed 
NS systems; from Di Salvo et al. (2002) and Maccarone \& Coppi (2003) for 
RLOF-fed NS systems; and from Miller et al. (2001) for BH systems.
These bolometric corrections will be applicable for the typical {\em
Chandra} observations of external galaxies. For deeper observations, where 
the lower luminosity cutoffs are below a few percent of the Eddington limit, 
the objects make spectral state transitions (see Maccarone 2003 and 
references within), and the bolometric corrections are much
larger\footnote{However, note that the quiescent X-ray luminosities are not
affected since they are adopted directly from the deep observations.}.

Combining all of the above information, we can calculate the X-ray 
luminosity of synthetic X-ray binaries from 
\begin{equation}
 L_{\rm x} = \left\{ \begin{array}{ll}
        10^{31} - 10^{32} & {\rm all\ quiescent\ NS\ transients} \\ 
        10^{30} - 10^{32} & {\rm 80\%\ quiescent\ BH\ transients} \\         
        10^{32} - 10^{33} & {\rm 20\%\ quiescent\ BH\ transients} \\
\eta_{\rm bol} \eta_{\rm out} L_{\rm edd} & {\rm outburst\ NS/BH\ transients} \\
\eta_{\rm bol} L_{\rm bol} & {\rm persistent\ (RLOF\ and\ wind)} \\
 \end{array} \right.
\label{lx05}
\end{equation}
where $L_{\rm x}$ is expressed in ${\rm erg\ s}^{-1}$ and $L_{\rm edd}$ represents 
the Eddington luminosity. 
Note that the X-ray luminosity is calculated directly from the mass transfer rate only for 
persistent sources. On the other hand, we adopt the above empirical description for 
transient sources since the relation between the quiescent, outburst, bolometric 
luminosities, and duty cycle are uncertain due to the mass loss from the 
system during the outburst state.  Evidence for such mass loss in the 
form of jets and/or wind have been observed in, for example, a Galactic BH transient  
GRS 1915-105 (Dhawan, Muno \& Remillard 2005; Truss \& Done 2006 ).

\subsection{High Mass X-ray Binaries: Be Star Transients}

\subsubsection{Observational Overview}

High mass X-ray binaries consist of a compact object (either
a NS or a BH) orbiting a massive star. Both galactic and extra-galactic
populations of HMXBs are known (Liu, van Paradijs \& van den Heuvel
2000, 2005). The majority of HMXBs (about $2/3$; see Liu et al.\ 2000, 
2005; Hayasaki \& Okazaki 2005) are so-called Be/X-ray binaries, in which 
the primary is a Be star, orbiting a magnetized NS. Orbits are generally wide 
with a moderate eccentricity. The compact star accretes from the wind of a
massive main sequence or subgiant Be (spectral types B3-O with Balmer 
emission lines; Zorec \& Briot 1997) companion. Many of these systems show 
transient behavior (see below). The remaining HMXBs are those in which the primary 
is a supergiant, so called SG/X-ray binaries (e.g., Liu et al.\ 2000). 
For these systems the compact object either accretes from the wind of 
the supergiant, or in brighter systems through RLOF (possibly atmospheric but not
always) via an accretion disk.

If indeed some HMXBs are confirmed to be evolving through stable RLOF, it
should pose a useful constraint on the development of a delayed dynamical
instability. In general, it is expected that mass transfer from a much more
massive donor to a low mass companion is dynamically unstable and leads to
the formation of a CE (see \S\,5) that ends HMXB phase.  
It has been shown that if a H-rich donor is $\sim 3$ times more massive 
than a compact star accretor (see \S\,5.1) the RLOF will lead to CE phase. 
For adopted the maximum NS mass adopted here ($2.5 \msun$) we predict that only stars of
spectral type later than B3 (masses smaller than $7.5 \msun$) could be in
dynamically stable RLOF with NS accretors. If a higher mass donor is found in a HMXB
with a solid case for ongoing RLOF, then either {\em (i)} compact object mass is
higher (e.g. BH), or {\em (ii)} the system is in the phase of short-lived
atmospheric RLOF and will soon end up in CE phase, or {\em (iii)} the
understanding of development of dynamical instability is incomplete and 
the observations could be used to set new limits.   
 
Some Be/X-ray binaries (B$e$ XRBs) are persistent sources (varying by
less than a factor of $\sim 10$) observed at low luminosity levels
$L_{\rm x} \sim 10^{32} - 10^{34}$ erg sec$^{-1}$ (e.g., Van Bever \& 
Vanbeveren 2000; Okazaki \& Negueruela 2001).
However, most B$e$ XRBs show periodic outbursts and are called
transient B$e$ XRBs. Transient B$e$ XRBs exhibit two different types of
outbursts (e.g., Bildsten et al. 1997; Okazaki \& Negueruela 2001;
Hayasaki \& Okazaki 2005; Baykal et al. 2005): \\
-- Type I (normal) outbursts, which are of moderate intensity ($L_{\rm x} 
\sim 10^{36} - 10^{37}$ erg sec$^{-1}$) and they appear to be related
to the orbital period. It is generally accepted that these outbursts
are associated with the periastron passage of a NS, and are explained
by the increased accretion from the B$e$ star wind at periastron. \\
-- Type II (giant) outbursts, with luminosities reaching $L_{\rm x} \gtrsim 
10^{37}$ erg sec$^{-1}$, are irregular, and although they seem to
appear shortly after the periastron passage, they do not exhibit any
other correlations with the orbital period.
Although the origin of the Type II outbursts remains unknown, it was
suggested that the outflow from the B$e$ star may lead to the formation of a 
transient accretion disk around the NS. Disk accretion results in higher 
X-ray luminosities than direct surface wind accretion (see Bildsten et 
al.\ 1997 for a discussion and references). Some systems show both types 
of outbursts, e.g., A 0535+262 (Motch et al.\ 1991; Finger, Wilson \& 
Harmon 1996), V0332+53 (Stella, White \& Rosner 1986) or 4U 0115+634 
(Baykal et al.\ 2005).

\subsubsection{Modeling}

Type I outbursts are averaged out of our calculations if we use the 
orbit-averaged wind accretion model (see \S\,4.2). 
In the general (arbitrary eccentricity) wind accretion model (see \S\,4.1) 
Type I outbursts are a natural outcome. However, it was noted (Avni \& Goldman 
1980) that the transient phenomenon may be difficult to explain. 

We construct a simple phenomenological model for Type II 
outbursts in order to be able to assess the influence of this transient 
activity on XRB population characteristics. 
For a system to be a potential Type II B$e$ XRB outburster we require:\\
-- a binary with  a NS or a BH accretor and a massive MS ($K_{\rm i}=1$) or 
   subgiant ($K_{\rm i}=2$) donor ($M \geq 8 \msun$, spectral type earlier than 
   B3),\\
-- that the system is tight enough so it appears as a HMXB with a persistent 
   (outside outbursts) wind accretion leading to an X-ray luminosity 
   greater than $L_{\rm x,Be}$. We allow $L_{\rm x,Be}$ to change within 
   the range $10^{32}-10^{34}$ erg sec$^{-1}$.

Furthermore, only a fraction ($f_{\rm Be}$) of donors in the above binaries 
are B$e$ stars (as opposed to a regular B stars), and can potentially trigger  
the Type II outbursts. To provide an upper limit on the contribution of 
bursting HMXBs to the XRB population one may choose $f_{\rm Be}=1$. 
For detail studies, the value of $f_{\rm Be}$ may be constrained based on the age of
a massive star (McSwain \& Gies 2005) or its spectral type and luminosity
class (Zorec \& Briot 1997).
Since little is known about the duty cycle of Type II outbursts, we 
allow the duty cycle to change within a wide range $DC_{\rm Be}=0.1-0.5$ 
and use Monte Carlo to decide whether the system is in outburst or in 
quiescence. Here, $DC_{\rm Be}$ gives the fraction of a time a given system 
spends in the outburst.
An orbit averaged X-ray luminosity (direct wind accretion) is used for 
quiescence ($\eta_{\rm bol}=0.15,0.8$  \S\,9.1), although thermal emission
from a NS is also observed in some systems. 
For systems in the Type II outburst the X-ray luminosity is taken to be uniformly 
distributed in the 
range $L_{\rm x} = 10^{37} - 10^{38}$ erg sec$^{-1}$. We adopt bolometric 
correction factors: $\eta_{\rm bol}=0.15,0.8$ for NS and BH accretors, 
respectively (see \S\,9.1).

The X-ray modeling will be further developed as we proceed with the studies of the 
Galactic and extragalactic X-ray binary populations (e.g., Belczynski et al.\ 2006, 
in preparation).

\section{Summary}

We have presented a detailed description of the updated {\tt StarTrack} 
population synthesis code. The code is being used to study populations 
of different varieties of binaries hosting compact objects. The code 
has been calibrated and tested against different sets of observations and 
detailed evolutionary calculations and the results are presented here. 
The updated version of {\tt StarTrack} was already used in several studies 
of compact object binaries and XRBs. {\tt StarTrack} allows for evolution 
of stellar systems with a wide variety of different initial conditions (IMF, 
metallicity, star formation history) and for a number of different 
evolutionary models, subject to the parametrization of the input physics.

The {\tt StarTrack} code can be compared to the {\tt BSE} population synthesis 
code (Hurley et al.\ 2002). {\tt StarTrack} incorporates the same single
star evolutionary formulas (Hurley et al. 2000) as the BSE code, however we 
extend the original formulas to 
{\em (i)} include wind mass loss rates from low- and intermediate-mass main
sequence stars (formation of pre-LMXBs, Belczynski \& Taam 2004b); 
{\em (ii)} account for the late evolution of low-mass helium stars
(important for formation of double neutron star systems, see Belczynski et al.
2007 and references therein); and 
{\em (iii)} calculate final masses of neutron stars and black holes, based on
recent hydrodynamical calculations (e.g., Belczynski et al. 2004b). 
For the treatment of tidal interactions we use the same equations (ODEs) as 
in Hurley's code, but we employ a fifth-order Runge-Kutta scheme with truncation 
error monitoring and adaptive step-size control integration of the ODEs instead 
of simple multiplication of the derivatives by the evolutionary timestep 
(Euler method). We also adopt convective tides that are more efficient 
(by a factor of 10-100) 
based on the observational calibration discussed in section 8.2. This will 
have a significant effect on the evolution of close binaries with low mass 
(convective) stars.
The calculation of X-ray luminosities for transient systems with NS and BH 
accretors is much more comprehensive in {\tt StarTrack}. We use both the recent
theoretical work and observations of low- and high-luminosity X-ray sources, 
to calibrate and test our approach (Belczynski \& Taam 2004b; Belczynski et al. 
2004a). The compact object masses formed in core collapse are calculated  
differently, and in particular we account for possibility of direct BH 
formation (no natal kick, no mass loss), with maximum BH masses formed reaching 
$\sim 10-20 \msun$ depending on metallicity and adopted wind mass loss rates 
(e.g., Belczynski et al. 2002; Belczynski et al. 2004b), a result that is 
consistent with maximum BH mass estimates in Galactic BH binaries (e.g., $\sim 
15 \msun$ for GRS 1915; $\sim 10-19 \msun$ for Cyg X-1; see Orosz 2003) 
In contrast, in the {\tt BSE} code all compact objects (including BH) 
have masses below $\sim 2.5 \msun$ for the entire spectrum of initial progenitor 
masses and different metallicities (see Fig.20 of Hurley et al.\ 2000), a
result that cannot be reconciled with the current estimates of BH masses.  
Also more recent
(Arzoumanian et al. 2002; Hobbs et al. 2005) natal kick distributions are 
used here as compared to Lyne \& Lorimer (1994) in {\tt BSE}. The above will
affect the post-SNa binary orbit, and subsequent evolution of massive
binaries. For example, the effect of natal kicks on population of double 
compact objects is rather dramatic and was quantified in Belczynski et al.\ 
(2002c). 
The treatment of nuclear mass transfer rate is different in the two codes. In 
{\tt BSE} it is calculated using a formula that keeps the donor star within 
its Roche lobe. The formula is calibrated to keep the mass transfer 
steady. In {\tt StarTrack}, we use the radius-mass exponents for the 
donor and its Roche lobe along with an estimate of the  
evolutionary donor radius change with time to calculate the 
mass transfer rate (see Sec. 5.1).  The calibration of the {\tt BSE} prescription 
is not discussed in detail by Hurley et al.\ (2002). 
Ruiter et al.\ (2006) find that for intermediate polars (low mass main sequence donors with WD 
accretors) the {\tt BSE} code results in mass transfer rates of about 2 orders 
magnitude  lower (as calculated by Liu \& Li 2006) than the rates predicted by 
{\tt StarTrack} and the observations of intermediate polars 
may indicate (Muno et al. 2006).
For low mass binaries, we use a different (less efficient) magnetic braking 
law in our standard model. As a result, binary orbits 
in our model will tend to take longer time to decay and initiate mass transfer, 
as compared to {\tt BSE} models. The {\tt StarTrack} prescription of mass accumulation onto white dwarfs is quite unique (see \S\,5.7). Related results of
calculations for accretion induced collapse and NS formation were presented 
by Belczynski \& Taam (2004a) and progenitor models of SN Ia  by Belczynski
et al. (2005b). On the other hand, the {\tt BSE} code is more fitted to work with dynamical codes, following in detail merger products (and their evolution) of 
various types of binary components. Also, the {\tt BSE} code is much faster 
than the {\tt StarTrack} code, and therefore may be used for simulations of 
larger stellar populations.

In a series of papers that will follow we will address the issues of modeling 
of XRBs, and will focus on the comparison of synthetic XRB populations 
with the observed X-ray point source populations in nearby galaxies. 
The code is also being used to study populations of binaries with NSs 
and BHs as potential source candidates for ground based interferometric 
gravitational radiation observatories (e.g., GEO, LIGO, VIRGO) as well 
as populations of less-massive WD binaries for space-based projects 
(e.g., LISA).   

Although a number of physical processes governing single and binary 
evolution remain highly uncertain, the advances in observational 
techniques and new results of massive surveys allow now various 
aspects of stellar evolution to be explored. We have incorporated several
different evolutionary models within {\tt StarTrack} (e.g., different  
magnetic braking laws or CE prescriptions) making possible tests of 
their validity. For example, one such test may be based on a comparison 
of synthetic and observed X-ray luminosity functions for nearby 
starburst galaxies. 

The {\tt StarTrack} code described in this paper may be used only for the
evolution of isolated stars and binaries, i.e., in stellar systems in  
which dynamical interactions are not important (e.g., field populations, 
open clusters). However, a number of interesting studies may be 
carried out for dense stellar environments, in which both stellar evolution
and dynamical interactions play an important role in the formation of compact
object binaries. In particular, {\tt StarTrack} was integrated with a dynamical 
code for these types of studies (for details see Ivanova et al.\ 2005).

\acknowledgements 
We would like to thank the anonymous referee for very detailed and
insightful reports that have helped us to improve our paper considerably. 
Also, we would like to thank R.Webbink, G.Nelemans, T. Di Salvo, J. 
Sepinsky and A.Ruiter for useful discussions and comments on the manuscript.
KB and TB acknowledge partial support through KBN Grant 1P03D02228 and
1P03D00530, and KB acknowledges support through NASA Chandra Theory Grant TM6-7006X. 
VK acknowledges support through a David and Lucile Packard Foundation  
Fellowship in Science and Engineering and through NASA grants  
NAG5-13056 and NAS8-03060. In addition, this research was supported in part 
by the NSF under Grant No. AST-0200876 to RT.

\pagebreak

\clearpage

\clearpage

\clearpage

\begin{figure}
\includegraphics[width=0.9\columnwidth]{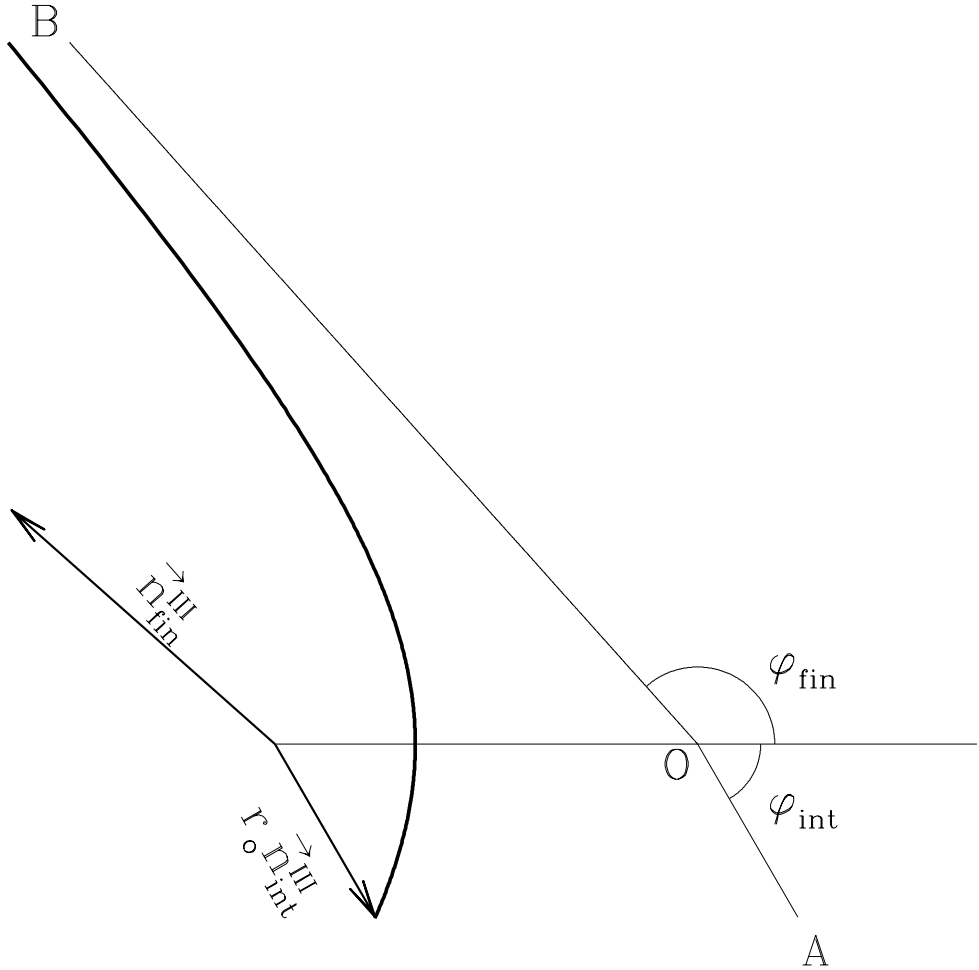}
\caption{ 
The case of a binary disrupted in a supernova explosion: we present the 
orbit in the coordinate system $III$ (for details see \S\,6). The line OA is 
parallel to the vector $\vec n^{III}_{\rm int}$, while the line OB, the 
asymptote of the hyperbola, is parallel to the vector $\vec n^{III}_{\rm fin}$. 
The point O is the focus of the hyperbola.
}
\label{fig02}
\end{figure}
\clearpage

\begin{figure}
\includegraphics[width=1.1\columnwidth]{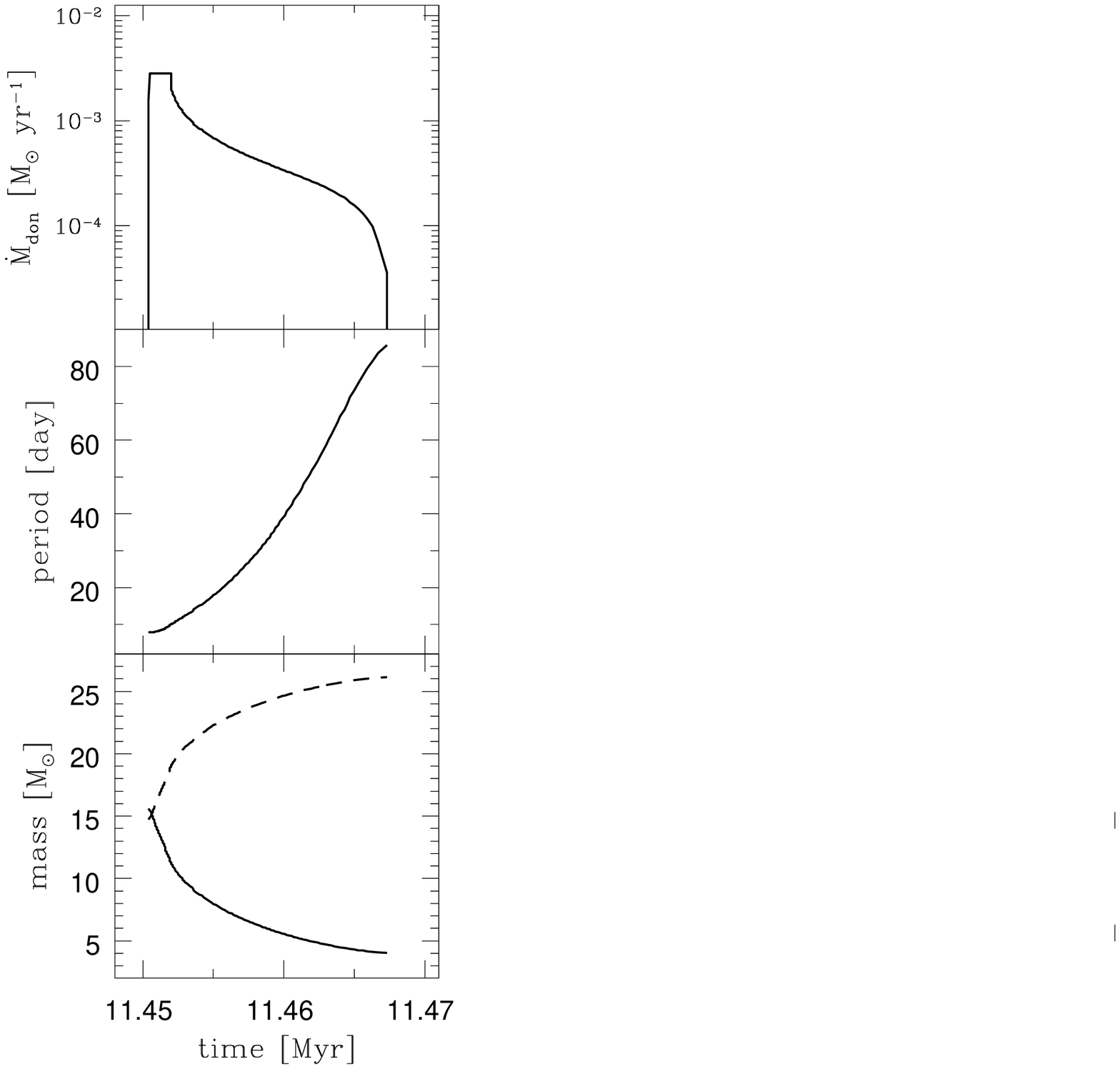}
\caption{RLOF sequence for $16 \msun$ HG + $15 \msun$ MS binary.
Top panel shows mass transfer rate, middle panel orbital period, while
bottom panel component mass evolution during the RLOF phase.  
}
\label{case1}
\end{figure}
\clearpage

\begin{figure}
\includegraphics[width=1.1\columnwidth]{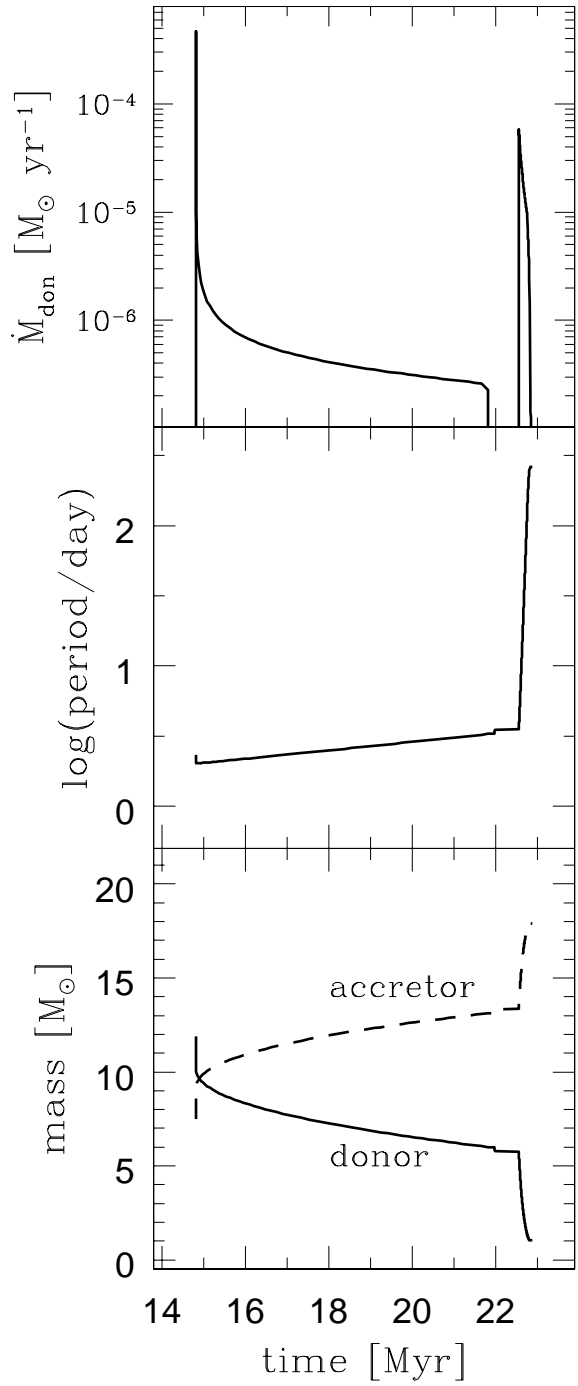}
\caption{RLOF sequence for $12 \msun$ MS + $7.5 \msun$ MS binary. 
Panels same as in Fig.~\ref{case1}.
}
\label{case2}
\end{figure}
\clearpage

\begin{figure}
\includegraphics[width=1.1\columnwidth]{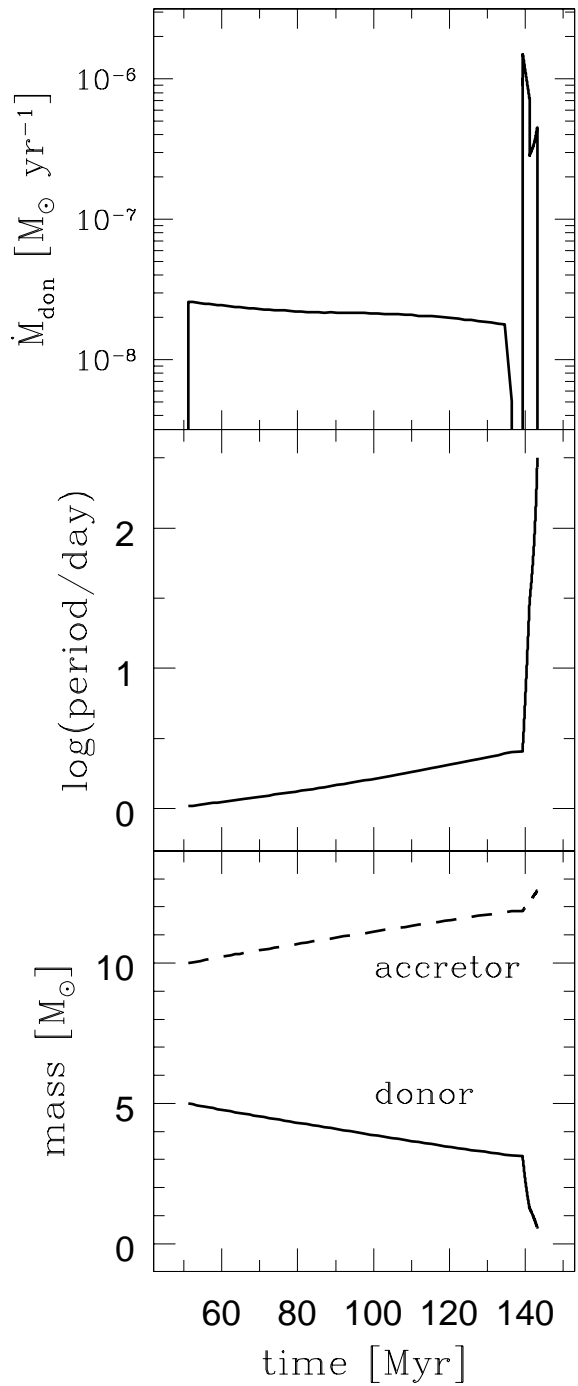}
\caption{RLOF sequence for $10 \msun$ BH + $5 \msun$ MS binary.
The critical Eddington mass accretion rate onto the BH is about $3.1-4 \times 10^{-7}
\mpy$. Panels same as in Fig.~\ref{case1}.
}
\label{case3}
\end{figure}
\clearpage

\begin{figure}
\includegraphics[width=1.1\columnwidth]{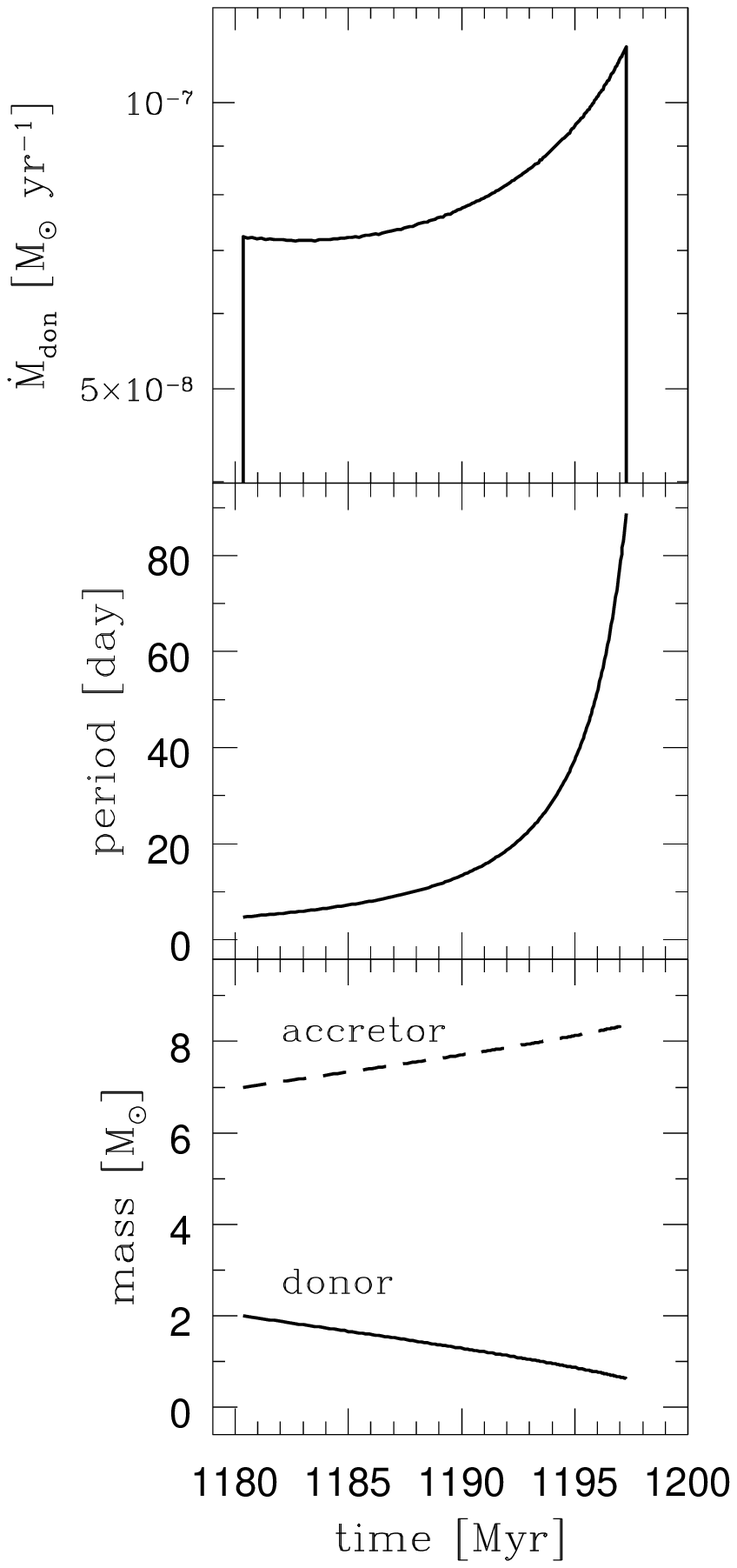}
\caption{RLOF sequence for $7 \msun$ BH + $2 \msun$ RG binary.
The critical Eddington mass accretion rate onto the BH is about $2.2-2.6 
\times 10^{-7} \mpy$. Panels same as in Fig.~\ref{case1}.
}
\label{case4}
\end{figure}
\clearpage

\begin{figure}
\includegraphics[width=1.1\columnwidth]{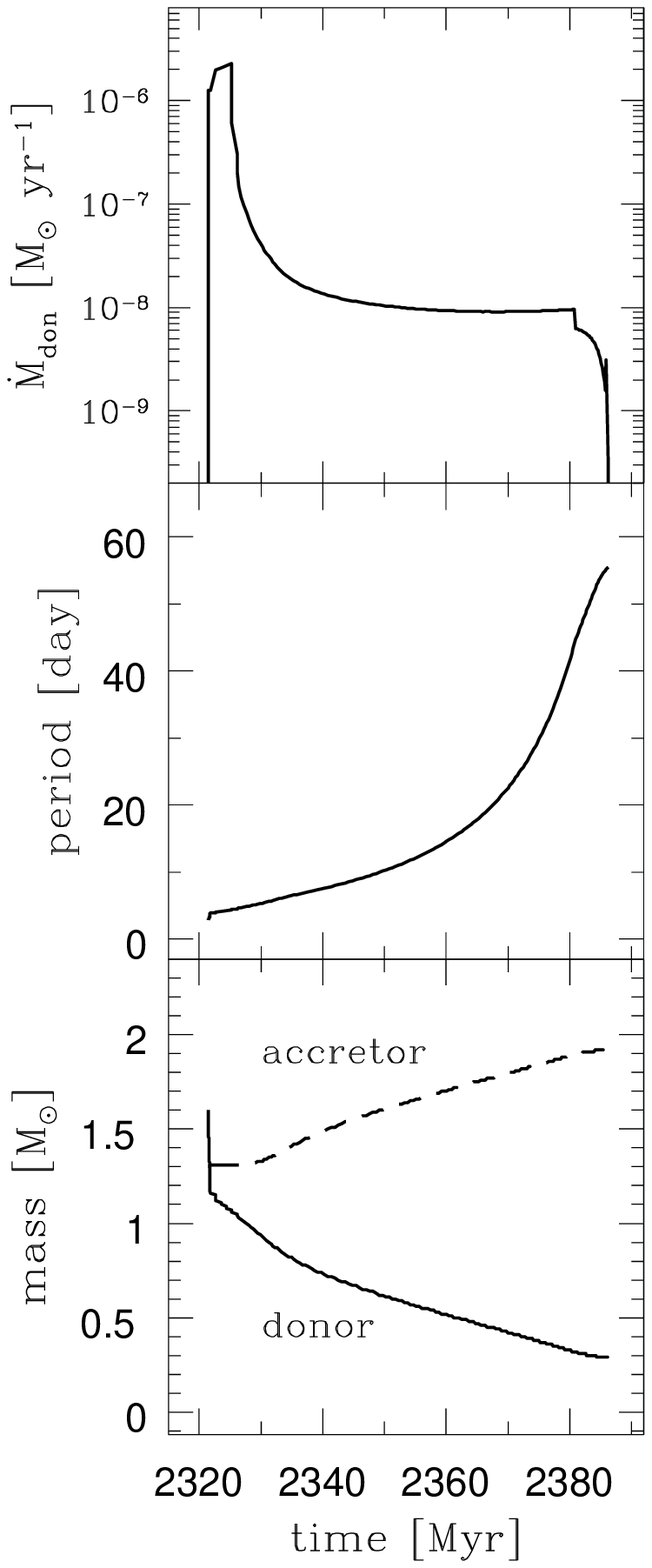}
\caption{RLOF sequence for $1.3 \msun$ NS + $1.6 \msun$ RG binary.
The critical Eddington mass accretion rate onto the NS is $\sim 1.7 
\times 10^{-8} \mpy$. Panels same as in Fig.~\ref{case1}.
}
\label{case5}
\end{figure}
\clearpage

\begin{figure}
\includegraphics[width=1.1\columnwidth]{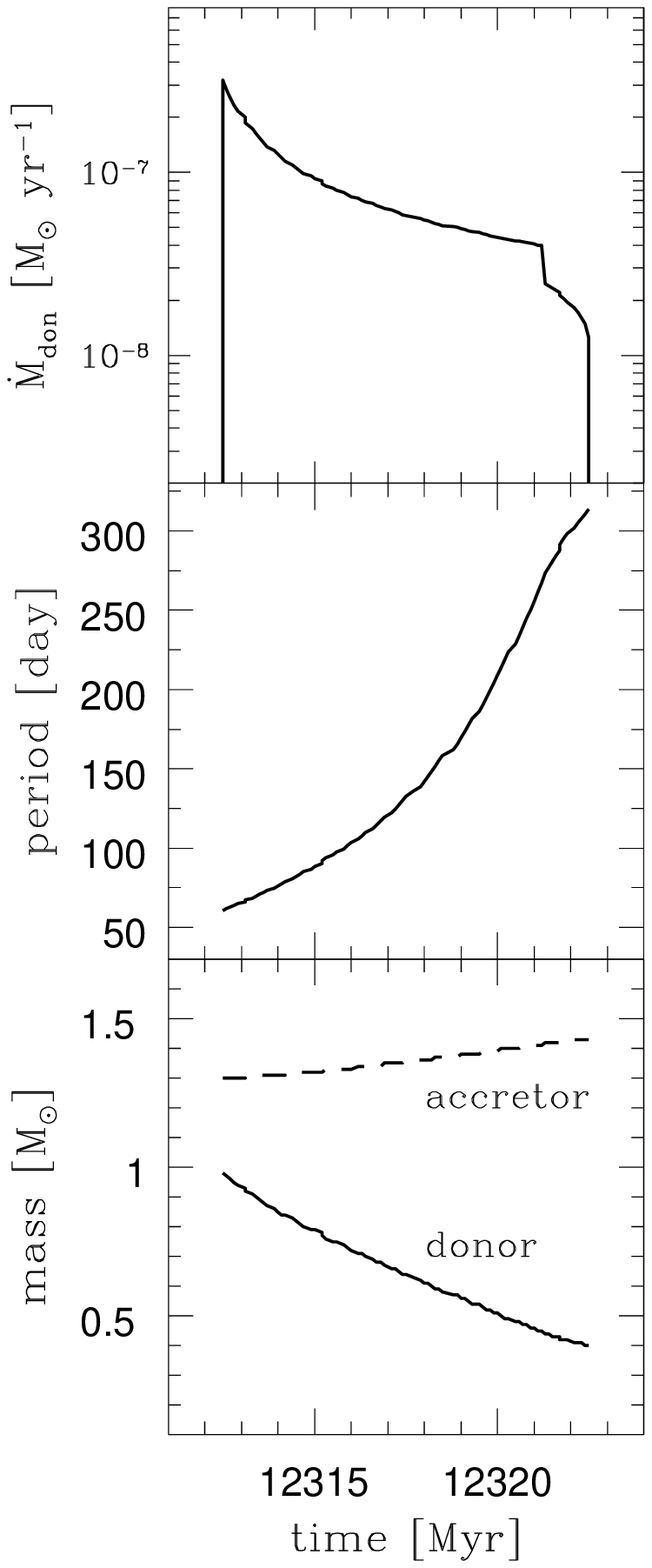}
\caption{RLOF sequence for $1.3 \msun$ NS + $1 \msun$ RG binary.
The critical Eddington mass accretion rate onto the NS is $\sim 1.7
\times 10^{-8} \mpy$. Panels same as in Fig.~\ref{case1}.
}
\label{case6}
\end{figure}
\clearpage

\begin{figure}
\includegraphics[width=1.1\columnwidth]{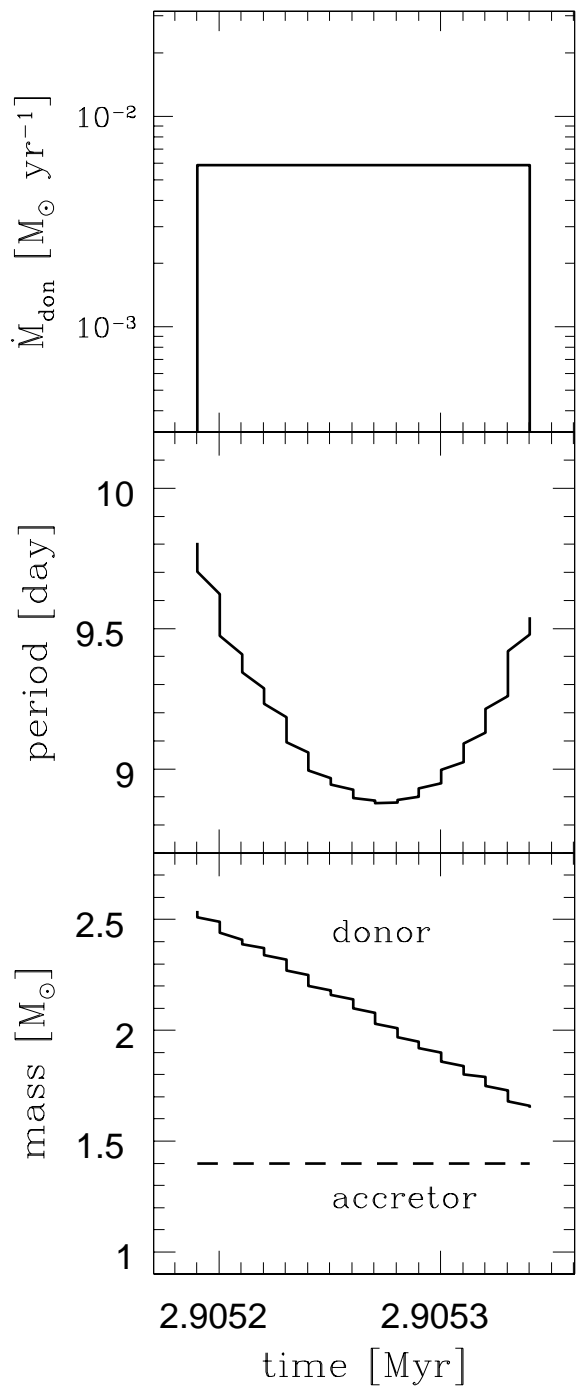}
\caption{RLOF sequence for $1.4 \msun$ NS + $2.8 \msun$ evolved 
He-star binary.
The critical Eddington mass accretion rate onto the NS is $\sim 2.9
\times 10^{-8} \mpy$. Panels same as in Fig.~\ref{case1}. Note the 
very short duration of this RLOF phase; the (finite) timesteps 
taken by the code may be seen through lines showing orbital period 
and donor mass. 
}
\label{case7}
\end{figure}
\clearpage

\begin{figure}
\includegraphics[width=1.1\columnwidth]{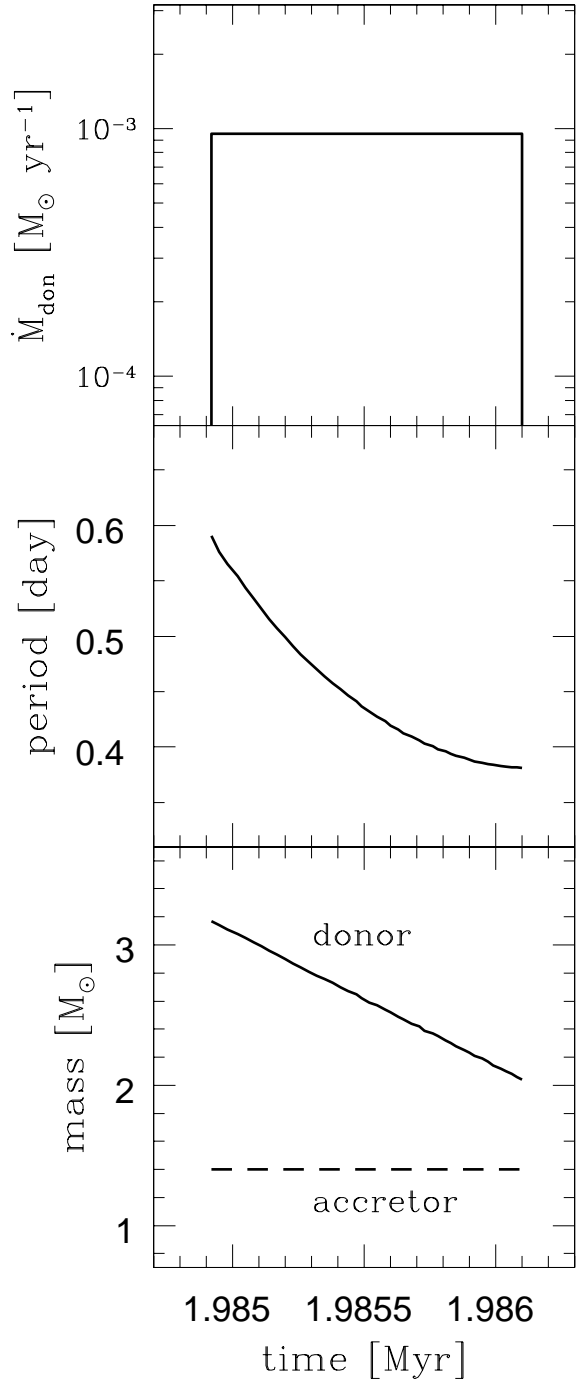}
\caption{RLOF sequence for $1.4 \msun$ NS + $3.6 \msun$  evolved
He-star binary.
The critical Eddington mass accretion rate onto the NS is $\sim 2.9
\times 10^{-8} \mpy$. Panels same as in Fig.~\ref{case1}.
}
\label{case8}
\end{figure}
\clearpage

\begin{figure}
\includegraphics[width=1.1\columnwidth]{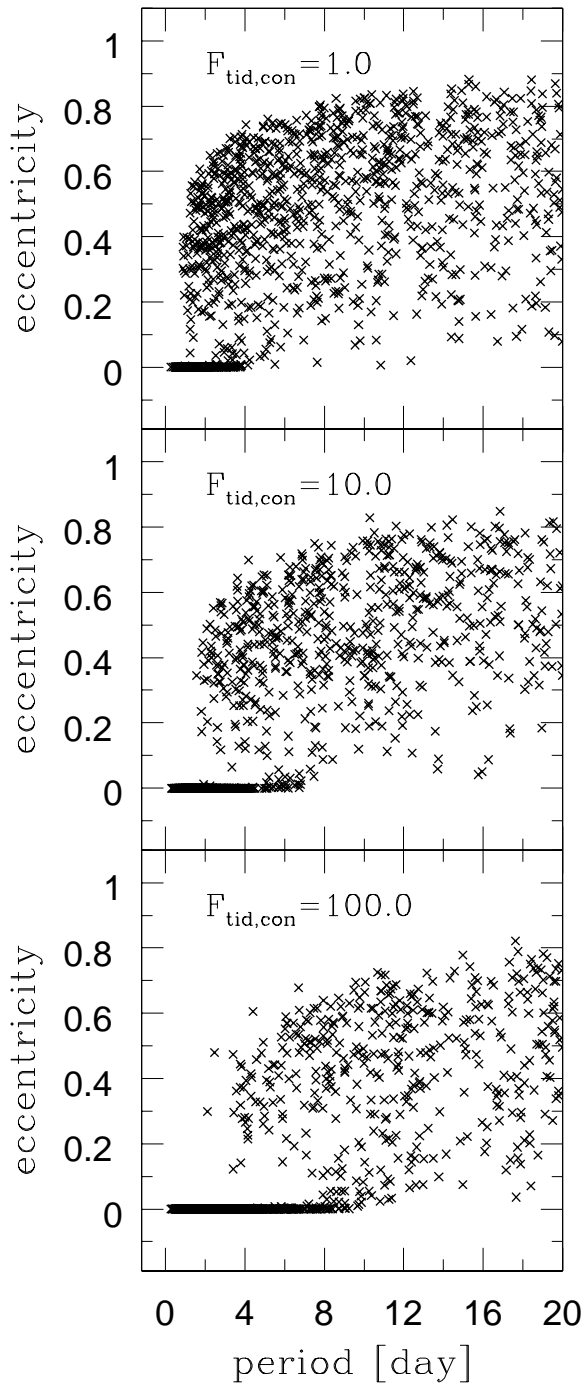}
\caption{Tidal calibration calculation for the open cluster M67. 
The figure shows the period--eccentricity plane with the population 
of main sequence binary stars at 3.98 Gyr, the current age 
of the cluster. Bottom and middle panels show the results of evolution 
with increased  tidal interactions ($F_{\rm tid,con}=100, 10$, 
respectively) as opposed to the standard prescription, 
presented on the top panel ($F_{\rm tid,con}=1$). 
Note the increase of cutoff period (the longest period circular 
binary in a given sample) with increasing $F_{\rm tid,con}$.
The observed cutoff period for M67 is $P_{\rm cut} \simeq 10-12$ days. 
For more details see \S\,8.2.1.
}
\label{m67}
\end{figure}
\clearpage

\begin{figure}
\includegraphics[width=1.1\columnwidth]{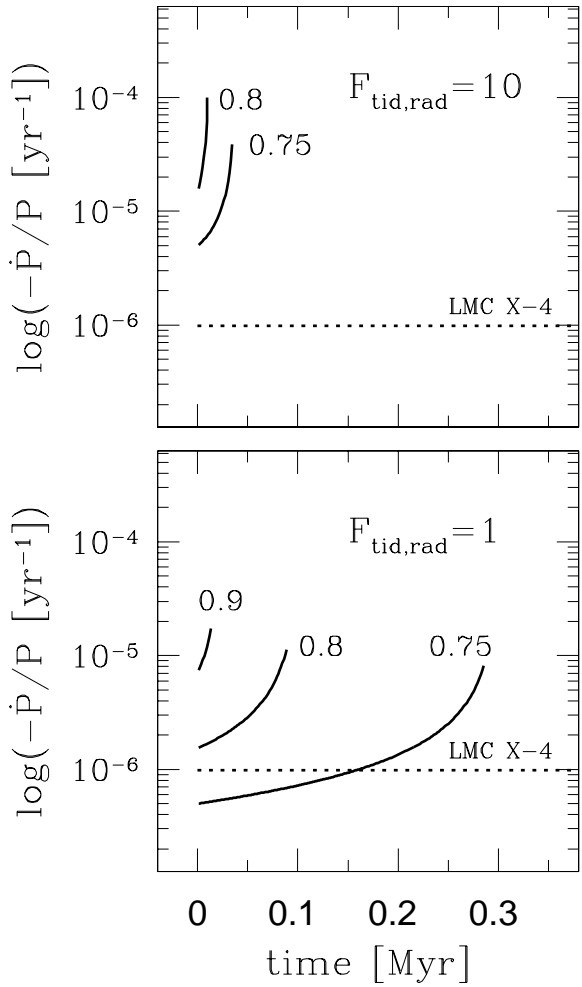}
\caption{Tidal calibration calculation for the high-mass X-ray 
binary LMC X-4. The observed orbital decay rate for LMC X-4 is 
$-9.8 \times 10^{-7} {\rm yr}^{-1}$ (marked with dotted line). 
Predicted decay rates for different radii of the main sequence 
secondary in respect to its Roche lobe ($R_2/R_{\rm 2,lob} = 
0.75,\ 0.8,\ 0.9$) are shown for $F_{\rm tid,rad}=1,\ 10$. 
Each track is displayed after the massive star reaches the 
relative radius corresponding to a given track and for all tracks 
the starting point is then set to time=0.
For more details see \S\,8.2.2.
}
\label{lmcx4}
\end{figure}
\clearpage

\end{document}